\documentclass[aps,prx,twocolumn,superscriptaddress,longbibliography]{revtex4-2}

\usepackage{mathrsfs}
\usepackage{tabularx}
\usepackage{slashed}
\usepackage{amsmath,amssymb,amsfonts,amsthm}
\usepackage{graphicx,color}
\usepackage[table]{colortbl}
\usepackage{times}
\usepackage{bbm}
\usepackage[caption=false]{subfig}
\usepackage{braket}
\usepackage{physics}
\usepackage{tikz-cd}
\usepackage{multirow}
\usepackage{comment}
\usepackage{array}
\usepackage{svg}
\newcolumntype{P}[1]{>{\centering\arraybackslash}p{#1}}

\graphicspath{{figs/}{}}

\newtheorem{theorem}{Theorem}

\newtheorem*{lemma*}{Lemma}

\theoremstyle{definition}
\newtheorem{definition}[theorem]{Definition}

\newtheorem*{claim*}{Claim}

\newcommand{\PT}{\ensuremath{\mathcal{PT}}}
\newcommand{\R}{\ensuremath{\mathbb{R}}}

\newcommand{\Z}{\ensuremath{\mathbb{Z}}}
\newcommand{\bk}{\ensuremath{\mathbf{k}}}
\newcommand{\B}{\ensuremath{\mathbb{B}}}

\newcommand{\specificthanks}[1]{\@fnsymbol{#1}}% Inserts a specific \thanks symbol

\usepackage[colorlinks,linkcolor=red,citecolor=blue]{hyperref}

\begin{document}

\title{Homotopy, Symmetry, and Non-Hermitian Band Topology}
\author{Kang Yang}
\thanks{Contributed equally. \href{mailto:kang.yang@fu-berlin.de}{kang.yang@fu-berlin.de}, \href{mailto:zli@perimeterinstitute.ca}{zli@perimeterinstitute.ca}}

\affiliation{Dahlem Center for Complex Quantum Systems and Fachbereich Physik, Freie Universit\"at Berlin, 14195 Berlin, Germany}
\author{Zhi Li}
\thanks{Contributed equally. \href{mailto:kang.yang@fu-berlin.de}{kang.yang@fu-berlin.de}, \href{mailto:zli@perimeterinstitute.ca}{zli@perimeterinstitute.ca}}
\affiliation{Perimeter Institute for Theoretical Physics, Waterloo, Ontario N2L 2Y5, Canada}
\author{J. Lukas K. K\"onig}
\affiliation{Department of Physics, Stockholm University, AlbaNova University Center, 106 91 Stockholm, Sweden}
\author{Lukas R\o dland}
\affiliation{Department of Physics, Stockholm University, AlbaNova University Center, 106 91 Stockholm, Sweden}
\author{Marcus St\aa lhammar}
\affiliation{Nordita, KTH Royal Institute of Technology and Stockholm University, Hannes Alfv\'ens v\"ag 12, SE-106 91 Stockholm, Sweden}
\author{Emil J. Bergholtz}
\email{emil.bergholtz@fysik.su.se}
\affiliation{Department of Physics, Stockholm University, AlbaNova University Center, 106 91 Stockholm, Sweden}
\date{\today}

\begin{abstract}

Non-Hermitian matrices are ubiquitous in the description of nature ranging from classical dissipative systems, including optical, electrical, and mechanical metamaterials, to scattering of waves and open quantum many-body systems. Seminal line-gap and point-gap classifications of non-Hermitian systems using K-theory have deepened the understanding of many physical phenomena. However, ample systems remain beyond this description; reference points and lines do not in general distinguish whether multiple non-Hermitian bands exhibit intriguing exceptional points, spectral braids and crossings. To address this we consider two different notions: non-Hermitian band gaps and separation gaps that crucially encompass a broad class of multi-band scenarios, enabling the description of generic band structures with symmetries. With these concepts, we provide a unified and comprehensive classification of both gapped and nodal systems in the presence of physically relevant parity-time ($\mathcal{PT}$) and pseudo-Hermitian symmetries using homotopy theory. This uncovers new stable topology stemming from both eigenvalues and wave functions, and remarkably also implies distinct fragile topological phases. In particular, we reveal different Abelian and non-Abelian phases in $\mathcal{PT}$-symmetric systems, described by frame and braid topology. The corresponding invariants are robust to symmetry-preserving perturbations that do not induce (exceptional) degeneracy, and they also predict the deformation rules of nodal phases. We further demonstrate that spontaneous $\mathcal{PT}$ symmetry breaking is captured by Chern-Euler and Chern-Stiefel-Whitney descriptions, a fingerprint of unprecedented non-Hermitian topology previously overlooked. These results open the door for theoretical and experimental exploration of a rich variety of novel topological phenomena in a wide range of physical platforms.

\end{abstract}

\maketitle

\makeatletter
\def\l@subsubsection#1#2{} %remove subsubsections in the table of contents
\makeatother
{
\hypersetup{linkcolor=darkgray}
  \tableofcontents
}

\section{Introduction}
Topology provides a powerful framework for describing robust and universal physical effects emerging in vastly different systems \cite{RevModPhys.89.040502,RevModPhys.89.041004}. 
A particularly important paradigm during the twenty-first century has been the enrichment of topological phases in presence of symmetries leading to topological insulators, superconductors and semimetals \cite{HasanKane,QiZhang,KaneMele2,BHZ,QiZhang,RevModPhys.88.035005,RevModPhys.90.015001,crystalline,beyond}. 
Already, this has revolutionized the understanding of quantum phases of materials, even though the theoretical description is entirely based on the idealized situation of (free fermions in) closed quantum systems. 

The current frontier of non-Hermitian topological phases \cite{RevModPhys.93.015005}, however, goes beyond this closed system perspective. Here, a rapid development in theory has been accompanied by an abundance of experimental realizations in classical metamaterials. These provide versatile platforms of diverse physical provenance for topological phenomena 
\cite{Raghu2008,Haldane2008,Huber2016,Albert2015,Lee2018,RevModPhys.91.015006,Lu_2014} in which non-Hermiticity is central: mechanical systems exhibit friction, electrical circuits have resistance and losses, optical systems feature gain and loss, and so on \cite{RevModPhys.93.015005,ashida2020non}. 
Although dissipation is arguably the most obvious source of non-Hermiticity, and correspondingly non-unitary dynamics, the range of applications of non-Hermitian topology in physics and beyond is much broader \cite{ashida2020non}. 
It includes quantum systems \cite{Rotter_2009,PhysRevX.8.041031,ep-optics}, for instance open Lindbladian systems where the non-Hermitian Liouvillian superoperator 
governs the dynamics   \cite{lindblad1976generators,dum1992monte,Prosen_2008,Daley2014,PhysRevLett.124.040401,PhysRevX.11.021037,PhysRevX.13.031019,PRXQuantum.4.030328,fei2019, ueda2021, emil2022l,kohei2023}, as well as the effective quasi-particle description of materials including effects of interactions, leads, phonons and disorder \cite{kozii2017,yoshidapeterskawakmi,PhysRevLett.125.227204,PhysRevResearch.4.L042025,PhysRevResearch.1.012003,PhysRevB.104.L121109}.

Symmetry plays a pivotal role also in non-Hermitian systems. 
Of special interest are non-Hermitian operators subject to parity-time (\PT) symmetry which were initially suggested as a fundamental amendment of Hermitian operators in quantum mechanics (of closed systems) due to the existence of extended parameter regions of real spectra \cite{Bender_2007,PhysRevLett.80.5243,Brody_2016}. 
More recently much interest has concerned applications whereby \PT symmetry implies a perfect balance between gain and loss in optics \cite{RevModPhys.91.015006,Lu_2014,ozdemir2019parity,el2018non}, and can be realized at remarkable precision in photonic crystals \cite{PhysRevLett.103.093902,ruter2010observation,PhysRevLett.106.213901,regensburger2012parity}, waveguides \cite{PhysRevLett.101.080402}, acoustics \cite{PhysRevX.4.031042}, optomechanics \cite{xu2016topological}, trapped ion systems \cite{PhysRevLett.126.083604} and microcavities \cite{peng2014parity,PhysRevLett.117.110802,poli2015selective}. 

The emergence of non-Hermitian topological phenomena has sparked the need to describe these within a unified and rigorous mathematical framework. A salient example is the K-theory classification of topological phases \cite{PhysRevX.9.041015,PhysRevB.99.235112} based on reference energies, which has provided several useful insights of physical importance, such as the 38-fold classification of non-Hermitian symmetry classes \cite{Bernard2002,PhysRevX.8.031079,PhysRevX.9.041015,PhysRevB.99.235112}, and illuminated the topological origin of the non-Hermitian skin effect \cite{PhysRevLett.77.570,PhysRevX.8.031079,lin2023,okuma2023}. 
While both elegant and useful, an apparent drawback of this classification scheme is that a lot of relevant band structures in physics cannot be detected by a reference point or a line. The reference classification is unable to detect either crossings of non-Hermitian bands (exceptional degeneracy), or non-Abelian topology sourced by their non-trivial eigenmode braiding \cite{PhysRevE.87.050101,wang2021topological,patil2022measuring,PhysRevLett.130.017201,PhysRevResearch.5.L022050}, since neither of these relate to the existence of a reference point or line. The insufficiency of the reference description is particularly severe when the spectrum is restricted by symmetry. In this situation only special reference lines are compatible with the symmetry and abundant systems elude the reference characterization.
From its Hermitian counterpart, a reference energy is most pertinent to many-body fermionic systems, whereas in experiments a very large portion are bosonic systems or a single-mode effective description.
Moreover, the gap width, and the corresponding energy cost of excitations, are missing in this framework, while for many experiments, the gap width serves as a prime diagnosis of phases.
Another question beyond the reference approach is how conventional Hermitian topology is modified by non-Hermiticity, especially the stability of Hermitian topology. For example, the sign of the topological Chern number can be changed by encircling an exceptional line or traversing through the Brillouin zone \cite{PhysRevResearch.2.023226,PhysRevB.103.155129,PhysRevB.101.205417}. This thus is highly relevant to the manipulation of conventional topological states through loss and gain. 

In the subsequent text, we develop a unified framework to classify non-Hermitian gapped and nodal systems subject to either \PT{} symmetry or pseudo-Hermitian \cite{MostafazadehPSH,doi:10.1142/S0219887810004816} symmetries by applying homotopy theory. 
To facilitate this we use two distinct concepts of gaps; the well-established band gaps, as well as a newly-formalized concept of separation gaps.
Band gaps characterize the energy cost of a single-mode excitation, which is relevant to bosonic systems and effective descriptions of dynamical processes. 
They provide a unified description of both single-body gapped phases and nodal phases. 
Separation gaps focus on excitations crossing bands of different static or dynamic properties, pertinent to many-body scenarios. 
They generalize the concepts of line gaps and many-body spectral gaps to situations where no natural reference energy exists. 
Notably, we do not restrict the number of gaps in these considerations and are therefore capable of answering questions that cannot be reduced to a single gap. 
In particular, this is the pertinent situation when \PT{} or pseudo-Hermitian symmetry is present, where the spectra on the upper and lower half complex planes are correlated. 
If the spectrum is restricted to the real axis, these geometry-independent gap concepts naturally reduce to those familiar from conventional Hermitian topological physics. 
As a result, the new gap formulations are able to shed light on how novel non-Hermitian topology arises in addition to existing Hermitian topological phases.

While we show that the implications of homotopy beyond the K-theory description have wider and arguably even more profound implications in the non-Hermitian realm, it is also known to be of relevance in the Hermitian limit. Notably, despite providing the 10-fold way classification of topological matter \cite{kitaevperio,ryu2010topological,RevModPhys.88.035005}, K-theory fails to capture certain physical phenomena of band insulators, such as Hopf insulators and Euler phases \cite{PhysRevLett.101.186805,PhysRevX.9.021013,PhysRevLett.125.053601,bouhon2022multi}. 
These phases open the door towards fragile and delicate topology \cite{PhysRevLett.121.126402,song2020twisted,PhysRevX.10.031001,brouwer2023homotopic,PhysRevB.100.195135}. 
Multiple-gapped phases constitute another example that is beyond K-theory. 
There, non-Abelian eigenvector topology attracts increasing interest and has been realized in recent experiments \cite{wu2019non,bouhon2020non,guo2021experimental,PhysRevX.13.021024,yang2023non}. 
Additional examples outside the scope of the tenfold way are gapless systems such as graphene, and Weyl semimetals, which host protected point-like band intersections \cite{Goerbig2017,RevModPhys.90.015001,annurev-conmatphys-031016-025458}. 
The most intriguing aspects of these gapless phases are the nodal structures concerning only a few bands and their crossings, sharply different from the single-gap many-band limit where K-theory is applicable. 
All of these systems require a general homotopy-theory description.

Due to the relaxation of the Hermiticity constraint, gapless nodal structures are much more abundant, as fewer parameters need to be tuned to achieve band intersections in non-Hermitian systems \cite{RevModPhys.93.015005,Heiss_2012,berry2004}.
Generically, non-Hermitian band crossings are comprised of exceptional points (EPs), at which eigenvalues and eigenvectors simultaneously coalesce \cite{kato2013perturbation}. 
These points appear generically already in two-dimensional systems \cite{Zhou2018,berry2004,Heiss_2012,kozii2017}, while they form potentially linked and knotted lines in three dimensions \cite{PhysRevLett.118.045701,PhysRevA.98.042114,Cerjan_2019,PhysRevB.99.161115,Stalhammar_2019,tidalsurfacestates,PhysRevLett.124.186402}.
The number of tuning parameters needed to render exceptional points generic can be further decreased by introducing certain discrete symmetries; important examples include the aforementioned pseudo-Hermitian and \PT{} symmetries \cite{PhysRevB.99.041406,PhysRevB.99.121101,PhysRevB.99.041202,Zhou2019,PhysRevA.84.021806,PhysRevB.100.115124,PhysRevB.104.L121109,PhysRevLett.123.066405}.
In such systems, symmetry-protected EPs appear stably already in one-dimensional systems, forming lines in two dimensions and surfaces of arbitrary genus in three dimensions \cite{PhysRevB.104.L201104,sayyad2022symmetryprotected}. 
The symmetry also stabilizes higher-order EPs, signifying the simultaneous coalescence of more than two eigenvalues and eigenvectors \cite{PhysRevLett.127.186601,PhysRevLett.127.186602,PhysRevResearch.4.023130,PhysRevB.104.L201104}, which have been experimentally realized in nonreciprocal circuits \cite{hu2023non} and single-photon setups \cite{doi:10.1126/sciadv.adi0732}. 
In the absence of symmetry, the topology protecting EPs comes from the complex eigenvalues around the EPs.
It is of the same origin as gradually rotating band gaps along a loop \cite{PhysRevLett.120.146402}. 
These types of topological phenomena are far beyond the scope of K-theory and reference points/lines. 
They are instead described by the mathematical structure of braid groups, which are captured within the framework of homotopy theory for non-degenerate matrices \cite{PhysRevB.103.155129,wang2021topological,PhysRevB.101.205417,PhysRevResearch.4.L022064,PhysRevB.106.L161401,zhong2023eigenenergy,PhysRevLett.126.010401,PhysRevLett.130.017201,PhysRevLett.130.157201}. 
This non-Abelian braid topology also permits exceptions to conventional doubling principles \cite{konig2022braid}.

A natural question to ask is how such new topology arising from eigenvalues will interact with the existing topology in Hermitian systems. This is even more relevant for \PT-symmetric systems, where winding numbers, non-Abelian frame topology, as well as stable $\Z_2$ topology \cite{PhysRevX.9.021013,wu2019non,bouhon2020non,PhysRevX.13.021024,yang2023non} already enter the stage for Hermitian systems in one dimension. All such Hermitian topology originates from the eigenvectors. Despite many recent advances including, e.g., the Zak phase \cite{PhysRevLett.126.215302}, the description of the interplay between this established eigenvector topology and the eigenvalue topology induced by non-Hermiticity is incomplete. A systematic way to answer this question will be provided by a general homotopy description of excitation gaps.

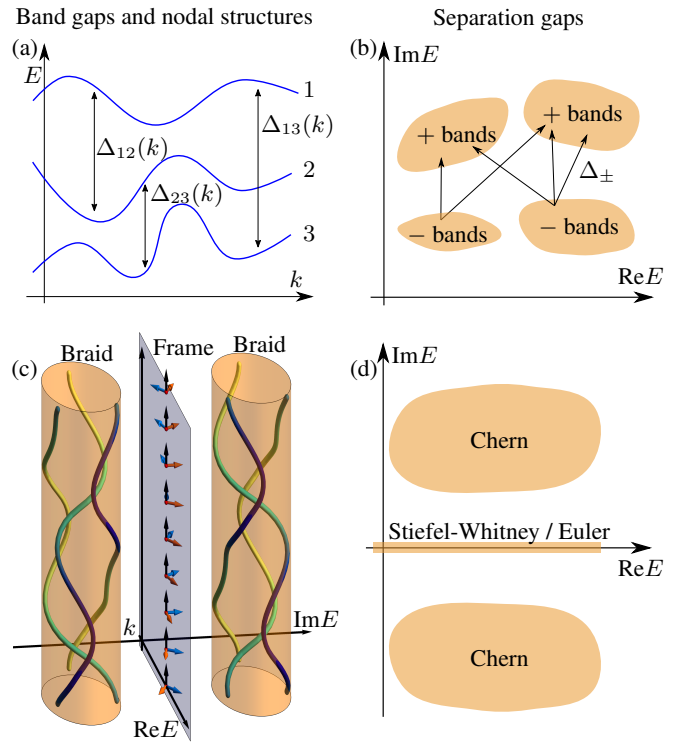
\begin{figure}[htbp!]
    \centering
    \def\svgwidth{\columnwidth}
    %--

\begingroup%
  \makeatletter%
  \providecommand\color[2][]{%
    \errmessage{(Inkscape) Color is used for the text in Inkscape, but the package 'color.sty' is not loaded}%
    \renewcommand\color[2][]{}%
  }%
  \providecommand\transparent[1]{%
    \errmessage{(Inkscape) Transparency is used (non-zero) for the text in Inkscape, but the package 'transparent.sty' is not loaded}%
    \renewcommand\transparent[1]{}%
  }%
  \providecommand\rotatebox[2]{#2}%
  \newcommand*\fsize{\dimexpr\f@size pt\relax}%
  \newcommand*\lineheight[1]{\fontsize{\fsize}{#1\fsize}\selectfont}%
  \ifx\svgwidth\undefined%
    \setlength{\unitlength}{264bp}%
    \ifx\svgscale\undefined%
      \relax%
    \else%
      \setlength{\unitlength}{\unitlength * \real{\svgscale}}%
    \fi%
  \else%
    \setlength{\unitlength}{\svgwidth}%
  \fi%
  \global\let\svgwidth\undefined%
  \global\let\svgscale\undefined%
  \makeatother%
  \begin{picture}(1,1.13636364)%
    \lineheight{1}%
    \setlength\tabcolsep{0pt}%
    \put(0,0){\includegraphics[width=\unitlength,page=1]{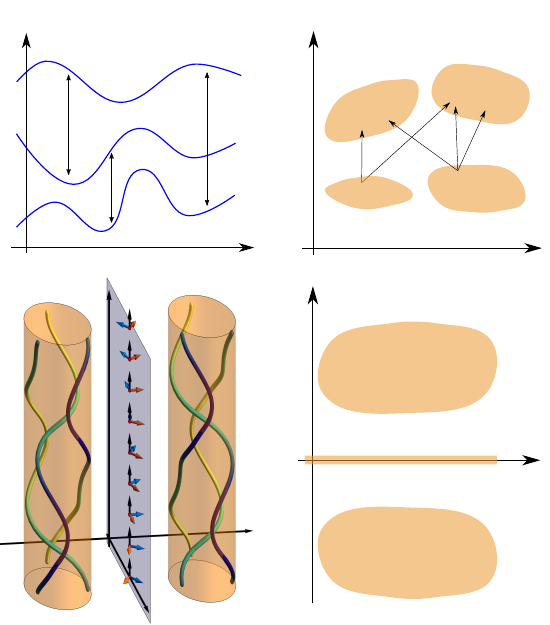}}%
    \put(0.003885,1.10626037){\color[rgb]{0,0,0}\makebox(0,0)[lt]{\lineheight{1.25}\smash{\begin{tabular}[t]{l}Band gaps and nodal structures\end{tabular}}}}%
    \put(0.7617221,1.10419667){\color[rgb]{0,0,0}\makebox(0,0)[t]{\lineheight{1.25}\smash{\begin{tabular}[t]{c}Separation gaps\end{tabular}}}}%
    \put(0.43524845,0.70030296){\color[rgb]{0,0,0}\makebox(0,0)[t]{\lineheight{1.25}\smash{\begin{tabular}[t]{c}\(k\)\end{tabular}}}}%
    \put(0.45522753,0.99306823){\color[rgb]{0,0,0}\makebox(0,0)[t]{\lineheight{1.25}\smash{\begin{tabular}[t]{c}\(1\)\end{tabular}}}}%
    \put(0.45522753,0.87090922){\color[rgb]{0,0,0}\makebox(0,0)[t]{\lineheight{1.25}\smash{\begin{tabular}[t]{c}\(2\)\end{tabular}}}}%
    \put(0.45522753,0.77147751){\color[rgb]{0,0,0}\makebox(0,0)[t]{\lineheight{1.25}\smash{\begin{tabular}[t]{c}\(3\)\end{tabular}}}}%
    \put(0.04504376,1.01452397){\color[rgb]{0,0,0}\makebox(0,0)[rt]{\lineheight{1.25}\smash{\begin{tabular}[t]{r}\(E\)\end{tabular}}}}%
    \put(0.12683108,0.90395062){\color[rgb]{0,0,0}\makebox(0,0)[lt]{\lineheight{1.25}\smash{\begin{tabular}[t]{l}\(\Delta_{12}(k)\)\end{tabular}}}}%
    \put(0.16853313,0.16507082){\color[rgb]{0,0,0}\makebox(0,0)[lt]{\lineheight{1.25}\smash{\begin{tabular}[t]{l}\(k\)\end{tabular}}}}%
    \put(0.21040654,0.83376434){\color[rgb]{0,0,0}\makebox(0,0)[lt]{\lineheight{1.25}\smash{\begin{tabular}[t]{l}\(\Delta_{23}(k)\)\end{tabular}}}}%
    \put(0.38105474,0.94026779){\color[rgb]{0,0,0}\makebox(0,0)[lt]{\lineheight{1.25}\smash{\begin{tabular}[t]{l}\(\Delta_{13}(k)\)\end{tabular}}}}%
    \put(0.58491064,1.04691722){\color[rgb]{0,0,0}\makebox(0,0)[lt]{\lineheight{1.25}\smash{\begin{tabular}[t]{l}Im\(E\)\end{tabular}}}}%
    \put(0.74799247,0.45305029){\color[rgb]{0,0,0}\makebox(0,0)[t]{\lineheight{1.25}\smash{\begin{tabular}[t]{c}Chern\end{tabular}}}}%
    \put(0.74799248,0.11833344){\color[rgb]{0,0,0}\makebox(0,0)[t]{\lineheight{1.25}\smash{\begin{tabular}[t]{c}Chern\end{tabular}}}}%
    \put(0.57701102,0.31301008){\color[rgb]{0,0,0}\makebox(0,0)[lt]{\lineheight{1.25}\smash{\begin{tabular}[t]{l}Stiefel-Whitney / Euler\end{tabular}}}}%
    \put(0.96519872,0.25616642){\color[rgb]{0,0,0}\makebox(0,0)[t]{\lineheight{1.25}\smash{\begin{tabular}[t]{c}Re\(E\)\end{tabular}}}}%
    \put(0.25513155,0.00879202){\color[rgb]{0,0,0}\makebox(0,0)[rt]{\lineheight{1.25}\smash{\begin{tabular}[t]{r}Re\(E\)\end{tabular}}}}%
    \put(0.43082581,0.17736392){\color[rgb]{0,0,0}\makebox(0,0)[lt]{\lineheight{1.25}\smash{\begin{tabular}[t]{l}Im\(E\)\end{tabular}}}}%
    \put(0.07257642,0.58780467){\color[rgb]{0,0,0}\makebox(0,0)[lt]{\lineheight{1.25}\smash{\begin{tabular}[t]{l}Braid\end{tabular}}}}%
    \put(0.33843881,0.6011796){\color[rgb]{0,0,0}\makebox(0,0)[lt]{\lineheight{1.25}\smash{\begin{tabular}[t]{l}Braid\end{tabular}}}}%
    \put(0.21509362,0.59589615){\color[rgb]{0,0,0}\makebox(0,0)[lt]{\lineheight{1.25}\smash{\begin{tabular}[t]{l}Frame\end{tabular}}}}%
    \put(0.58237213,0.58463637){\color[rgb]{0,0,0}\makebox(0,0)[lt]{\lineheight{1.25}\smash{\begin{tabular}[t]{l}Im\(E\)\end{tabular}}}}%
    \put(0.96519887,0.70613658){\color[rgb]{0,0,0}\makebox(0,0)[t]{\lineheight{1.25}\smash{\begin{tabular}[t]{c}Re\(E\)\end{tabular}}}}%
    \put(0.68176648,0.92457587){\color[rgb]{0,0,0}\makebox(0,0)[t]{\lineheight{1.25}\smash{\begin{tabular}[t]{c}\(+\) bands\end{tabular}}}}%
    \put(0.87554895,0.95613666){\color[rgb]{0,0,0}\makebox(0,0)[t]{\lineheight{1.25}\smash{\begin{tabular}[t]{c}\(+\) bands\end{tabular}}}}%
    \put(0.66801489,0.77055363){\color[rgb]{0,0,0}\makebox(0,0)[t]{\lineheight{1.25}\smash{\begin{tabular}[t]{c}\(-\) bands\end{tabular}}}}%
    \put(0.86992111,0.78184519){\color[rgb]{0,0,0}\makebox(0,0)[t]{\lineheight{1.25}\smash{\begin{tabular}[t]{c}\(-\) bands\end{tabular}}}}%
    \put(0.86974131,0.86815287){\color[rgb]{0,0,0}\makebox(0,0)[lt]{\lineheight{1.25}\smash{\begin{tabular}[t]{l}\(\Delta_ {\pm}\)\end{tabular}}}}%
    \put(0.56001848,1.05754711){\color[rgb]{0,0,0}\makebox(0,0)[rt]{\lineheight{1.25}\smash{\begin{tabular}[t]{r}(b)\end{tabular}}}}%
    \put(0.03842761,1.05754711){\color[rgb]{0,0,0}\makebox(0,0)[rt]{\lineheight{1.25}\smash{\begin{tabular}[t]{r}(a)\end{tabular}}}}%
    \put(0.03780263,0.56409097){\color[rgb]{0,0,0}\makebox(0,0)[rt]{\lineheight{1.25}\smash{\begin{tabular}[t]{r}(c)\end{tabular}}}}%
    \put(0.56052985,0.56409097){\color[rgb]{0,0,0}\makebox(0,0)[rt]{\lineheight{1.25}\smash{\begin{tabular}[t]{r}(d)\end{tabular}}}}%
  \end{picture}%
\endgroup%

    %--
    \caption{
    Gap structures and their accompanying topological charges.
    (a) Band gaps indicate that there is a non-zero energy cost for single-particle excitations preserving the momentum (or other parameters), \(\Delta k=0\).
    (b) Separation gaps consider excitations across distinct types of bands, labeled here by $\pm$. These types are generalizations of occupied and empty bands. In Hermitian fermionic systems, we usually separate occupied bands ($-$) from empty bands ($+$). In non-Hermitian systems, $\pm$ can be bands carrying different dynamic properties, e.g., the steady states and the evanescent states for the Liouvillian gap.
    (c) Band-gapped systems are described by braid invariants, the braids of eigenvalues, for \PT-breaking bands, and frame invariants, the frames spanned by eigenvectors, for \PT-preserving bands.
    (d) The separation gap between real and non-real eigenvalues is described by the Stiefel-Whitney class / Euler number for the \PT-preserving bands and the Chern number for \PT-breaking bands. 
    } 
    \label{fig_nphs}
\end{figure}

Motivated by the high physical significance of \PT-symmetric and pseudo-Hermitian systems, the topological homotopy classification scheme developed in this work provides an important missing piece in recent physics research. 
Homotopy theory provides a simple topological explanation for the abundance of nodal structures such as EPs in non-Hermitian systems with symmetries.  
We use this classification to explain the previously unknown interplay between eigenvalue and eigenvector in \PT-symmetric systems. 
This interplay leads us to posit a very different point of view on stable and fragile topological phenomena.
These are comprised of novel Abelian and non-Abelian phases, for which we further derive explicit stable topological invariants related to frame and braid topology, as well as winding numbers, which we explicitly exemplify in illustrative two-band models. 
Moreover, we show that the spontaneous breaking of \PT{} symmetry, which appears as a separation between states with different dynamic properties, can be described using a combined notion of Chern and Euler numbers. The results are briefly summarized in Fig.~\ref{fig_nphs}. All of these novel topologies echo the unforeseen exceptional structures of non-Hermitian matrices \cite{PhysRevA.102.032216,PhysRevB.107.144304}.
Importantly, we identify some novel mathematical spaces beyond the symmetric spaces \cite{kitaevperio} and flag manifolds \cite{wu2019non,PhysRevB.102.115135} that often arise in describing Hermitian systems \cite{PhysRevLett.51.51}.
They bring us to a much broader realm of topology beyond existing examples. 

We note that our results also apply to charge-conjugation parity (CP) and pseudo-chiral symmetries \cite{PhysRevLett.127.186602}. These can be obtained from \PT{} and pseudo-Hermitian symmetry by multiplying a factor of $i$ to the corresponding matrices. Another highly relevant situation is $C_2\mathcal{T}$ symmetry, which is equivalent to $\mathcal{PT}$ symmetry for two-dimensional spinless particles. 
We anticipate ample further topological aspects to emerge from the homotopy description of other symmetry classes. Note this work is also qualitatively different from previous homotopy description \cite{PhysRevB.103.155129,PhysRevB.101.205417} for systems without symmetries. The presence of symmetry essentially induces coexistence of eigenvalue and wave-function topology from the lowest dimension. This involves several new algebraic structures calling for more advanced tools to resolve the overall topological structure. Moreover, the separation gap notion is for the first time considered in the non-Hermitian setting. This concept much broadens the scope of band structures and turns out to be indispensable when the spectra of the system are correlated by anti-linear symmetries. The methods and concepts introduced here can be subsequently generalized to any other symmetries.

The remainder of the article is organized as follows. 
We will use \PT{} symmetry as the main example to illustrate the framework and method of the homotopy classification, as well as the interpretations of different topologies in detail.
Subsequently, we provide a list of analogous results for the case of pseudo-Hermitian symmetry. 
We set the physical context in Sec.~\ref{sec:preliminaries}. 
We introduce the notion of \PT{} symmetry, the concepts of band gaps and separation gaps.
We give a brief introduction on how homotopy theory captures all topological invariants that might appear for both gapped phases and nodal structures.
In Sec.~\ref{sec_2bd}, we start from the two-band models, where the topology can be visualized easily and directly read out without involving complicated mathematical tools. 
We apply the results to a series of models, showing how the nontrivial zeroth homotopy set explains the abundant existence of nodal structures.
We then illustrate how the winding number in the \PT{} symmetry-preserving sector emerges as an invariant from the fundamental group.
This predicts protected non-trivial band-gapped phases and induces rules governing exceptional rings and surfaces. 
The results for band-gapped phases and nodal structures for an arbitrary number of bands are presented in Sec.~\ref{sc_ggp}, including a detailed derivation. 
We turn to the separation gap protecting the bands of different symmetry and dynamic properties in Sec.~\ref{sc_spphs}, 
where we emphasize the intrinsic non-Hermitian nature of this band structure thoroughly. 
In Sec.~\ref{sc_expti}, we express all the topological invariants found in the preceding Sections~\ref{sc_ggp}~and~\ref{sc_spphs} using concepts familiar to most physicists, and answer how these expressions must be modified to account for the non-orthogonal eigenvectors of non-Hermitian operators.
The issue of fragile and stable topology follows in Sec.~\ref{sc_fgstbto}. 
There we unveil that some Hermitian topology remains stable under non-Hermiticity, while some becomes fragile in a very different way compared to Hermitian models, accompanied by new stable topology emerges from complex eigenvalues. 
The analogous pseudo-Hermitian results are given in Sec.~\ref{sc_psedh}. Different from previous research, we reveal that pseudo-Hermitian systems can contain an exponentially large zeroth homotopy set, making nodal structures even more abundant. 
Finally, we summarize our main results in Sec.~\ref{sc_cocloutl}, and outline a range of exciting directions for future research building on the extensive framework established here. 
Supporting mathematical details and generalizations are provided in Appendices~\ref{ap_PTgU}~to~\ref{ap_cherneuler}.

\section{Symmetry, models, and methods} \label{sec:preliminaries}

We begin by introducing the questions, concepts, and tools used throughout the paper. We are interested in systems where the properties are captured by general $N\times N$ non-Hermitian matrices. Such matrices prominently appear in the equations of motion as effective Hamiltonians but their applications in physics are in fact much more general \cite{ashida2020non}. Following the convention in the literature, we denote these matrices as $H$, while keeping in mind their more general applications beyond effective Hamiltonians. The matrix $H$ acts on an $N$-dimensional Hilbert space, which represents the degrees of freedom of the system. In the absence of any constraints, the entries of these matrices are complex numbers. All such matrices form an $N^2$-dimensional complex vector space $\mathrm{Mat}(N,\mathbb C)$. 
In reality, this matrix depends on a series of tunable parameters in experiments. So it can viewed as a matrix-valued function. We denote this dependence as $H(\mathbf k)$, where $\mathbf k$ is a $d$-dimensional real vector, representing the control parameters in experiments. 
Such a description also appears as a Bloch Hamiltonian describing some tight-binding models on a lattice. The $N$ indices of the matrix $H$ correspond to the $N$ orbitals inside a unit cell, and it may also include spin degrees of freedom. The $d$ real parameters $\mathbf k=(k_1,k_2,\ldots,k_d)$ are interpreted as the momentum of the Bloch Hamiltonians. The momenta form a Brillouin zone in the form of a $d$-dimensional torus.

The spectrum of matrix $H(\mathbf k)$ reflects key properties of the corresponding system. When the matrix is Hermitian, the spectrum is real. This usually represents that the modes are oscillating and dissipationless. If loss and gain are included, $H(\mathbf k)$ can be non-Hermitian, where the modes have nonzero imaginary parts and decay or grow with respect to time. A systematic way to balance these oscillating modes and evanescent modes is to impose the matrix $H$ to be real, $H\in \mathrm{Mat}(N,\mathbb R)$. Such systems are called $\mathcal{PT}$-symmetric systems \cite{PhysRevLett.101.080402,PhysRevLett.103.093902,ruter2010observation,PhysRevLett.106.213901,regensburger2012parity,PhysRevX.4.031042,peng2014parity,PhysRevLett.112.143903,xu2016topological,el2018non,ozdemir2019parity,PhysRevLett.126.083604,arkhipov2023dynamically}. The oscillating modes and decaying/growing modes follow different patterns under the symmetry transformation. Their appearance is related to the spontaneous symmetry breaking, as we recapitulate below. 

In addition to these real matrices, other forms of \PT symmetry also appear naturally in reality.
For example, consider Bloch Hamiltonians where the parity transformation $\mathcal P$ may also bring a unitary transformation $U_{\mathcal P}$ on the space of orbitals besides reversing the momentum $\mathbf k\to -\mathbf k$. 
The time-reversal transformation reverses the momentum, brings a complex conjugate on all operators: $\mathcal T H(\mathbf k)\mathcal T^{-1}=U_{\mathcal T}H^\ast(-\mathbf k)U^{-1}_{\mathcal T}$ where $U_{\mathcal T}$ is unitary.
The joint action of the parity and time reversal transformation in general takes the form \(H(\mathbf k)\to U_{\PT} H^\ast(\mathbf k) U^{-1}_{\PT}\). The forms of the unitary matrix $ U_{\PT}$ are decided by the crystal structure \cite{PhysRevX.10.031001,brouwer2023homotopic} and the spins of the system.

When including all possible forms of unitary transformations, there are three classes of $U_\mathcal{PT}$, details in Appendix~\ref{ap_PTgU}. Interestingly, they correspond to the three types of associative division algebra over the real numbers: real, complex, and quaternion. A most relevant and representative class is $U_{\PT}=I$, the identity matrix. Intuitively, this class includes all systems where the orbitals inside each unit cell can be adiabatically deformed to the lattice position while preserving the crystalline symmetry and the spectrum. For this class, the lattice $\mathcal{PT}$ symmetry requires the Bloch Hamiltonian to be a real $N\times N$ matrix, exactly the form familiar in photonics, lasers, and other metamaterials as mentioned above. The real $N\times N$ matrices $H(\mathbf k)$ also apply to fermions, including $\mathcal{PT}$-symmetric systems without spin-orbit coupling and $C_2\mathcal T$-symmetric systems \cite{PhysRevLett.118.056401,PhysRevX.9.021013,bouhon2020non}. The second class corresponds to a direct sum of an arbitrary complex matrix and its complex conjugation, whose structure has already been covered in Ref.~\cite{PhysRevB.103.155129}, the system without any symmetry. In the third class, $H$ is equivalent to a matrix whose entries are quaternions. It includes more complicated geometries of orbitals and couplings with spins. This class requires more delicate experimental setups. We will not concentrate on this case here, but for completeness, we summarize the outcome of the theoretical analysis in Appendix~\ref{ap_PTgU}.

Eigenvalues of a real matrix can be classified into two types: real and complex.
Eigenvectors corresponding to real eigenvalues can also be chosen to have real components. We denote such eigenvalues as $\mathcal{PT}$-preserving eigenvalues, as they are invariant under the $\mathcal{PT}$ transformation. 
Complex eigenvalues must appear as complex conjugate pairs $E, E^\ast\in \mathbb C\backslash \mathbb R$. The corresponding eigenvectors also appear in complex conjugate pairs. Under the $\mathcal{PT}$ transformation these two eigenvalues and eigenvectors switch with each other. They are denoted as $\mathcal{PT}$-symmetry breaking eigenvalues \cite{PhysRevLett.103.093902}.

In the rest of this article, we will not distinguish whether $\mathbf k$ represents control parameters or momenta. We will always use the name Brillouin zone (BZ) for the domain of $\mathbf k$, albeit we have to keep in mind that when $\mathbf k$ refers to control parameters, it could live on an open non-compact space $\mathbb R^d$ instead of the torus.

The key object that we are studying is this matrix-valued function $H(\mathbf k)$. It is a continuous map from the BZ to the space of $N\times N$ matrices ($\mathrm{Mat}(N,\mathbb R)$). 
This map gives the eigenvalues as a function of $\mathbf k$. These eigenvalues forms a set of \emph{bands}: \(E_1(\mathbf k),\ldots,E_N(\mathbf k) \), with the corresponding eigenvectors given by $|u_1(\mathbf k)\rangle,\ldots, |u_N(\mathbf k)\rangle$. 
(Note: here and in the following, the subscripts are for distinguishing different eigenvalues only; if the eigenvalues can braid, each $E_i$ might not be a periodic function of $\mathbf k$.)
We are interested in certain band structures, especially nodal structures and gaps, as shown in Fig.~\ref{fig_nphs}. In the following, we are going to introduce these band structures and comment on their physical significance. After transforming the band structures into mathematical notions and concepts, we will be able to apply homotopy theory to describe and classify them.

\subsection{Nodal phases and structures}

We begin by examining how physical properties are determined by band structures, which motivates us to define the central concepts to study. Let us first consider single-body physics, which is highly relevant to bosonic systems and dynamic evolution captured by an effective non-Hermitian matrix.

In these situations, the state of the system is described by a single mode in one of the bands, which we denote as $E_i(\mathbf k)$. To excite the system to another mode $E_j(\mathbf k)$, we need to overcome the difference $\Delta_{ij}(\bk) := E_i(\bk)-E_j(\bk) $, depicted as Fig.~\ref{fig_nphs} (a). As long as this difference is non-vanishing, the properties of the state are continuous with respect to different $\mathbf k$. But when degeneracy $\Delta_{ij}(\bk_\ast)=0$ takes place at some $\mathbf k_\ast$, the system bifurcates, and even exhibits singular behaviors since $E_i(\bk)$ can become non-analytic at $\mathbf k_\ast$. This singularity is further enhanced for non-Hermitian matrices, due to the coalescence of the corresponding eigenvectors, known as EPs  \cite{kato2013perturbation,berry2004,RevModPhys.93.015005}. The regions of $\mathbf k$ where degeneracy happens play an important role as the jump or transition of physical properties. This leads us to the first type of band structure:
\begin{definition}[Nodal Structure]
    A connected region $\mathcal K$ in the momentum space is called a \emph{nodal structure} if the $N\times N$ matrix  \(H(\bk)\) is degenerate for any point \(\bk\) in the region. More rigorously, there exists $i\ne j$ such that $E_i(\mathbf k)=E_j(\mathbf k),\, \forall\, \mathbf k\in \mathcal K$. 
    Depending on its dimension, we also call this region nodal point, line, surface, etc.
\end{definition}

In the above definition, we take into account all degeneracy between every pair of bands. This is because in many situations there is no preference for which band the single-particle/mode may occupy.
At nodal points \(\bk_*\), some bands must touch, $ E_i(\bk_*)-E_j(\bk_*) = 0$ for some \(i,j\). So we also call nodal points \emph{band crossings}. A phase with nodal points somewhere in the BZ is said to be a \emph{nodal phase}.

The dimension of the nodal structures tells us how large the singularities are in the system. The singular behaviors are, instead captured by how the system looks like in the vicinity of the nodal structures. These can be dispersion of the eigenvalues, coalescence of the eigenvectors, or evolution of the eigenvalue and eigenstate encircling the nodal structure. For this purpose, it is sometimes useful to consider the codimension of nodal structures. If a nodal structure has dimension $d'$ inside a $d$-dimensional BZ, then its codimension $\bar d$ is defined as $\bar d=d-d'$. This codimension characterizes how many parameters in $\mathbf k$ we need to tune to reach the degeneracy. 

The codimension allows us to unify the characterization of nodal structures that appear in BZs of different dimensions. For example, in non-Hermitian systems, degeneracies are generically of codimension 2 \cite{berry2004,Heiss_2012,RevModPhys.93.015005}, independently of the momentum space dimension \(d\). They occur as points in two-dimensional momentum space, as lines in three dimensions, and so forth. 
Hermitian nodal structures in contrast generally occur with codimension 3 \cite{RevModPhys.90.015001,annurev-conmatphys-031016-025458}.
Discrete symmetries, such as \PT{} symmetry or pseudo-Hermiticity considered in this text, decrease the generic codimension of nodal structures further to 1, making them even more abundant \cite{PhysRevB.99.041406,PhysRevLett.127.186602,PhysRevLett.127.026404,PhysRevB.104.L201104,PhysRevB.99.121101,PhysRevB.99.041202,Zhou2019}.

Despite we mostly talk about how the nodal structure affects single-particle behaviors, it also impacts many-particle physics, for example through the singular density of states. The detail will depend on how the bands are occupied in a many-particle system.  We will not discuss this aspect in this work. Rather, we will use another concept (the separation gap) to systematically describe many-particle scenarios.

\subsection{Band gaps and separation gaps}

\begin{figure}
    \centering
    \def\svgwidth{\columnwidth}
    %--

    \begingroup%
  \makeatletter%
  \providecommand\color[2][]{%
    \errmessage{(Inkscape) Color is used for the text in Inkscape, but the package 'color.sty' is not loaded}%
    \renewcommand\color[2][]{}%
  }%
  \providecommand\transparent[1]{%
    \errmessage{(Inkscape) Transparency is used (non-zero) for the text in Inkscape, but the package 'transparent.sty' is not loaded}%
    \renewcommand\transparent[1]{}%
  }%
  \providecommand\rotatebox[2]{#2}%
  \newcommand*\fsize{\dimexpr\f@size pt\relax}%
  \newcommand*\lineheight[1]{\fontsize{\fsize}{#1\fsize}\selectfont}%
  \ifx\svgwidth\undefined%
    \setlength{\unitlength}{246bp}%
    \ifx\svgscale\undefined%
      \relax%
    \else%
      \setlength{\unitlength}{\unitlength * \real{\svgscale}}%
    \fi%
  \else%
    \setlength{\unitlength}{\svgwidth}%
  \fi%
  \global\let\svgwidth\undefined%
  \global\let\svgscale\undefined%
  \makeatother%
  \begin{picture}(1,0.79674803)%
    \lineheight{1}%
    \setlength\tabcolsep{0pt}%
    \put(0,0){\includegraphics[width=\unitlength,page=1]{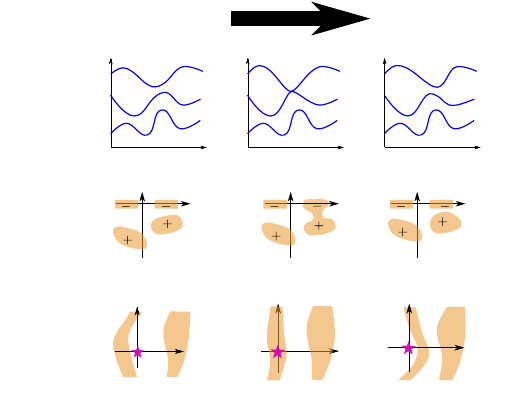}}%
    \put(0.0003511,0.66128723){\color[rgb]{0,0,0}\makebox(0,0)[lt]{\lineheight{1.25}\smash{\begin{tabular}[t]{l}(a)\end{tabular}}}}%
    \put(-0.00078711,0.42566406){\color[rgb]{0,0,0}\makebox(0,0)[lt]{\lineheight{1.25}\smash{\begin{tabular}[t]{l}(b)\end{tabular}}}}%
    \put(0.0014893,0.19004086){\color[rgb]{0,0,0}\makebox(0,0)[lt]{\lineheight{1.25}\smash{\begin{tabular}[t]{l}(c)\end{tabular}}}}%
    \put(0.31186519,0.7669298){\color[rgb]{0,0,0}\makebox(0,0)[t]{\lineheight{1.25}\smash{\begin{tabular}[t]{c}Topological\\invariant \(\mu\)\end{tabular}}}}%
    \put(0.84710407,0.7669298){\color[rgb]{0,0,0}\makebox(0,0)[t]{\lineheight{1.25}\smash{\begin{tabular}[t]{c}Topological\\invariant \(\mu'\)\end{tabular}}}}%
    \put(0.12735193,0.62451696){\color[rgb]{0,0,0}\makebox(0,0)[t]{\lineheight{1.25}\smash{\begin{tabular}[t]{c}Band\\gap\end{tabular}}}}%
    \put(0.12747387,0.37967772){\color[rgb]{0,0,0}\makebox(0,0)[t]{\lineheight{1.25}\smash{\begin{tabular}[t]{c}Separation\\gap\end{tabular}}}}%
    \put(0.12806332,0.13083344){\color[rgb]{0,0,0}\makebox(0,0)[t]{\lineheight{1.25}\smash{\begin{tabular}[t]{c}Reference\\point\end{tabular}}}}%
    \put(0.39093693,0.51930911){\color[rgb]{0,0,0}\makebox(0,0)[lt]{\lineheight{1.25}\smash{\begin{tabular}[t]{l}\(k\)\end{tabular}}}}%
    \put(0.21220068,0.66128723){\color[rgb]{0,0,0}\makebox(0,0)[rt]{\lineheight{1.25}\smash{\begin{tabular}[t]{r}\(E\)\end{tabular}}}}%
    \put(0.47878029,0.66128724){\color[rgb]{0,0,0}\makebox(0,0)[rt]{\lineheight{1.25}\smash{\begin{tabular}[t]{r}\(E\)\end{tabular}}}}%
    \put(0.74402412,0.66128724){\color[rgb]{0,0,0}\makebox(0,0)[rt]{\lineheight{1.25}\smash{\begin{tabular}[t]{r}\(E\)\end{tabular}}}}%
    \put(0.65321138,0.51930911){\color[rgb]{0,0,0}\makebox(0,0)[lt]{\lineheight{1.25}\smash{\begin{tabular}[t]{l}\(k\)\end{tabular}}}}%
    \put(0.92150405,0.51930911){\color[rgb]{0,0,0}\makebox(0,0)[lt]{\lineheight{1.25}\smash{\begin{tabular}[t]{l}\(k\)\end{tabular}}}}%
    \put(0.57297588,0.47436165){\color[rgb]{0,0,0}\makebox(0,0)[t]{\lineheight{1.25}\smash{\begin{tabular}[t]{c}\textcolor{red}{band crossing}\end{tabular}}}}%
    \put(0.57297588,0.25413502){\color[rgb]{0,0,0}\makebox(0,0)[t]{\lineheight{1.25}\smash{\begin{tabular}[t]{c}\textcolor{red}{spectral crossing}\end{tabular}}}}%
    \put(0.57297587,0.01241119){\color[rgb]{0,0,0}\makebox(0,0)[t]{\lineheight{1.25}\smash{\begin{tabular}[t]{c}\textcolor{red}{reference crossing}\end{tabular}}}}%
    \put(0.2796485,0.41658192){\color[rgb]{0,0,0}\makebox(0,0)[lt]{\lineheight{1.25}\smash{\begin{tabular}[t]{l}Im\(E\)\end{tabular}}}}%
    \put(0.55601866,0.41658192){\color[rgb]{0,0,0}\makebox(0,0)[lt]{\lineheight{1.25}\smash{\begin{tabular}[t]{l}Im\(E\)\end{tabular}}}}%
    \put(0.8242493,0.41658192){\color[rgb]{0,0,0}\makebox(0,0)[lt]{\lineheight{1.25}\smash{\begin{tabular}[t]{l}Im\(E\)\end{tabular}}}}%
    \put(0.27325197,0.19004086){\color[rgb]{0,0,0}\makebox(0,0)[lt]{\lineheight{1.25}\smash{\begin{tabular}[t]{l}Im\(E\)\end{tabular}}}}%
    \put(0.5499211,0.19613842){\color[rgb]{0,0,0}\makebox(0,0)[lt]{\lineheight{1.25}\smash{\begin{tabular}[t]{l}Im\(E\)\end{tabular}}}}%
    \put(0.80595662,0.19613842){\color[rgb]{0,0,0}\makebox(0,0)[lt]{\lineheight{1.25}\smash{\begin{tabular}[t]{l}Im\(E\)\end{tabular}}}}%
    \put(0.36141266,0.36328864){\color[rgb]{0,0,0}\makebox(0,0)[lt]{\lineheight{1.25}\smash{\begin{tabular}[t]{l}Re\(E\)\end{tabular}}}}%
    \put(0.65207062,0.36328864){\color[rgb]{0,0,0}\makebox(0,0)[lt]{\lineheight{1.25}\smash{\begin{tabular}[t]{l}Re\(E\)\end{tabular}}}}%
    \put(0.91248343,0.36328864){\color[rgb]{0,0,0}\makebox(0,0)[lt]{\lineheight{1.25}\smash{\begin{tabular}[t]{l}Re\(E\)\end{tabular}}}}%
    \put(0.36955564,0.11335223){\color[rgb]{0,0,0}\makebox(0,0)[lt]{\lineheight{1.25}\smash{\begin{tabular}[t]{l}Re\(E\)\end{tabular}}}}%
    \put(0.66064315,0.11335223){\color[rgb]{0,0,0}\makebox(0,0)[lt]{\lineheight{1.25}\smash{\begin{tabular}[t]{l}Re\(E\)\end{tabular}}}}%
    \put(0.91007118,0.11335223){\color[rgb]{0,0,0}\makebox(0,0)[lt]{\lineheight{1.25}\smash{\begin{tabular}[t]{l}Re\(E\)\end{tabular}}}}%
    \put(0,0){\includegraphics[width=\unitlength,page=2]{fig2.pdf}}%
  \end{picture}%
\endgroup%
    
    %--
    \caption{Manifestation of different types of topology under deformation of a system. 
    (a) The band-gap topology can only be modified via a band crossing. 
    (b) The separation-gap topology is robust until there is a spectral crossing violating the partition of bands. 
    (c) The reference topology detects when a reference point or line is crossed by the spectrum of the system.}
    \label{fig_phstrans}
\end{figure}

Outside the nodal region, the eigenvalues of $H(\mathbf k)$ are pairwise distinct, \(i\neq j \implies E_i\neq E_j\), then we call such matrices \emph{non-degenerate}. For non-degenerate matrices, there is non-zero cost $\Delta_{ij}(\mathbf k)$ to create excitation that preserves the momentum or the control parameters, illustrated in Fig.~\ref{fig_nphs} (a). This is reminiscent of a single-particle gap in usual Hermitian physics. When we deform the system, many physical properties remain unchanged whenever the costs $\Delta_{ij}(\mathbf k)$ remain non-vanishing. Some of these properties originate from topological invariants associated with the system. Based on these physical considerations, we give the following definition of a band-gapped phase. It formulates some earlier ideas (cf. \cite{PhysRevLett.118.040401,PhysRevLett.120.146402}) into a systematic description: 
\begin{definition}[Band-gapped phases]
    We call a system characterized by an $N\times N$ operator $H(\bk)$ \emph{band-gapped}, or simply \emph{gapped}, if and only if its spectrum is non-degenerate for all \(\bk\). 
    Namely, $\left|E_i(\mathbf k)- E_j(\mathbf k)\right|>0,\, \forall\, \mathbf k, i\ne j$.
\end{definition}
Thus, a system is gapped if and only if there are no nodal structures throughout the BZ, or equivalently, if and only if all pairs of its bands are band-gapped. The gap width between the bands $i,j$ is defined as $\Delta^b_{ij}=\textrm{min}_{\mathbf{k}}\left|E_i(\mathbf k)- E_j(\mathbf k)\right|$. This type of band structure, together with the nodal structures, naturally emerges in systems described by a single mode/particle. There, a finite number $N$ degrees of freedom in experiments are relevant to our interests. All physical properties are encoded in how these $N$ bands cross or repel from each other, as depicted in Fig.~\ref{fig_phstrans} (a). We want to know what topological property protects our gaps and nodal structures. This comprises one branch of topology that we will study in this paper. We denote the space of all \PT-symmetric non-degenerate \(N\times N\) matrices as \(X_N\).

By now we have discussed many physical properties and band structures related to single-particle physics or bosonic systems. In Hermitian fermionic many-body systems, we usually have some bands occupied and other bands empty. If these occupied bands have a clear separation from the empty bands, we have a gap in the many-body spectrum. This kind of many-particle gap can also be generalized to non-Hermitian systems. The challenge is that the spectrum of non-Hermitian systems is lying on a two-dimensional complex plane. The geometry of the spectrum is much more complicated than the Hermitian realm. In order to do the generalization, we are obliged to resort to some more abstract languages. Nevertheless, the physical meaning of this generalized many-particle gap will be clear in specific realizations, as we explain and depict in Fig.~\ref{fig_nphs} (b).

Let us start by extracting the key ingredients from the Hermitian insulating gap in a many-body free fermionic system. Assume we have $N$ bands in total, labeled by band indices $j=1,2,\dots, N$. The bands are filled according to the value of their energy. The system is insulating if any excitation from the occupied states to an empty state requires non-zero energy. This happens when there is no band that is partially filled (every band is either completely filled or empty). So an insulator gives us a partition of all bands into occupied and empty bands. All indices $j$ correspond to completely occupied bands form a set $J_-$. And all bands that are completely empty form another set $J_+$. Clearly, there is no band that possesses occupied states and empty states simultaneously for an insulator, so $J_+\cap J_-=\varnothing$, and every band is either filled or empty, $\{1,2,\dots,N\}=J_+\cup J_-$. Thus an insulator with a many-body gap is equivalent to a partition of band indices into two disjoint subsets $\{1,2,\dots,N\}=J_+\sqcup J_-$, where $\sqcup $ denotes disjoint set union. The two subsets $J_-$ and $J_+$ are energetically \emph{separate}. The insulating behavior comes from the absence of energy overlap between the occupied and the empty bands. 
With this motivation, we now generalize such many-body gaps to non-Hermitian systems. To make our definition cover broader physical phenomena, we do not restrict the number of sets or bands in the partition (cf. \cite{PhysRevLett.120.146402}):
\begin{definition}[Separation-gapped phases]
    A system has \emph{separation gaps} if and only if there exists a partition of bands $\{1,2,\dots,N\}=\sqcup_\alpha J_\alpha$ according to certain rules on their spectrum.
    The partition is featured by the nonzero cost of excitation crossing different sets: $\textrm{min}_{\mathbf k, \mathbf k'}\,\left|E_j(\mathbf k)- E_{j'}(\mathbf k')\right|>0$ when $j$ and $j'$ belonging to different $J_\alpha$'s. The rule of the partition $J_\alpha$ must be chosen and fixed according to the physical context.
\end{definition}
In the presence of separation gaps, all excitation across different types of bands has to pay a price, as shown in Fig.~\ref{fig_nphs} (b). The most important ingredient of the separation gap is the partition rule based on the location of the eigenvalues $E_i$ on the complex plane. Different partition rules give different physical properties. This is why we stress in its definition that the partition rule must be fixed by the physical question. To understand the partition rule, let us look at a few concrete examples. 

As we have already shown, in Hermitian systems, the partition rule is given by sorting the spectrum on the real axis. The occupied bands $J_-$ are entirely below the chemical potential and the empty bands $J_+$ are above the chemical potential. The gap width is the distance from the top of the valence band to the bottom of the conduction band: $\Delta^s_{+-}=\textrm{min}\, \left[E_j(\mathbf k)- E_{j'}(\mathbf k')\right]$ with respect to all $j$ in the empty bands and all $j'$ in the occupied bands. In this situation, the separation gap coincides with a reference gap at the chemical potential \cite{PhysRevX.9.041015}, as the partition is given by the chemical potential.

Another common example of partition rule comes from the dynamical properties of the system, namely the Liouvillian gap \cite{PhysRevA.98.042118}. The system has loss and dissipation. The eigenvalue is either on the real axis, those steady states, or on the lower half complex plane, those evanescent states. The physical properties in the long-time limit are controlled by the steady states. The Liouvillian gap is defined as $\Delta^s_{+-}=\textrm{min}\,\left| \textrm{Im }E_{j}(\mathbf k)\right|$, for all bands $j$ living on the lower half complex plane. It tells the time scale at which all evanescent states decay. From this definition, we formulate it into a separation gap. The partition is to put the bands on the real axis as $J_-$, and put the bands on the lower half complex plane in $J_+$. Whenever the Liouvillian gap exists, there is no band mixing between $J_-$ and $J_+$. The nonvanishing Liouvillian gap guarantees the separation of spectra: $ \textrm{min}_{\mathbf k, \mathbf k'}\,\left|E_{j\in J_+}(\mathbf k)- E_{j'\in J_-}(\mathbf k')\right|\ge \Delta^s_{+-}>0$.

The independence of reference and flexibility in the number of bands in the definition of separation gap will play a crucial role in \PT-symmetric and pseudo-Hermitian systems. We are now using this language to describe the typical separation there.

As we have mentioned before, in \PT-symmetric or pseudo-Hermitian systems, there are two types of eigenvalues, those spontaneous symmetry-breaking eigenvalues as a mirror pair on the upper and the lower half complex planes, and those symmetry-preserving eigenvalues on the real axis. The symmetry-breaking eigenvalues are either decaying or growing, depending on which half of the complex plane they belong to, while the symmetry-preserving eigenvalues are oscillating. Due to these differences in symmetry and dynamics, we can do a natural partition of bands into three types: the bands on the real axis, the bands on the upper half complex plane and the bands on the lower half complex plane.
This partition is always possible whenever no band is crossing the real axis and connecting the lower or upper complex plane. We will show that this separation is intrinsically multi-separation-gapped in Sec.~\ref{sc_spphs}. It is an example that cannot be captured by a single, fixed reference. We leave more more discussion contrasting references and our gap definitions to Appendix~\ref{ap_dfgs}. 

The separation gap is closed when it is impossible to establish the partition of bands. Often this is caused by some band $j$ that cannot be assigned to either $J_-$ or $J_+$.  A typical example is that the spectrum of a band is connecting the spectrum in $J_-$ to the spectrum in $J_+$, as shown in Fig.~\ref{fig_phstrans} (b). This is similar to how the many-body gap closes in free fermionic Hermitian systems.

In the last part of this section, we elaborate on the connections and the differences between band gaps and separation gaps. 
The band gap describes the repulsion of bands locally in the BZ. As a complex number, the band gap $\Delta_{ij}(\mathbf k)$ can rotate on the complex plane, bringing the eigenmode braids \cite{PhysRevB.103.155129,PhysRevB.101.205417,wang2021topological,patil2022measuring,PhysRevLett.130.017201,PhysRevLett.130.157201}. The separation gap requires the repulsion of band spectrum in the complex plane. The gap itself has fewer degrees of freedom. Yet, in separation-gapped systems, there can be band crossings between bands $j,j'$ belonging to the same subset $J_+$ or $J_-$.

We summarize the typical features of our band gaps and separation gaps below:
\begin{enumerate}
    \item Both types of gaps are based on the cost of excitations. They are not influenced by any constant that can be added to $H$. The band gap focuses on excitation for a fixed control parameter or carrying zero momentum. The separation gap is a generalization of the many-body spectrum gap.
    \item The closing of a band gap is associated with band crossings, while the closing of a separation gap is often accompanied by a spectral crossing. In comparison, the reference approach detects when the reference is crossed, as reflected in Fig.~\ref{fig_phstrans}. 
    \item In the definitions of band gaps and separation gaps, we do not exactly specify whether the spectrum lives on the entire complex plane or is restricted to the real axis. When Hermiticity is imposed, the two concepts naturally transform back to their familiar counterparts in Hermitian physics. Hence this is a unified framework for both Hermitian and non-Hermitian physics. This will help us understand how conventional Hermitian topology behaves in the presence of non-Hermiticity.
\end{enumerate}

Having established the notions of gaps and nodal structures, we now turn to the study of their topological properties with the help of homotopy theory. We explain why this is fruitful, and show how homotopy groups provide invariants under gap-preserving deformations of the system. 
Importantly, we emphasize that homotopy groups give us the topological characters of both gapped phases and nodal structures.

\subsection{Classification of nodal structures}

In this section, we introduce how to apply the idea of homotopy theory to classify degeneracies. Concrete examples will be given later in Sec.~\ref{sec_2bd}. The formalism is similar to the classification of defects in ordered media. A thorough review of this topic can be found in Ref.~\onlinecite{RevModPhys.51.591}.

\begin{figure}[htbp!]
    \centering
    \def\svgwidth{\columnwidth}
    %--
    
    \begingroup%
		  \makeatletter%
		  \providecommand\color[2][]{%
		    \errmessage{(Inkscape) Color is used for the text in Inkscape, but the package 'color.sty' is not loaded}%
		    \renewcommand\color[2][]{}%
		  }%
		  \providecommand\transparent[1]{%
		    \errmessage{(Inkscape) Transparency is used (non-zero) for the text in Inkscape, but the package 'transparent.sty' is not loaded}%
		    \renewcommand\transparent[1]{}%
		  }%
		  \providecommand\rotatebox[2]{#2}%
		  \newcommand*\fsize{\dimexpr\f@size pt\relax}%
		  \newcommand*\lineheight[1]{\fontsize{\fsize}{#1\fsize}\selectfont}%
		  \ifx\svgwidth\undefined%
		    \setlength{\unitlength}{246bp}%
		    \ifx\svgscale\undefined%
		      \relax%
		    \else%
		      \setlength{\unitlength}{\unitlength * \real{\svgscale}}%
		    \fi%
		  \else%
		    \setlength{\unitlength}{\svgwidth}%
		  \fi%
		  \global\let\svgwidth\undefined%
		  \global\let\svgscale\undefined%
		  \makeatother%
		  \begin{picture}(1,1.42682927)%
		    \lineheight{1}%
		    \setlength\tabcolsep{0pt}%
		    \put(0,0){\includegraphics[width=\unitlength,page=1]{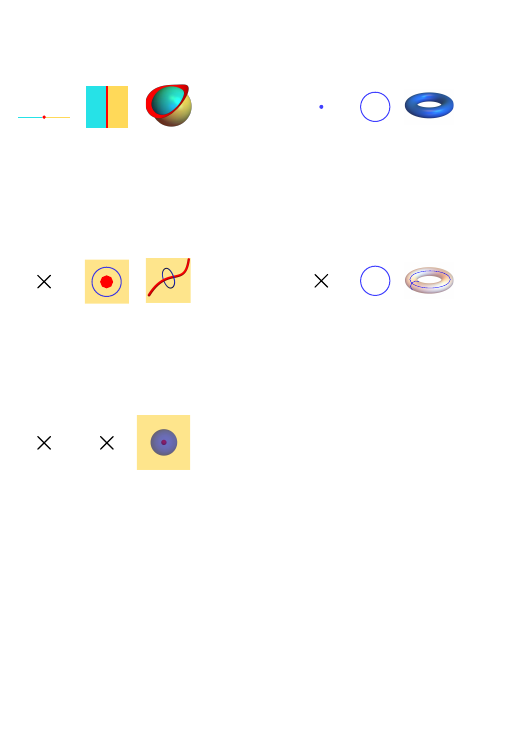}}%
		    \put(0.20868499,1.26579841){\color[rgb]{0,0,0}\makebox(0,0)[t]{\lineheight{1.25}\smash{\begin{tabular}[t]{c}d=2\end{tabular}}}}%
		    \put(0.20943717,1.21251881){\color[rgb]{0,0,0}\makebox(0,0)[t]{\lineheight{1.25}\smash{\begin{tabular}[t]{c}\(a\)~~\(a'\)\end{tabular}}}}%
		    \put(0.21334318,1.02405315){\color[rgb]{0,0,0}\makebox(0,0)[t]{\lineheight{1.25}\smash{\begin{tabular}[t]{c}\(\pi_1\): Can be encircled\\by a loop\end{tabular}}}}%
		    \put(0,0){\includegraphics[width=\unitlength,page=2]{fig3.pdf}}%
		    \put(0.22671597,1.39545046){\color[rgb]{0,0,0}\makebox(0,0)[t]{\lineheight{1.25}\smash{\begin{tabular}[t]{c}Nodal Structure\end{tabular}}}}%
		    \put(0.79963705,1.39545046){\color[rgb]{0,0,0}\makebox(0,0)[t]{\lineheight{1.25}\smash{\begin{tabular}[t]{c}Gapped Phases\end{tabular}}}}%
		    \put(0.21409164,1.33551652){\color[rgb]{0,0,0}\makebox(0,0)[t]{\lineheight{1.25}\smash{\begin{tabular}[t]{c}\(\pi_0\): divides the space\end{tabular}}}}%
		    \put(0.32873318,0.92625478){\color[rgb]{0,0,0}\makebox(0,0)[t]{\lineheight{1.25}\smash{\begin{tabular}[t]{c}d=3\end{tabular}}}}%
		    \put(0.62736424,1.26579841){\color[rgb]{0,0,0}\makebox(0,0)[t]{\lineheight{1.25}\smash{\begin{tabular}[t]{c}d=0\end{tabular}}}}%
		    \put(0.08541003,1.26579841){\color[rgb]{0,0,0}\makebox(0,0)[t]{\lineheight{1.25}\smash{\begin{tabular}[t]{c}d=1\end{tabular}}}}%
		    \put(0.73153075,0.92625478){\color[rgb]{0,0,0}\makebox(0,0)[t]{\lineheight{1.25}\smash{\begin{tabular}[t]{c}d=1\end{tabular}}}}%
		    \put(0.73153075,1.26579841){\color[rgb]{0,0,0}\makebox(0,0)[t]{\lineheight{1.25}\smash{\begin{tabular}[t]{c}d=1\end{tabular}}}}%
		    \put(0.83728286,1.26579841){\color[rgb]{0,0,0}\makebox(0,0)[t]{\lineheight{1.25}\smash{\begin{tabular}[t]{c}d=2\end{tabular}}}}%
		    \put(0.83728286,0.92625478){\color[rgb]{0,0,0}\makebox(0,0)[t]{\lineheight{1.25}\smash{\begin{tabular}[t]{c}d=2\end{tabular}}}}%
		    \put(0.20868499,0.92625478){\color[rgb]{0,0,0}\makebox(0,0)[t]{\lineheight{1.25}\smash{\begin{tabular}[t]{c}d=2\end{tabular}}}}%
		    \put(0.32873318,1.26579841){\color[rgb]{0,0,0}\makebox(0,0)[t]{\lineheight{1.25}\smash{\begin{tabular}[t]{c}d=3\end{tabular}}}}%
		    \put(0.32065185,0.61930221){\color[rgb]{0,0,0}\makebox(0,0)[t]{\lineheight{1.25}\smash{\begin{tabular}[t]{c}d=3\end{tabular}}}}%
		    \put(0.21553446,1.13555423){\color[rgb]{0,0,0}\makebox(0,0)[t]{\lineheight{1.25}\smash{\begin{tabular}[t]{c}\(a\neq a'\in\pi_0\)\end{tabular}}}}%
		    \put(0.21047094,0.70820465){\color[rgb]{0,0,0}\makebox(0,0)[t]{\lineheight{1.25}\smash{\begin{tabular}[t]{c}\(\pi_2\): Can be enclosed\\by a sphere\end{tabular}}}}%
		    \put(0.78371142,1.13555423){\color[rgb]{0,0,0}\makebox(0,0)[t]{\lineheight{1.25}\smash{\begin{tabular}[t]{c}fix \(\pi_0\)-element\end{tabular}}}}%
		    \put(0.78494185,0.82066069){\color[rgb]{0,0,0}\makebox(0,0)[t]{\lineheight{1.25}\smash{\begin{tabular}[t]{c}fix \(\pi_1\)-elements\end{tabular}}}}%
		    \put(0.78482018,1.33551652){\color[rgb]{0,0,0}\makebox(0,0)[t]{\lineheight{1.25}\smash{\begin{tabular}[t]{c}entire parameter space\end{tabular}}}}%
		    \put(0.93746128,1.2081895){\color[rgb]{0,0,0}\makebox(0,0)[t]{\lineheight{1.25}\smash{\begin{tabular}[t]{c}\(\cdots\)\end{tabular}}}}%
		    \put(0.93746128,0.86874776){\color[rgb]{0,0,0}\makebox(0,0)[t]{\lineheight{1.25}\smash{\begin{tabular}[t]{c}\(\cdots\)\end{tabular}}}}%
		    \put(0.78471855,1.02329593){\color[rgb]{0,0,0}\makebox(0,0)[t]{\lineheight{1.25}\smash{\begin{tabular}[t]{c}loops in parameter space\end{tabular}}}}%
		    \put(0.60567466,0.671335){\color[rgb]{0,0,0}\makebox(0,0)[t]{\lineheight{1.25}\smash{\begin{tabular}[t]{c}\(\vdots\)\end{tabular}}}}%
		    \put(0.08695484,1.21251881){\color[rgb]{0,0,0}\makebox(0,0)[t]{\lineheight{1.25}\smash{\begin{tabular}[t]{c}\(a\)~~\(a'\)\end{tabular}}}}%
		    \put(0.02347628,1.39545046){\color[rgb]{0,0,0}\makebox(0,0)[t]{\lineheight{1.25}\smash{\begin{tabular}[t]{c}(a)\end{tabular}}}}%
		    \put(0.58783603,1.39545046){\color[rgb]{0,0,0}\makebox(0,0)[t]{\lineheight{1.25}\smash{\begin{tabular}[t]{c}(b)\end{tabular}}}}%
		    \put(0.02233799,0.41403525){\color[rgb]{0,0,0}\makebox(0,0)[t]{\lineheight{1.25}\smash{\begin{tabular}[t]{c}(c)\end{tabular}}}}%
		    \put(0.40110116,1.21251881){\color[rgb]{0,0,0}\makebox(0,0)[t]{\lineheight{1.25}\smash{\begin{tabular}[t]{c}\(\cdots\)\end{tabular}}}}%
		    \put(0.14317669,0.39276519){\color[rgb]{0,0,0}\makebox(0,0)[lt]{\lineheight{1.25}\smash{\begin{tabular}[t]{l}Homotopy\\groups\end{tabular}}}}%
		    \put(0.36906104,0.36689131){\color[rgb]{0,0,0}\makebox(0,0)[t]{\lineheight{1.25}\smash{\begin{tabular}[t]{c}\(d=0\)\end{tabular}}}}%
		    \put(0.49578571,0.36689131){\color[rgb]{0,0,0}\makebox(0,0)[t]{\lineheight{1.25}\smash{\begin{tabular}[t]{c}\(d=1\)\end{tabular}}}}%
		    \put(0.62101937,0.36689131){\color[rgb]{0,0,0}\makebox(0,0)[t]{\lineheight{1.25}\smash{\begin{tabular}[t]{c}\(d=2\)\end{tabular}}}}%
		    \put(0.74592143,0.36689131){\color[rgb]{0,0,0}\makebox(0,0)[t]{\lineheight{1.25}\smash{\begin{tabular}[t]{c}\(d=3\)\end{tabular}}}}%
		    \put(0.2362658,0.25688859){\color[rgb]{0,0,0}\makebox(0,0)[t]{\lineheight{1.25}\smash{\begin{tabular}[t]{c}\(\pi_0\)\end{tabular}}}}%
		    \put(0.2362658,0.15252727){\color[rgb]{0,0,0}\makebox(0,0)[t]{\lineheight{1.25}\smash{\begin{tabular}[t]{c}\(\pi_1\)\end{tabular}}}}%
		    \put(0.2362658,0.05296037){\color[rgb]{0,0,0}\makebox(0,0)[t]{\lineheight{1.25}\smash{\begin{tabular}[t]{c}\(\pi_2\)\end{tabular}}}}%
		    \put(0.32269601,0.28231536){\color[rgb]{0,0,0}\makebox(0,0)[lt]{\lineheight{1.25}\smash{\begin{tabular}[t]{l}frame orientation, number\\of real/complex eigenvalues\end{tabular}}}}%
		    \put(0.44254306,0.17703938){\color[rgb]{0,0,0}\makebox(0,0)[lt]{\lineheight{1.25}\smash{\begin{tabular}[t]{l}winding number / frame\\rotation, eigenvalue braid\end{tabular}}}}%
		    \put(0.56948399,0.07487087){\color[rgb]{0,0,0}\makebox(0,0)[lt]{\lineheight{1.25}\smash{\begin{tabular}[t]{l}Euler number,\\Chern number\end{tabular}}}}%
		  \end{picture}%
		\endgroup%
    
    %--
    \caption{
    Classification algorithm for nodal structures and gapped phases using the homotopy groups. 
    (a) To characterize a nodal structure, we observe its neighborhood. 
    Structures of codimension $1$ usually separate the parameter space into several components; 
    each component is labeled by a unique element in $\pi_0$. 
    If the nodal structure can be enclosed by a loop, for example, a nodal point or circle in two dimensions, it can be protected by an element $\pi_1$ on this loop. 
    (b) For a gapped phase, we begin by describing the topological character with its unique $\pi_0$-element. 
    Subsequently, the phase can be further characterized by higher homotopy groups. 
    Using the fundamental group $\pi_1$, we may assign a character along each non-contractible loop in the BZ or the space of control parameters, for instance, the meridian and the longitude of a two-dimensional BZ (which forms a torus).  
    (c) Summary of the classification and the results. Here $d$ is the minimal system dimension where the topology appears. The topological invariant for a phase/nodal structure is a list $\mu=(\pi_0,\pi_1,\dots)$.
    }
    \label{fig_hptiv}
\end{figure}

Topological properties of nodal structures are studied by how the system behaves in their vicinity. Away from the nodal structure in the BZ, the system is band-gapped. We assume that our nodal structure can be enclosed by a sphere $S^{n}$, as shown in Fig.~\ref{fig_hptiv} (a). The matrix function $H(\mathbf k)$ on this sphere defines a map from the sphere to the space of non-degenerate matrices, $S^{n}\to X_N$. 
When system parameters are continuously varied, the nodal structure will deform and move. It is regarded topologically invariant, as long as the variation does not cause collisions with other nodal structures in the BZ.
This is equivalent to say that $H(\mathbf k)$ stays gapped on the surrounding sphere during the variation. To classify nodal structures,  we define that two nodal structures are of the same type if and only if the maps $S^{n}\to X_N$ on their enclosing spheres can be continuously deformed into each other. 
A caution is that in order to compare two nodal structures, we need to make sure that the corresponding enclosing spheres have a common point. This guarantees that we can travel around the nodal structures in a specified manner starting from the common point \cite{RevModPhys.51.591}, which is important when the topology is non-Abelian. 
We call the nodal structure trivial if it can be gapped out without crossing this sphere. In contrast, a non-trivial nodal structure cannot be gapped out until another nodal structure crosses its enclosing sphere and annihilates it.

The above idea has a direct mathematical correspondence to the homotopy theory.
There, two maps are considered homotopic, if they may be continuously deformed into one another. The homotopy defines equivalent classes of maps. Especially, when the maps are defined on spheres, the equivalent classes can be endowed with a group structure.
The homotopy groups $\pi_n(X)$ consist of such equivalence classes of maps from \(n\)-dimensional spheres to a given space \(X\), $S^n\to X$.
For the classification, we simply assign an element $a\in \pi_n(X_N)$ to each nodal structure, representing the map on its enclosing sphere. Moreover, homotopy groups also tell us what happens when two nodal structures meet. 
As mentioned above, when two nodal structures are present, we choose their individual enclosing spheres to share a common point. 
Via this common point, the two enclosing spheres can be joined into a larger sphere enclosing both nodal structures. The total property of the two nodal structures is carried by this larger sphere. The merging of nodal structures is then given by their product in the homotopy group \cite{RevModPhys.51.591}.

As illustrative examples of homotopy groups, consider the cases when $n=0$ and $n=1$. 
The \emph{set} $\pi_0(X_N)$ classifies maps from a point to $X_N$. 
Two such maps can be deformed into each other if and only if the images of the point under the two maps can be joined by a path in $X_N$, 
so elements $\pi_0(X_N)$ correspond to the path-connected components \footnote{We do not distinguish the terminology ``connected'' and ``path-connected'', as they are equivalent for the questions we are interested in. } of the space $X_N$. 
It is in general not a group, except for special situations, for example, when the space $X_N$ itself is a Lie group. Unlike higher homotopy groups, the zero-dimensional sphere, a pair of points, does not always completely enclose a nodal structure.
Rather, it labels the different gapped regions into which the nodal structure divides the BZ. 
Each such region is an element of $\pi_0(X_N)$ representing the non-degenerate matrices that can continuously transform into each other without gap closing. 
The nodal structure described by $\pi_0(X_N)$ is then the boundary between different gapped regions, illustrated in Fig.~\ref{fig_hptiv} (a). 
It must divide the Brillouin zone and thus has codimension one, which can be thought as a domain wall. 
A nontrivial set $\pi_0(X_N)$ explains why nodal structures are more common in symmetry-protected systems \cite{PhysRevB.104.L201104,PhysRevLett.127.026404}. 

The so-called \emph{fundamental group} $\pi_1(X_N)$ classifies continuous maps from a circle $S^1$ to the target space $X_N$ with one fixed point $p_0\in S^1$ always mapped to another fixed point $x_0\in X_N$. 
Each element in $\pi_1 $ represents a class of maps that cannot be continuously interpolated to the others. 
Intuitively, such a map can be treated as a loop in $X_N$ with a fixed starting and ending point $x_0$. 
In other words, $\pi_1(X_N)$ gives the classes of loops that can be deformed continuously to each other while keeping the start and end point $x_0$ fixed.
Thus studying the fundamental group is equally relevant to point degeneracy in two dimensions and nodal lines in three dimensions since we can encircle both along a loop in their neighborhoods. Note that the image of $S^1$ is also connected in $X_N$, so each element of  $\pi_1(X_N)$ must belong to a specific connected component of $X_N$, with the latter being exactly labeled by $\pi_0(X_N)$. So this dictates the order of applying homotopy sets/groups, as in Fig.~\ref{fig_hptiv}.

In practice, the nodal region may be enclosed by several types of spheres of different dimensions. 
In such cases, they have to be classified starting from the lowest homotopy group (set), as illustrated in Fig.~\ref{fig_hptiv}(a). 
If the nodal structure divides the Brillouin zone into different connected components, its first character comes from $\pi_0$.
If it can be further enclosed with a loop, for example,  a nodal line that does not cut through a two-dimensional Brillouin zone, we may use an element in $\pi_1$ to describe it. 
The procedure continues until we cannot find a higher-dimensional sphere to enclose the nodal region.

\subsection{Description of gapped systems}
\label{sec:cdg}
To classify gapped phases, we switch our objects from the spheres enclosing the nodal structures to the whole BZ. By doing so, homotopy theory induces a natural classification of gapped systems in a closely related manner \cite{PhysRevLett.51.51,PhysRevLett.101.186805}. We use the band-gapped case as the example, where the homotopy groups $\pi_n(X_N)$ provide a unified description of both nodal structures and band-gapped systems. The method carries over to separation gaps in a straightforward way, by studying the corresponding homotopy groups of separation-gapped matrices.

A phase with a band gap completely lacks any band crossings. We are interested in properties robust against deformations of the system as long as the band gaps (and symmetries) are preserved. Band-gapped phases that cannot be adiabatically joined in this way are separated by phase transitions or gapless phases. These robust properties are topological invariants. 

The band-gapped phase $H(\mathbf k)$ defines a map from the $d$-dimensional torus $T^d$ to the space of $N\times N$ non-degenerate matrices \(X_N\). A continuous gap-preserving deformation of the system brings a continuous deformation of this map. Mathematically, two maps can be continuously deformed into each other if and only if they are homotopic. Quantities that are conserved under homotopy are called homotopy invariants. 
Thus, topological invariants in physics correspond to homotopy invariants mathematically.

The classification of band-gapped phases now transforms to the homotopy classes of maps $T^d\to X_N$ \cite{PhysRevLett.51.51,PhysRevLett.101.186805,1995JETPL6149M}. This is not the homotopy group but has close connections to the homotopy group. An illustrative observation is that there are many natural nontrivial loops on a $d$-dimensional torus, corresponding to traveling through the BZ by varying one component of the momentum $k_a\to k_a+2\pi$, [see  Fig.~\ref{fig_hptiv} (b)]. The matrix function $H(\mathbf k)$ restricted to these loops induces maps from a circle $S^1$ to non-degenerate matrices $X_N$. This is an element in the fundamental group $\pi_1(X_N)$ (with the caveat that conjugacy classes should be considered when the group is non-Abelian, see Sec.~\ref{sc_fdPTbk}). So a first type of topological invariants comes from the fundamental groups along different nontrivial loops in the BZ. A systematic way of using homotopy groups to find out all homotopy classes on the BZ is to decompose the $d$-dimensional BZ into spheres of different dimensions and study the maps induced on the spheres \cite{PhysRevLett.51.51,PhysRevB.103.155129,PhysRevB.101.205417}.
This procedure is known as a decomposition of the $d$-dimensional torus as a CW complex \cite{bott1982differential,hatcherAlgebraicTopology2002,brouwer2023homotopic,bouhon2023quantum}. Usually, for bands on a $d$-torus BZ, all homotopy groups $\pi_{n\le d}$ play a role. An explicit example is the Hopf insulator with nontrivial Chern numbers on the boundary surfaces of the $3$-dimensional BZ \cite{PhysRevLett.101.186805}. The system is characterized by three Chern numbers on the BZ boundaries, originating from $\pi_2$, and one Hopf invariant in the bulk determined by both $\pi_2$ and $\pi_3$.
Although this procedure can be very complicated for higher dimensional tori, it turns out to be rather straightforward in lower dimensions. 
We illustrate the cases when $d=0,d=1$ and $d=2$, as shown in Fig.~\ref{fig_hptiv}.

A $d=0$ the BZ is a single point \(p_0\), and a gapped phase is simply represented by a non-degenerate matrix \(H_0\) at this point. 
We only need to identify the non-degenerate matrices that cannot be connected by continuous interpolation. 
This information is provided by $\pi_0(X_N)$, and all gapped $d=0$ systems are therefore classified by the set
$\pi_0(X_N)$. 

For $d=1$, the Brillouin zone is a circle $S^1$. A gapped system defines a map from $S^1$ to $X_N$. As $S^1$ is connected, its image is also connected in $X_N$; the matrices $H(\mathbf k)$ must belong to the same element of $\pi_0(X_N)$ for all $\mathbf k$. 
This gives the first character of a $1$d gapped phase. 
Besides $\pi_0(X_N)$, the map $S^1\to X_N$ also belongs to an element of $\pi_1(X_N)$. 
This gives a second topological character of the gapped phase. 

The $d=2$ situation is similar to $d=1$. 
First, the gapped phase must correspond to a unique element in $\pi_0$. 
Then notice that in a two-dimensional Brillouin zone, there are two types of non-contractible loops, the meridian, $k_x=\text{constant}$, and the longitude, $k_y=$ constant, of the torus. 
Therefore two elements of $\pi_1$ are required to describe the topology along these two loops. 
Furthermore, the map on the two-dimensional torus can be extended to a map from a two-dimensional sphere to $X_N$ \cite{PhysRevLett.51.51,PhysRevB.103.155129}. 
Thus, $\pi_2$ gives a third and final topological character of a two-dimensional gapped phase.

As a conclusion of the classification scheme, we comment on two subtleties in the previous description that have to be dealt with carefully in practice. 
First, higher homotopy groups differ for different connected components of $X_N$, so $\pi_{n>0}(X_N)$ is not a rigorous notation. 
It depends on which $\pi_0$ is fixed in the first step of the classification, emphasized in  Fig.~\ref{fig_hptiv}. This will become relevant in the classification laid out in this text, as the degree of spontaneous \PT{} symmetry breaking of a Hamiltonian (given by \(\pi_0(X_N)\)) affects the higher homotopy groups. This will in the end lead to sector-dependent topology in our later results, which provides different levels of topology compared to Hermitian examples.
Second, the homotopy groups describe maps with a fixed base point. 
There can be maps that can be continuously deformed into each other with a moving base point. 
Allowing such deformations will lead to further equivalent classes in invariants from $\pi_1$ and $\pi_2$ \cite{RevModPhys.51.591,PhysRevB.103.155129,PhysRevB.101.205417}, corresponding to how the fundamental group acts on homotopy groups. Such problems can arise when $\pi_1$ is not trivial, giving a much more sophisticated classification.
We will briefly introduce an example in Sec.~\ref{sc_fdPTbk}. 

In Fig.~\ref{fig_hptiv} (c), we list the appearance of the homotopy groups according to the dimensions $d$ of the BZ. We give a summary of what topology these homotopy invariants represent. The details of working out these invariants will be introduced in Sec.~\ref{sec_2bd} and Sec.~\ref{sc_expti}. From all the discussion in this section, the topological indices of a gapped phase or a nodal structure take the form of a \emph{sequence} $(\pi_0,\pi_1,\dots,\pi_d)$, where $d$ is the dimension of the system. We have to fix them one by one following the order in Fig.~\ref{fig_hptiv}. The element $\pi_0$ is the \emph{first} character to be specified for any phases. It determines the types of $\pi_{n>0}$ that appear subsequently.

We briefly introduce our conventions here such that readers can take a quick look through our results. We use $\mathbb Z$ to denote the additive group of integers and $\mathbb Z_2=\mathbb Z/2\mathbb Z$ is the additive group only distinguishing odd and even integers. 
We use $\{1\}$ to denote the trivial group; when the group is Abelian, we also use $0$ to represent the trivial group. 
The direct product of two groups is represented by the Cartesian product $\times$; for two Abelian groups, it is the same as the direct product $\oplus$.

\section{The case of two bands}\label{sec_2bd}
Before deriving the general classification scheme, we discuss the special case of two-band systems to build useful intuition. 
We find that the band gap classification corresponds to a pictorially simple geometry, while imposing a separation between symmetry-breaking and symmetry-preserving bands leads to trivial classification for only two bands [see Sec.~\ref{sc_spphs}]. A qualitative difference from Hermitian topological phases is that the topological invariants depend on which sector the system is in, i.e., whether bands spontaneously break $\mathcal{PT}$ symmetry or not. 
This is due to the connection between the spontaneous symmetry breaking and the zeroth homotopy set $\pi_0$. 
As a result, one should be cautious not to mix different topological invariants from different sectors. The analysis must follow the sequence laid out in Fig.~\ref{fig_hptiv}.

\subsection{Parameterization and topology}\label{sec-PTtwoband}

We can express an arbitrary \PT-symmetric (real valued) \(2\times 2\) matrix in the basis of Pauli matrices \(\sigma_x,\sigma_y,\sigma_z\) and the identity \(\sigma_0\), with real coefficients depending on \(\bk\):
\begin{equation}
    H(\mathbf k)=
        d_0(\mathbf k) \sigma_0+
        d_x(\mathbf k) \sigma_x + 
        i \cdot d_y(\mathbf k) \sigma_y +
        d_z(\mathbf k) \sigma_z.
\end{equation}
The inclusion of a factor of \(i\) in the \(\sigma_y\) term ensures $H(\mathbf k)$ to be \PT-symmetric, i.e., real, for real components \(d_{x,y,z}\).
The operator's eigenvalues are given by $d_0\pm\sqrt{d_x^2-d^2_y+d^2_z}$. 
We are interested in the gap between the two bands, which is unaffected by \(d_0\), so we may set \(d_0=0\) and consider only traceless matrices without loss of generality.
In this way we identify the space of traceless real matrices with \(\R^3\).
The coefficients define a function from the BZ to the 3-dimensional vector space: \(\mathbf{d}(\bk)=(d_x(\bk),d_y(\bk),d_z(\bk))\in \R^3\).

\begin{figure}[htbp]
    \centering
    \def\svgwidth{\columnwidth}
    %--
    
    %%
\begingroup%
  \makeatletter%
  \providecommand\color[2][]{%
    \errmessage{(Inkscape) Color is used for the text in Inkscape, but the package 'color.sty' is not loaded}%
    \renewcommand\color[2][]{}%
  }%
  \providecommand\transparent[1]{%
    \errmessage{(Inkscape) Transparency is used (non-zero) for the text in Inkscape, but the package 'transparent.sty' is not loaded}%
    \renewcommand\transparent[1]{}%
  }%
  \providecommand\rotatebox[2]{#2}%
  \newcommand*\fsize{\dimexpr\f@size pt\relax}%
  \newcommand*\lineheight[1]{\fontsize{\fsize}{#1\fsize}\selectfont}%
  \ifx\svgwidth\undefined%
    \setlength{\unitlength}{246bp}%
    \ifx\svgscale\undefined%
      \relax%
    \else%
      \setlength{\unitlength}{\unitlength * \real{\svgscale}}%
    \fi%
  \else%
    \setlength{\unitlength}{\svgwidth}%
  \fi%
  \global\let\svgwidth\undefined%
  \global\let\svgscale\undefined%
  \makeatother%
  \begin{picture}(1,1)%
    \lineheight{1}%
    \setlength\tabcolsep{0pt}%
    \put(0,0){\includegraphics[width=\unitlength,page=1]{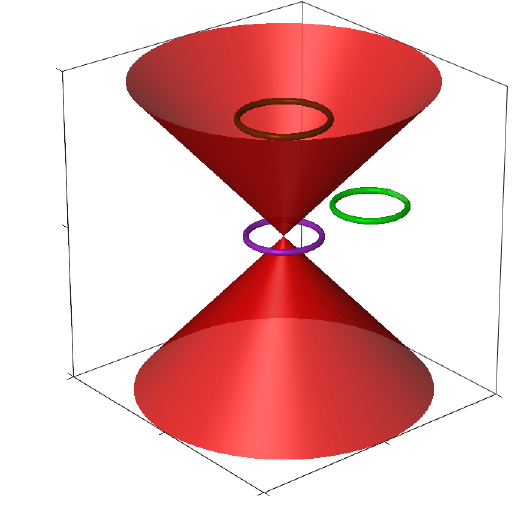}}%
    \put(0.092553,0.86966007){\color[rgb]{0,0,0}\transparent{0.99684298}\makebox(0,0)[t]{\lineheight{1.25}\smash{\begin{tabular}[t]{c}\(1\)\end{tabular}}}}%
    \put(0.10694949,0.27841066){\color[rgb]{0,0,0}\transparent{0.99684298}\makebox(0,0)[t]{\lineheight{1.25}\smash{\begin{tabular}[t]{c}\(-1\)\end{tabular}}}}%
    \put(0.98951578,0.18778881){\color[rgb]{0,0,0}\transparent{0.99684298}\makebox(0,0)[t]{\lineheight{1.25}\smash{\begin{tabular}[t]{c}\(1\)\end{tabular}}}}%
    \put(0.5428936,0.01691726){\color[rgb]{0,0,0}\transparent{0.99684298}\makebox(0,0)[t]{\lineheight{1.25}\smash{\begin{tabular}[t]{c}\(-1\)\end{tabular}}}}%
    \put(0.47446309,0.02134157){\color[rgb]{0,0,0}\transparent{0.99684298}\makebox(0,0)[t]{\lineheight{1.25}\smash{\begin{tabular}[t]{c}\(1\)\end{tabular}}}}%
    \put(0.11065896,0.23251925){\color[rgb]{0,0,0}\transparent{0.99684298}\makebox(0,0)[t]{\lineheight{1.25}\smash{\begin{tabular}[t]{c}\(-1\)\end{tabular}}}}%
    \put(0.37897285,0.07262347){\color[rgb]{0,0,0}\transparent{0.99684298}\makebox(0,0)[t]{\lineheight{1.25}\smash{\begin{tabular}[t]{c}\(d_x\)\end{tabular}}}}%
    \put(0.88211364,0.13719523){\color[rgb]{0,0,0}\transparent{0.99684298}\makebox(0,0)[t]{\lineheight{1.25}\smash{\begin{tabular}[t]{c}\(d_z\)\end{tabular}}}}%
    \put(0.08433113,0.71036593){\color[rgb]{0,0,0}\transparent{0.99684298}\makebox(0,0)[t]{\lineheight{1.25}\smash{\begin{tabular}[t]{c}\(d_y\)\end{tabular}}}}%
    \put(0.29458923,0.12507482){\color[rgb]{0,0,0}\transparent{0.99684298}\makebox(0,0)[t]{\lineheight{1.25}\smash{\begin{tabular}[t]{c}\(0\)\end{tabular}}}}%
    \put(0.77573157,0.10365864){\color[rgb]{0,0,0}\transparent{0.99684298}\makebox(0,0)[t]{\lineheight{1.25}\smash{\begin{tabular}[t]{c}\(0\)\end{tabular}}}}%
    \put(0.09712031,0.54878059){\color[rgb]{0,0,0}\transparent{0.99684298}\makebox(0,0)[t]{\lineheight{1.25}\smash{\begin{tabular}[t]{c}\(0\)\end{tabular}}}}%
    \put(0.6667823,0.85260949){\color[rgb]{0,0,0}\transparent{0.99684298}\makebox(0,0)[t]{\lineheight{1.25}\smash{\begin{tabular}[t]{c}\(X_{+}^{(1,0)}\)\end{tabular}}}}%
    \put(0.56132072,0.23780493){\color[rgb]{0,0,0}\transparent{0.99684298}\makebox(0,0)[t]{\lineheight{1.25}\smash{\begin{tabular}[t]{c}\(X_{-}^{(1,0)}\)\end{tabular}}}}%
    \put(0.81751998,0.44207327){\color[rgb]{0,0,0}\transparent{0.99684298}\makebox(0,0)[t]{\lineheight{1.25}\smash{\begin{tabular}[t]{c}\(X^{(0,2)}\)\end{tabular}}}}%
    \put(0.46463148,0.5310863){\color[rgb]{0,0,0}\transparent{0.99684298}\makebox(0,0)[rt]{\lineheight{1.25}\smash{\begin{tabular}[t]{r}\(\pi_1\left(X^{(0,2)}\right)\)\end{tabular}}}}%
  \end{picture}%
\endgroup%
    
    %--
    \caption{The space $X_2$ of two-band gapped Hamiltonians is the complement of the double cone in $\mathbb R^3$. 
    Note that the cones extend to infinity and separate the space $\mathbb R^3$ into three regions, representing the three elements of the zeroth homotopy set $\pi_0(X_2)$. 
    In the region $X^{(0,2)}$, the eigenvalues are real and there is a non-trivial loop (purple) generating its fundamental group $\pi_1(X^{(0,2)})=\mathbb Z$. In the other two regions $X^{(1,0)}_-$ and $ X^{(1,0)}_+$, the eigenvalues are complex. 
    These two regions are distinguished by a $\mathbb Z_2$ invariant, and are both contractible.}
    \label{fig_2bspace}
\end{figure}
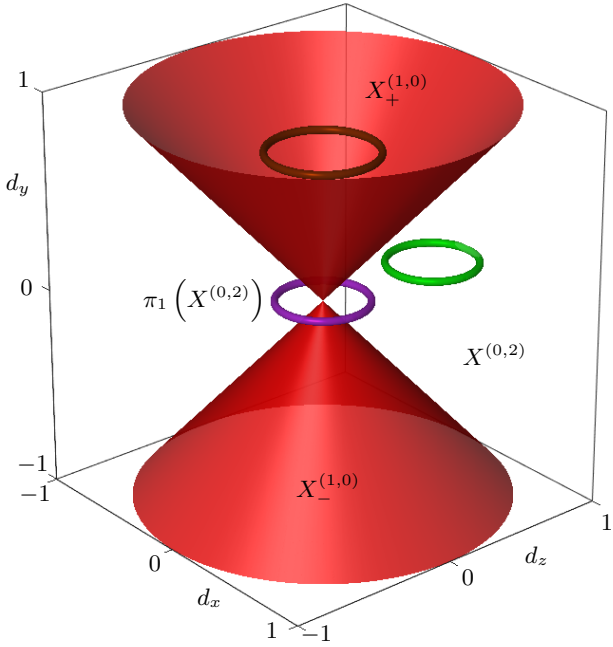

The key object that we study is the space $X_2$ made up of all $2\times 2$ \textit{non-degenerate} real matrices. 
To understand which coefficients \(\mathbf d\) lead to non-degenerate matrices, observe that the gap closes and the eigenvalues become equal at \(d_x^2+d_z^2 = d_y^2\), which describes a double cone in \(\R^3\).
This makes it easy to describe \(X_2\) as
\begin{equation}
    X_2 = \{\mathbf{d}\in\R^3 | d_x^2+d_z^2 \neq d_y^2\},
\end{equation}
which is the complement of this double cone.
We illustrate this space and its homotopy structure in Fig.~\ref{fig_2bspace}. 

The space \(X_2\) consists of three disjoint connected components, which we will index according to their eigenvalue structure.
\begin{itemize}
    \item \(X^{(0,2)}\): the component `around' the cone, where $d_x^2+d^2_z> d^2_y$. In this region, $H(\bk)$ has two distinct real eigenvalues. In other words, \PT{} symmetry is satisfied by individual eigenstates. Hence this region is also called the \emph{\PT-preserving} region.
\item \(X^{(1,0)}_{+}\) and \(X^{(1,0)}_{-}\): the other two components of \(X_2\), where $d_x^2+d^2_z<d^2_y, \ d_y\lessgtr 0$.
The two eigenvalues form a single pair of non-real complex numbers that are each other's complex conjugate.
Therefore, \PT{} symmetry is spontaneously broken, and we call this region \emph{\PT-breaking}.
We denote their union by \(X^{(1,0)}\).
\end{itemize}

The zeroth homotopy set \(\pi_0\), which lists the connected components of a space, is therefore made up of these three sets, % $X^{(0,2)},X^{(1,0)}_-$ and $X^{(1,0)}_+$ together form the set $\pi_0(X_2)$, the zeroth homotopy set of the total space, 
\begin{equation}
    \pi_0(X_2)=\{X^{(0,2)},X^{(1,0)}_-,X^{(1,0)}_+\}\label{eq_2bdpi0}.
\end{equation}
Following Fig.~\ref{fig_hptiv}, when we talk about topology, we have to fix $\pi_0$ first. This exactly corresponds to fixing the extent of spontaneous symmetry breaking. To make this degree of \PT-breaking explicit, we introduce the order parameter \(m\), counting the number of pairs of complex conjugate eigenvalues, i.e. \(m=0\) for \(X^{(0,2)}\), and \(m=1\) for \(X^{(1,0)}_\pm\). This is the most important order parameter in describing the phases. 

Let us look at the topology of the subspaces with a given $m$. 
We start from $m=0$, the \PT-preserving regime. 
The homotopy groups \(\pi_0, \pi_1\) describe the connected components and nonequivalent loops in $X^{(0,2)}$, respectively. 
From Fig.~\ref{fig_2bspace} we can read off the homotopy group structure directly:
\begin{equation}
    \pi_0(X^{(0,2)})=\{1\}, \,\pi_1(X^{(0,2)})=\mathbb Z.
\end{equation}
This result can be obtained rigorously using the concept of deformation retraction \cite{hatcherAlgebraicTopology2002}. 
It means that the space \(X^{(0,2)}\) consists of just one connected component but hosts non-trivial loops. 
These loops are characterized by the number \(W\) of times they wind around the cone; exemplary loops are shown in Fig.~\ref{fig_2bspace}.

We now give a physical consequence of these homotopy invariants. The trivial $\pi_0(X^{(0,2)})$ shows that the region is connected. This
means that all zero-dimensional \PT-preserving gapped phases can be continuously deformed into each other, without closing the band gap.

The same is not true for one-dimensional gapped phases. These can be thought of as the embedding of a circle \(\mathbf{d}:S^1\to X^{(0,2)}\) in this space, and not all such loops can be deformed into one another.
The corresponding $\mathbb Z$ invariant in the fundamental group $\pi_1(X^{(0,2)})$ is precisely the winding number of $(d_x,d_z)$ around the cone.
It can be calculated by an integral:
\begin{align}\label{eq_2BandWinding}
    W=\frac{1}{2\pi i}\oint_{\mathcal C} \partial_{\bk}\ln\left[d_x(\mathbf k)+id_z(\mathbf k)\right]\cdot d\mathbf k\in \mathbb Z.
\end{align}
The integrand represents the rotation of the $\mathbf d$-vector projected in the $d_x$-$d_z$ plane. 
The resulting winding number \(W\) counts how many times the parameterized operator $H(\mathbf k)$ encircles the double cone when following the loop $\mathcal C$. 
Under a $2\pi$ rotation of $(d_x,d_z)$, or $W=1$ equivalently, the corresponding real eigenvectors undergo a $\pi$ rotation.
The winding number can thus be equivalently obtained from the eigenvector geometry.
We note that when using Eq.~\eqref{eq_2BandWinding}, we must restrict the contour $\mathcal C$ inside regions $d_x^2(\mathbf k)+d^2_z(\mathbf k)> d^2_y(\mathbf k)$, i.e., inside $X^{(0,2)}$.
The result computed outside this region will \emph{not} be an invariant, as we will see in the next subsection \ref{sc_2bdmodels}. This is in accordance with the scheme presented in Fig.~\ref{fig_hptiv}: the connected component must be fixed in the first step.
Note that this winding of the eigenvectors is different from the spectral winding number that also features prominently in non-Hermitian systems \cite{PhysRevX.8.031079}.

For the $m=0$ gapped systems, the entire BZ is chosen as the domain of integration $\mathcal C$. 
Phases with different winding numbers cannot be adiabatically deformed into each other. 
This means that such a deformation must instead leave the gapped regime and close the band gap, which amounts to a phase transition. We label these distinct phases as  \((m=0, W)\).

The winding number in Eq.~\eqref{eq_2BandWinding} is also able to classify nodal structures that can be enclosed by a loop. 
For example, an isolated nodal point at \(\bk=0\) on a two-dimensional parameter space can be encircled by a loop $(k_x,k_y)=(\cos\phi,\sin\phi)$. Integrating over $\phi\in\mathcal C = [0,2\pi]$ produces a winding number. 
If it is non-zero, the nodal point cannot be removed unless it touches another one carrying the opposite winding number. 

Note that this winding number generalizes the Hermitian gapped phase invariant for systems with chiral or \(\mathcal{C}_2\mathcal{T}\) symmetry \cite{bouhon2020non,PhysRevLett.89.077002,asboth2016short}. 
This can be seen in the Hermitian limit \(d_y\to0\), in which \PT-symmetric models become chiral symmetric and the winding number becomes the established winding number of such systems. 
The standard representation of chiral symmetry can be obtained via the basis transformation \(\exp(i\pi \sigma_x /2)\).
When non-Hermitian couplings are turned on, \(d_y\neq 0\), the winding number remains a good topological invariant, so long as the symmetry is preserved by the eigenvalues.
This scenario changes only when transitioning to the symmetry-breaking phase, where $m$ jumps from $0$ to $1$.

Now we turn to $m=1$, the \PT-breaking phase. From Fig.~\ref{fig_2bspace} we can read out the homotopy groups as
\begin{equation}
    \pi_0(X^{(1,0)})=\mathbb Z_2,\quad
    \pi_1(X^{(1,0)})=\{1\}.
\end{equation}
We see that \(X^{(1,0)}\) consists of two connected components, which we have already labeled by an index \(\nu\in\pm1\), but it only contains trivial loops.

The non-trivial \(\pi_0\) of the \PT-breaking phase distinguishes the two components lying inside the double cone. This can be used to classify zero-dimensional gapped phases where \PT{} symmetry is spontaneously broken. A deformation from one of the components to the other necessarily closes a gap.
The origin of this topological invariant is as follows.
Due to \PT symmetry, two eigenvectors must also form a complex conjugate pair, i.e., \(\ket{u}, \ket{u^\ast}\) are the eigenvectors to \(E, E^*\).
We pick the eigenvector corresponding to \(\mathrm{Im} E>0\), and decompose it as \(|u\rangle = |u_R\rangle + i |u_I\rangle \) 
where $|u_R\rangle$ and $|u_I\rangle$ are 2-dimensional real vectors.
Since $|u\rangle$ and $|u^\ast\rangle$ are linearly independent, so are $|u_R\rangle$ and $|u_I\rangle$.
The topological invariant is then the relative orientation of these vectors, and can be calculated as:
\begin{equation}
    \nu=\frac{|u_R\rangle\times |u_I\rangle}{\norm{|u_R\rangle\times\ |u_I\rangle}}\in \mathbb \pi_0(X^{(1,0)}) = \mathbb Z_2.\label{eq_2bdort}
\end{equation}
Note that a complex scalar multiplication on $\ket{u}$ induces a real linear transformation with positive determinant on $(|u_R\rangle,|u_I\rangle)$, so the orientation is independent of the gauge of $\ket{u}$.
The invariant \(\nu\) matches the sign of \(d_y\) in the space of all matrices; 
it is precisely the sign we use to label the upper and lower in-cone regions in Fig.~\ref{fig_2bspace}: $X^{(1,0)}=X^{(1,0)}_-\sqcup X^{(1,0)}_+$.
A direct transition between these two phases can only happen through the tip of the cone \(\mathbf{d}=0\), so they are joined by a singular degeneracy called a non-defective degeneracy (at which the Jordan decomposition is unstable) \cite{sayyad2022symmetryprotected}. Again, we need to remember that Eq.~\eqref{eq_2bdort} is a good topological quantity when we are in regions $d_x^2(\mathbf k)+d^2_z(\mathbf k)< d^2_y(\mathbf k)$, the symmetry-breaking regime. 
It loses its meaning in the symmetry-preserving regime.
The topological index is then most accurately written as a pair \((m=1,\nu)\).

When plotting the band structure of a general $2$-band system on the BZ, each region where $H(\bk)$ is gapped corresponds to one of these three elements in Eq.~\eqref{eq_2bdpi0}. 
They are usually separated by nodal points, lines, and surfaces in one, two, and three dimensions, respectively \cite{PhysRevB.104.L201104,PhysRevLett.127.186602,PhysRevLett.127.186601,PhysRevB.99.041406}. 
Hence we can use $\pi_0(X_2)$ to characterize these codimension-1 degeneracies, which are the boundaries between gapped regions labeled by different $\pi_0(X_2)$. 

All higher homotopy groups are trivial: $\pi_n(X^{(0,2)})=\pi_n(X^{(1,0)})=\{1\}$ for $ n\geq 2$. They do not give new topology for two-band models.

\subsection{Models: gapped phases and nodal structures}\label{sc_2bdmodels}

\begin{figure}
    \centering
    \def\svgwidth{\columnwidth}
    %--

    \begingroup%
  \makeatletter%
  \providecommand\color[2][]{%
    \errmessage{(Inkscape) Color is used for the text in Inkscape, but the package 'color.sty' is not loaded}%
    \renewcommand\color[2][]{}%
  }%
  \providecommand\transparent[1]{%
    \errmessage{(Inkscape) Transparency is used (non-zero) for the text in Inkscape, but the package 'transparent.sty' is not loaded}%
    \renewcommand\transparent[1]{}%
  }%
  \providecommand\rotatebox[2]{#2}%
  \newcommand*\fsize{\dimexpr\f@size pt\relax}%
  \newcommand*\lineheight[1]{\fontsize{\fsize}{#1\fsize}\selectfont}%
  \ifx\svgwidth\undefined%
    \setlength{\unitlength}{246bp}%
    \ifx\svgscale\undefined%
      \relax%
    \else%
      \setlength{\unitlength}{\unitlength * \real{\svgscale}}%
    \fi%
  \else%
    \setlength{\unitlength}{\svgwidth}%
  \fi%
  \global\let\svgwidth\undefined%
  \global\let\svgscale\undefined%
  \makeatother%
  \begin{picture}(1,0.75203252)%
    \lineheight{1}%
    \setlength\tabcolsep{0pt}%
    \put(0,0){\includegraphics[width=\unitlength,page=1]{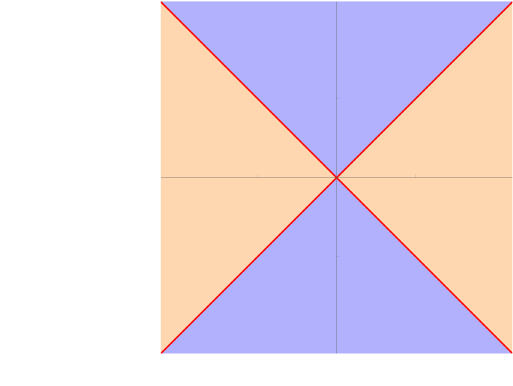}}%
    \put(0.99184772,0.42589397){\color[rgb]{0,0,0}\transparent{0.99684298}\makebox(0,0)[rt]{\lineheight{1.25}\smash{\begin{tabular}[t]{r}\(m_R\)\end{tabular}}}}%
    \put(0.66164108,0.71975491){\color[rgb]{0,0,0}\transparent{0.99684298}\makebox(0,0)[lt]{\lineheight{1.25}\smash{\begin{tabular}[t]{l}\(m_I\)\end{tabular}}}}%
    \put(0.80967308,0.42852462){\color[rgb]{0,0,0}\transparent{0.99684298}\makebox(0,0)[t]{\lineheight{1.25}\smash{\begin{tabular}[t]{c}\(1\)\end{tabular}}}}%
    \put(0.6737512,0.54989869){\color[rgb]{0,0,0}\transparent{0.99684298}\makebox(0,0)[lt]{\lineheight{1.25}\smash{\begin{tabular}[t]{l}\(1\)\end{tabular}}}}%
    \put(0,0){\includegraphics[width=\unitlength,page=2]{fig5.pdf}}%
    \put(0.79455377,0.67663299){\color[rgb]{0,0,0}\transparent{0.99684298}\makebox(0,0)[t]{\lineheight{1.25}\smash{\begin{tabular}[t]{c}\(\nu=+1\)\end{tabular}}}}%
    \put(0.79455377,0.09654466){\color[rgb]{0,0,0}\transparent{0.99684298}\makebox(0,0)[t]{\lineheight{1.25}\smash{\begin{tabular}[t]{c}\(\nu=-1\)\end{tabular}}}}%
    \put(0.33625426,0.10659801){\color[rgb]{0,0,0}\transparent{0.99684298}\rotatebox{45}{\makebox(0,0)[lt]{\lineheight{1.25}\smash{\begin{tabular}[t]{l}\PT-preserving\end{tabular}}}}}%
    \put(0.36674224,0.06759618){\color[rgb]{0,0,0}\transparent{0.99684298}\rotatebox{45}{\makebox(0,0)[lt]{\lineheight{1.25}\smash{\begin{tabular}[t]{l}\PT-breaking\end{tabular}}}}}%
    \put(0.82191387,0.52364259){\color[rgb]{0,0,0}\transparent{0.99684298}\rotatebox{45}{\makebox(0,0)[lt]{\lineheight{1.25}\smash{\begin{tabular}[t]{l}\PT-preserving\end{tabular}}}}}%
    \put(0.78938502,0.55979759){\color[rgb]{0,0,0}\transparent{0.99684298}\rotatebox{45}{\makebox(0,0)[lt]{\lineheight{1.25}\smash{\begin{tabular}[t]{l}\PT-breaking\end{tabular}}}}}%
    \put(0,0){\includegraphics[width=\unitlength,page=3]{fig5.pdf}}%
    \put(-0.00308234,0.72117192){\color[rgb]{0,0,0}\transparent{0.9940688}\makebox(0,0)[lt]{\lineheight{1.25}\smash{\begin{tabular}[t]{l}(a)\end{tabular}}}}%
    \put(0.31222715,0.72117192){\color[rgb]{0,0,0}\transparent{0.9940688}\makebox(0,0)[rt]{\lineheight{1.25}\smash{\begin{tabular}[t]{r}(b)\end{tabular}}}}%
    \put(0,0){\includegraphics[width=\unitlength,page=4]{fig5.pdf}}%
    \put(0.08032733,0.50674713){\color[rgb]{0,0,0}\transparent{0.9940688}\makebox(0,0)[rt]{\lineheight{1.25}\smash{\begin{tabular}[t]{r}\(m_I\)\end{tabular}}}}%
    \put(0.08032734,0.05420393){\color[rgb]{0,0,0}\transparent{0.9940688}\makebox(0,0)[rt]{\lineheight{1.25}\smash{\begin{tabular}[t]{r}\(-1\)\end{tabular}}}}%
    \put(0.08032733,0.39966302){\color[rgb]{0,0,0}\transparent{0.9940688}\makebox(0,0)[rt]{\lineheight{1.25}\smash{\begin{tabular}[t]{r}\(0\)\end{tabular}}}}%
    \put(0.24964384,0.02308607){\color[rgb]{0,0,0}\transparent{0.9940688}\makebox(0,0)[lt]{\lineheight{1.25}\smash{\begin{tabular}[t]{l}\(m_R\)\end{tabular}}}}%
    \put(0.11215273,0.02308607){\color[rgb]{0,0,0}\transparent{0.9940688}\makebox(0,0)[t]{\lineheight{1.25}\smash{\begin{tabular}[t]{c}\(-\frac12\)\end{tabular}}}}%
    \put(0.15655745,0.02308607){\color[rgb]{0,0,0}\transparent{0.9940688}\makebox(0,0)[t]{\lineheight{1.25}\smash{\begin{tabular}[t]{c}\(0\)\end{tabular}}}}%
    \put(0.20638895,0.02308607){\color[rgb]{0,0,0}\transparent{0.9940688}\makebox(0,0)[t]{\lineheight{1.25}\smash{\begin{tabular}[t]{c}\(\frac32\)\end{tabular}}}}%
    \put(0.08032733,0.72117192){\color[rgb]{0,0,0}\transparent{0.9940688}\makebox(0,0)[rt]{\lineheight{1.25}\smash{\begin{tabular}[t]{r}\(1\)\end{tabular}}}}%
    \put(0,0){\includegraphics[width=\unitlength,page=5]{fig5.pdf}}%
    \put(0.2524118,0.48685501){\color[rgb]{0,0,0}\transparent{0.9940688}\makebox(0,0)[lt]{\lineheight{1.25}\smash{\begin{tabular}[t]{l}\(E^2\)\end{tabular}}}}%
    \put(0.2524118,0.23965593){\color[rgb]{0,0,0}\transparent{0.9940688}\makebox(0,0)[lt]{\lineheight{1.25}\smash{\begin{tabular}[t]{l}\(-5\)\end{tabular}}}}%
    \put(0.2524118,0.55738932){\color[rgb]{0,0,0}\transparent{0.9940688}\makebox(0,0)[lt]{\lineheight{1.25}\smash{\begin{tabular}[t]{l}\(5\)\end{tabular}}}}%
  \end{picture}%
\endgroup%
    
    %--
    \caption{
   Phase diagram of the zero-dimensional model Eq.~\eqref{eq:2band-0D}.
    (a)
    Plot of the squared eigenvalues when sweeping \(m_I\) from \(-1\) to \(1\) for \(m_R\) taking fixed values \(-\frac12, 0, \frac32\).
    For the first two sweeps, the initial and final operators lie in the \PT-broken phase (blue) in which they have \(m=1\) pair of complex conjugate eigenvalues.
    The parametric sweep deforms them from the \(\nu=-1\) oriented phase to the \(\nu=+1\) phase. 
    For \(m_R=-\frac12\), the transition happens via two exceptional points shown in red, and traversing the \PT-preserving region (orange) in between them. This is the generic case.
    Tuning \(m_R=0\), the transition happens without entering the \PT-preserving phase, via a single degenerate point which is a non-defective EP at which the operator is zero \(\mathbf0_2\).
    The sweep for \(m_R=\frac32\) lies entirely in the gapped \PT-preserving phase. 
    This shows that the sign change in \(m_I\) does not correspond to a change in topological invariant \(\nu\), since this invariant is only defined in the \PT-broken region. 
    (b)
    As (a), for general \(m_R, m_I\), which produces the phase diagram of \(H^{\text{0D}}\). This is essentially a section of the cone shown in Fig.~\ref{fig_2bspace}. 
    Cyan arrows indicate the parameter sweeps in (a).
    }
    \label{fig:2band-0d}
\end{figure}

We apply the classification of two-band models derived in the previous section to explicit \PT-symmetric two-band models in zero, one, two and three dimensions. Generically,  \PT-symmetric models host nodal structures of codimension 1, meaning that there are symmetry-protected nodal points, lines, and surfaces in one, two, and three-dimensional models, respectively.

We begin with a simple zero-dimensional \PT-symmetric two-level system described by 
\begin{equation}
    H^{\text{0D}} = m_R \sigma_z + i m_I \sigma_y,
    \label{eq:2band-0D}
\end{equation}
corresponding to staggered potentials leading to a real mass term \(m_R\) and an imaginary mass term \(i m_I\) originating from orbital dependent gain and loss (of equal magnitude).  

This model is well-suited to illustrate the topology stemming from the zeroth homotopy set $\pi_0(X_2)$, the different connected components of \(X_2\): for \(\abs{m_R}>\abs{m_I}\), the operator \(H^{\text{0D}}\) lies in \(X^{(0,2)}\) and has two real eigenvalues, while for \(\abs{m_R}<\abs{m_I}\), it lies in \(X^{(1,0)}\) and has one pair of complex conjugate eigenvalues.
This space  \(X^{(1,0)}\) is divided into regions \( m_I\lessgtr0\), which are distinguished by the orientation invariant \(\nu\in\Z_2\) given in Eq.~\eqref{eq_2bdort}. Restricted to \(\abs{m_R}<\abs{m_I}\) in this model $\nu$ is equal to \(\operatorname{sign}(m_I)\).

The role of the orientation invariant becomes relevant when, starting from a phase with $\nu=-1$ and \(\abs{m_R}<\abs{m_I}\), we tune the model to $\nu=+1,\abs{m_R}<\abs{m_I}$. 
The change in \(\nu\) along this deformation is necessarily accompanied by a closing of the band gap.
This can happen in two slightly different ways, which we show in Fig.~\ref{fig:2band-0d} by sweeping $m_I$ for different fixed values of $m_R$. In the first way, the operator crosses the border between the symmetry-breaking phase and the symmetry-preserving phase, where an exceptional point occurs. 
The sign change of $\nu$ is accompanied by entering the regime  \(\abs{m_R}>\abs{m_I}\) where $\nu$ is not defined (see remarks under Eq.~\eqref{eq_2bdort} and Fig.~\ref{fig:2band-0d}). 
This is the generic shape of such a deformation.
One may also tune the deformation to take the operator directly from the sector \(\nu=-1\) to \(\nu=+1\) without crossing through the \PT-preserving sector.
This happens only by deforming through \(H^{\text{0D}}=\mathbf0_2\),  where this operator constitutes a non-defective exceptional point. 
We reiterate that a deformation between these two regions with different topological invariants is associated with a closing of the band gap in either case.

\begin{figure*}[htbp]
    \centering
    \def\svgwidth{\linewidth}
    %--
    
    \begingroup%
  \makeatletter%
  \providecommand\color[2][]{%
    \errmessage{(Inkscape) Color is used for the text in Inkscape, but the package 'color.sty' is not loaded}%
    \renewcommand\color[2][]{}%
  }%
  \providecommand\transparent[1]{%
    \errmessage{(Inkscape) Transparency is used (non-zero) for the text in Inkscape, but the package 'transparent.sty' is not loaded}%
    \renewcommand\transparent[1]{}%
  }%
  \providecommand\rotatebox[2]{#2}%
  \newcommand*\fsize{\dimexpr\f@size pt\relax}%
  \newcommand*\lineheight[1]{\fontsize{\fsize}{#1\fsize}\selectfont}%
  \ifx\svgwidth\undefined%
    \setlength{\unitlength}{510bp}%
    \ifx\svgscale\undefined%
      \relax%
    \else%
      \setlength{\unitlength}{\unitlength * \real{\svgscale}}%
    \fi%
  \else%
    \setlength{\unitlength}{\svgwidth}%
  \fi%
  \global\let\svgwidth\undefined%
  \global\let\svgscale\undefined%
  \makeatother%
  \begin{picture}(1,0.48235294)%
    \lineheight{1}%
    \setlength\tabcolsep{0pt}%
    \put(0,0){\includegraphics[width=\unitlength,page=1]{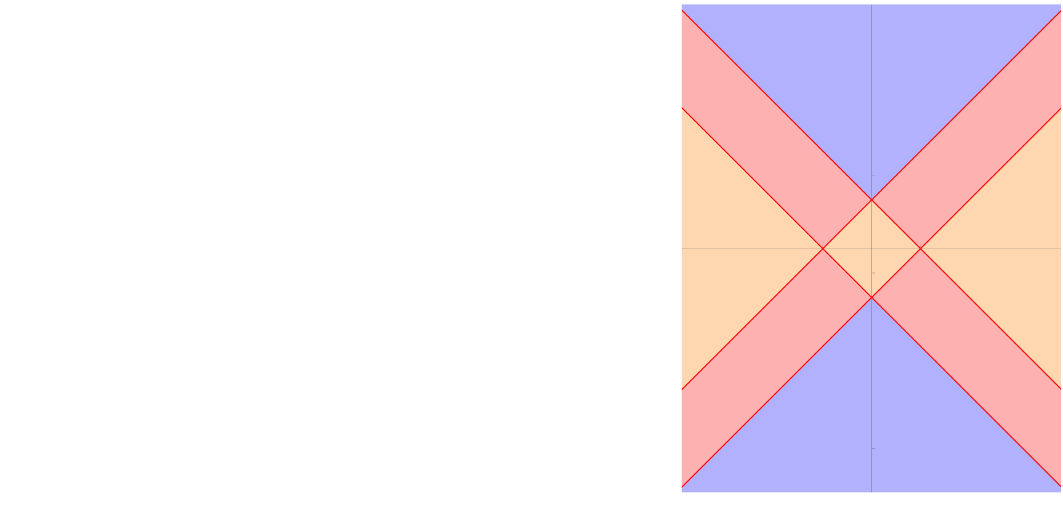}}%
    \put(0.99753332,0.23115813){\color[rgb]{0,0,0}\transparent{0.99684298}\makebox(0,0)[rt]{\lineheight{1.25}\smash{\begin{tabular}[t]{r}\(m_R\)\end{tabular}}}}%
    \put(0.82400302,0.46341127){\color[rgb]{0,0,0}\transparent{0.99684298}\makebox(0,0)[lt]{\lineheight{1.25}\smash{\begin{tabular}[t]{l}\(m_I\)\end{tabular}}}}%
    \put(0.82698529,0.28939054){\color[rgb]{0,0,0}\transparent{0.99684298}\makebox(0,0)[lt]{\lineheight{1.25}\smash{\begin{tabular}[t]{l}\(t\)\end{tabular}}}}%
    \put(0.86574216,0.25647958){\color[rgb]{0,0,0}\transparent{0.99684298}\makebox(0,0)[t]{\lineheight{1.25}\smash{\begin{tabular}[t]{c}\(t\)\end{tabular}}}}%
    \put(0,0){\includegraphics[width=\unitlength,page=2]{fig6.pdf}}%
    \put(0.64450492,0.24913755){\color[rgb]{0,0,0}\transparent{0.99684298}\makebox(0,0)[lt]{\lineheight{1.25}\smash{\begin{tabular}[t]{l}\(W=0\)\end{tabular}}}}%
    \put(0.90773998,0.24913755){\color[rgb]{0,0,0}\transparent{0.99684298}\makebox(0,0)[lt]{\lineheight{1.25}\smash{\begin{tabular}[t]{l}\(W=0\)\end{tabular}}}}%
    \put(0.7597408,0.43167549){\color[rgb]{0,0,0}\transparent{0.99684298}\makebox(0,0)[t]{\lineheight{1.25}\smash{\begin{tabular}[t]{c}\(\nu=+1\)\end{tabular}}}}%
    \put(0.7597408,0.04638141){\color[rgb]{0,0,0}\transparent{0.99684298}\makebox(0,0)[t]{\lineheight{1.25}\smash{\begin{tabular}[t]{c}\(\nu=-1\)\end{tabular}}}}%
    \put(0,0){\includegraphics[width=\unitlength,page=3]{fig6.pdf}}%
    \put(0.81977336,0.24913755){\color[rgb]{0,0,0}\transparent{0.99684298}\makebox(0,0)[t]{\lineheight{1.25}\smash{\begin{tabular}[t]{c}\(W=1\)\end{tabular}}}}%
    \put(0,0){\includegraphics[width=\unitlength,page=4]{fig6.pdf}}%
    \put(0.52543111,0.33823534){\color[rgb]{0,0,0}\transparent{0.99684298}\makebox(0,0)[rt]{\lineheight{1.25}\smash{\begin{tabular}[t]{r}\(m_I=1.5\)\end{tabular}}}}%
    \put(0.52543111,0.18528974){\color[rgb]{0,0,0}\transparent{0.99684298}\makebox(0,0)[rt]{\lineheight{1.25}\smash{\begin{tabular}[t]{r}\(m_I=-0.5\)\end{tabular}}}}%
    \put(0,0){\includegraphics[width=\unitlength,page=5]{fig6.pdf}}%
    \put(0.02236651,0.10195286){\color[rgb]{0,0,0}\transparent{0.99684298}\makebox(0,0)[rt]{\lineheight{1.25}\smash{\begin{tabular}[t]{r}\(-4\)\end{tabular}}}}%
    \put(0.01876418,0.33571147){\color[rgb]{0,0,0}\transparent{0.99684298}\makebox(0,0)[rt]{\lineheight{1.25}\smash{\begin{tabular}[t]{r}\(4\)\end{tabular}}}}%
    \put(0.02150977,0.21633193){\color[rgb]{0,0,0}\transparent{0.99684298}\makebox(0,0)[rt]{\lineheight{1.25}\smash{\begin{tabular}[t]{r}\(0\)\end{tabular}}}}%
    \put(0.04272784,0.01102987){\color[rgb]{0,0,0}\transparent{0.99684298}\makebox(0,0)[t]{\lineheight{1.25}\smash{\begin{tabular}[t]{c}\(-4\)\end{tabular}}}}%
    \put(0.25096829,0.00613929){\color[rgb]{0,0,0}\transparent{0.99684298}\makebox(0,0)[t]{\lineheight{1.25}\smash{\begin{tabular}[t]{c}\(4\)\end{tabular}}}}%
    \put(0.28637038,0.10841074){\color[rgb]{0,0,0}\transparent{0.99684298}\makebox(0,0)[lt]{\lineheight{1.25}\smash{\begin{tabular}[t]{l}\(4\)\end{tabular}}}}%
    \put(0.14914388,0.00828082){\color[rgb]{0,0,0}\transparent{0.99684298}\makebox(0,0)[t]{\lineheight{1.25}\smash{\begin{tabular}[t]{c}\(0\)\end{tabular}}}}%
    \put(0.38544498,0.00882358){\color[rgb]{0,0,0}\transparent{0.99684298}\makebox(0,0)[t]{\lineheight{1.25}\smash{\begin{tabular}[t]{c}\(0\)\end{tabular}}}}%
    \put(0.45970968,0.00882358){\color[rgb]{0,0,0}\transparent{0.99684298}\makebox(0,0)[t]{\lineheight{1.25}\smash{\begin{tabular}[t]{c}\(m_R\)\end{tabular}}}}%
    \put(0.53397438,0.00882358){\color[rgb]{0,0,0}\transparent{0.99684298}\makebox(0,0)[t]{\lineheight{1.25}\smash{\begin{tabular}[t]{c}\(3\)\end{tabular}}}}%
    \put(0.38134346,0.02959079){\color[rgb]{0,0,0}\transparent{0.99684298}\makebox(0,0)[rt]{\lineheight{1.25}\smash{\begin{tabular}[t]{r}\(0\)\end{tabular}}}}%
    \put(0.38134346,0.09366936){\color[rgb]{0,0,0}\transparent{0.99684298}\makebox(0,0)[rt]{\lineheight{1.25}\smash{\begin{tabular}[t]{r}\(k\)\end{tabular}}}}%
    \put(0.38134346,0.15774795){\color[rgb]{0,0,0}\transparent{0.99684298}\makebox(0,0)[rt]{\lineheight{1.25}\smash{\begin{tabular}[t]{r}\(2\pi\)\end{tabular}}}}%
    \put(0.38134346,0.18213699){\color[rgb]{0,0,0}\transparent{0.99684298}\makebox(0,0)[rt]{\lineheight{1.25}\smash{\begin{tabular}[t]{r}\(0\)\end{tabular}}}}%
    \put(0.38134346,0.2434628){\color[rgb]{0,0,0}\transparent{0.99684298}\makebox(0,0)[rt]{\lineheight{1.25}\smash{\begin{tabular}[t]{r}\(k\)\end{tabular}}}}%
    \put(0.38134346,0.31029415){\color[rgb]{0,0,0}\transparent{0.99684298}\makebox(0,0)[rt]{\lineheight{1.25}\smash{\begin{tabular}[t]{r}\(2\pi\)\end{tabular}}}}%
    \put(0.38134346,0.33507817){\color[rgb]{0,0,0}\transparent{0.99684298}\makebox(0,0)[rt]{\lineheight{1.25}\smash{\begin{tabular}[t]{r}\(0\)\end{tabular}}}}%
    \put(0.38134346,0.39915673){\color[rgb]{0,0,0}\transparent{0.99684298}\makebox(0,0)[rt]{\lineheight{1.25}\smash{\begin{tabular}[t]{r}\(k\)\end{tabular}}}}%
    \put(0.38134346,0.46323532){\color[rgb]{0,0,0}\transparent{0.99684298}\makebox(0,0)[rt]{\lineheight{1.25}\smash{\begin{tabular}[t]{r}\(2\pi\)\end{tabular}}}}%
    \put(0.27731977,0.02765373){\color[rgb]{0,0,0}\transparent{0.99684298}\makebox(0,0)[lt]{\lineheight{1.25}\smash{\begin{tabular}[t]{l}\(-4\)\end{tabular}}}}%
    \put(0.28333801,0.07074419){\color[rgb]{0,0,0}\transparent{0.99684298}\makebox(0,0)[lt]{\lineheight{1.25}\smash{\begin{tabular}[t]{l}\(0\)\end{tabular}}}}%
    \put(0.01251095,0.46647634){\color[rgb]{0,0,0}\transparent{0.99684298}\makebox(0,0)[t]{\lineheight{1.25}\smash{\begin{tabular}[t]{c}(a)\end{tabular}}}}%
    \put(0.16101289,0.3833931){\color[rgb]{0,0,0}\transparent{0.99684298}\makebox(0,0)[t]{\lineheight{1.25}\smash{\begin{tabular}[t]{c}\(W=1\)\end{tabular}}}}%
    \put(0.23739255,0.3833931){\color[rgb]{0,0,0}\transparent{0.99684298}\makebox(0,0)[t]{\lineheight{1.25}\smash{\begin{tabular}[t]{c}\(W=0\)\end{tabular}}}}%
    \put(0.26606516,0.05562251){\color[rgb]{0,0,0}\transparent{0.99684298}\makebox(0,0)[rt]{\lineheight{1.25}\smash{\begin{tabular}[t]{r}\(d_x\)\end{tabular}}}}%
    \put(0.1197253,0.04117652){\color[rgb]{0,0,0}\transparent{0.99684298}\makebox(0,0)[rt]{\lineheight{1.25}\smash{\begin{tabular}[t]{r}\(d_z\)\end{tabular}}}}%
    \put(0.03872546,0.27058828){\color[rgb]{0,0,0}\transparent{0.99684298}\makebox(0,0)[lt]{\lineheight{1.25}\smash{\begin{tabular}[t]{l}\(d_y\)\end{tabular}}}}%
    \put(0.3428463,0.46653184){\color[rgb]{0,0,0}\transparent{0.99684298}\makebox(0,0)[rt]{\lineheight{1.25}\smash{\begin{tabular}[t]{r}(b)\end{tabular}}}}%
    \put(0.3428463,0.31029415){\color[rgb]{0,0,0}\transparent{0.99684298}\makebox(0,0)[rt]{\lineheight{1.25}\smash{\begin{tabular}[t]{r}(c)\end{tabular}}}}%
    \put(0.3428463,0.15774795){\color[rgb]{0,0,0}\transparent{0.99684298}\makebox(0,0)[rt]{\lineheight{1.25}\smash{\begin{tabular}[t]{r}(d)\end{tabular}}}}%
    \put(0,0){\includegraphics[width=\unitlength,page=6]{fig6.pdf}}%
    \put(0.62957069,0.46653184){\color[rgb]{0,0,0}\transparent{0.99684298}\makebox(0,0)[rt]{\lineheight{1.25}\smash{\begin{tabular}[t]{r}(e)\end{tabular}}}}%
    \put(0.56949025,0.46738339){\color[rgb]{0,0,0}\transparent{0.99684298}\makebox(0,0)[lt]{\lineheight{1.25}\smash{\begin{tabular}[t]{l}\(10\)\end{tabular}}}}%
    \put(0.56949025,0.24705887){\color[rgb]{0,0,0}\transparent{0.99684298}\makebox(0,0)[lt]{\lineheight{1.25}\smash{\begin{tabular}[t]{l}\(0\)\end{tabular}}}}%
    \put(0.64931182,0.13163809){\color[rgb]{0,0,0}\transparent{0.99684298}\rotatebox{45.00000008}{\makebox(0,0)[lt]{\lineheight{1.25}\smash{\begin{tabular}[t]{l}\PT-preserving\end{tabular}}}}}%
    \put(0.70544798,0.06169606){\color[rgb]{0,0,0}\transparent{0.99684298}\rotatebox{45.00000008}{\makebox(0,0)[lt]{\lineheight{1.25}\smash{\begin{tabular}[t]{l}\PT-breaking\end{tabular}}}}}%
    \put(0.67737989,0.09666709){\color[rgb]{0,0,0}\transparent{0.99684298}\rotatebox{45.00000008}{\makebox(0,0)[lt]{\lineheight{1.25}\smash{\begin{tabular}[t]{l}nodal\end{tabular}}}}}%
    \put(0.56949025,0.02056549){\color[rgb]{0,0,0}\transparent{0.99684298}\makebox(0,0)[lt]{\lineheight{1.25}\smash{\begin{tabular}[t]{l}\(-10\)\end{tabular}}}}%
    \put(0.56949025,0.34986819){\color[rgb]{0,0,0}\transparent{0.99684298}\makebox(0,0)[lt]{\lineheight{1.25}\smash{\begin{tabular}[t]{l}\(E^2\)\end{tabular}}}}%
    \put(0,0){\includegraphics[width=\unitlength,page=7]{fig6.pdf}}%
    \put(0.52543111,0.03186129){\color[rgb]{0,0,0}\transparent{0.99684298}\makebox(0,0)[rt]{\lineheight{1.25}\smash{\begin{tabular}[t]{r}\(m_I=-4.1\)\end{tabular}}}}%
  \end{picture}%
\endgroup%
    
    %--
    \caption{
    Illustration on the one-dimensional model Eq.~\eqref{eq:2band-1D} for fixed \(t=1\).
    (a)
    For fixed \(t,m_I,m_R\), the model corresponds to a loop \(\mathbf{d}(k)\in\R^3\).
    The winding number according to Eq.~\eqref{eq_2BandWinding} is \(1\) for \(\abs{m_R}<\abs{t}\) and \(0\) otherwise.
    As a topological invariant it is defined only in the region outside the double cone.
    Each pair of rings corresponds to the start and endpoint of a parametric sweep of \(m_R\) between \(0\) and \(3\), shown for \(m_I\) fixed at \(1.5,-0.5,-4.1\) shown in green, purple, and brown, respectively.
    The second sweep starts and ends in the \PT-preserved region, and deforms the operator from the \(W=1\) to the \(W=0\) phase.
    The first and third sweeps start in the \PT-broken phase, where the winding number is not defined. 
    A change of the integral in Eq.~\eqref{eq_2BandWinding} need thus not accompany a gap closing.
    (b)-(d) Squared eigenvalues $E^2$ of the model in Eq.~\eqref{eq:2band-1D} for the deformations discussed in (a), for (b) $m_I=1.5$, (c) $m_I=-0.5$, and (d) $m_I=-4.1$. 
    Blue (orange) identifies a gap between two complex conjugate (real) eigenvalues. Red signifies a gap closing.
    The solid (dashed) lines at $m_R=0 (3)$ correspond to the solid (dashed) loops in (c). 
    (b) The system transitions from the \PT-breaking to the \PT-preserving phase via a pair-creation of exceptional points that move through parameter space before annihilating pairwise.
    (c) The system transitions from the gapped \PT-preserving phase to itself, with a change in winding number \(W=1\to W=0\). The phase transition happens via pair-creation and annihilation of exceptional points that is local in parameter space.
    (d) The system remains gapped in the \PT-breaking phase when increasing \(m_R\) to \(3\) at \(m_I=-4.1\). 
    (e)
    The phase diagram of the one-dimensional model for general values of the mass parameters \(m_R,m_I\). 
    Blue (orange) signifies the model is band-gapped in the \PT-broken (\PT-preserving) phase, which is further divided into regions with different orientation \(\nu\) (different winding \(W\)).
    Light red signifies the nodal region in which the band gap closes at two exceptional points (EPs). In its dark red boundary the two EPs merge into a single degeneracy and can pairwise annihilate.
    At the boundary intersection demarcated by red markers, the single degeneracy is a non-defective exceptional point. 
    At the boundary intersection without markers, the band gap closes for the entire band simultaneously.
    The colored arrows correspond to the parameter sweeps discussed in (a)-(d).
    }
    \label{fig:2band-1D-mockup}
\end{figure*}

We extend this model into one dimension and obtain periodic operator
\begin{align}
    H^{\text{1D}}(k) = t\cos(k)\sigma_x + \left[t\sin(k)+m_R\right]\sigma_z+ im_{I} \sigma_y,
    \label{eq:2band-1D}
\end{align}
which can be thought of as the tight-binding Bloch Hamiltonian of a one-dimensional chain with a two-level system \(H^{\text{0D}}\) per unit cell. 
The unit cells are coupled by Hermitian nearest neighbor hopping with strength \(t\). 
Mathematically this model may be thought of as an embedding of a circle \(S^1\) into the space of $2\times 2$ \PT-symmetric operators.

In the following, we probe relevant phases of this model by varying the mass terms \(m_R, m_I \in\R\). With \(m_R=0\) and increasing \(m_I\) transitions the system between the different gapped phases when \(m_I=\pm t\), starting from \(X^{(1,0)}_-\), through \(X^{(0,2)}\) to \(X^{(1,0)}_+\).
Explicitly, the \(\pi_0\)-invariant \(\nu\in\Z_2\) can be calculated and is \(\nu=-1\) for \(m_I<-t\), \(\nu=+1\) for \(m_I>+t\).
This can be seen in Fig.~\ref{fig:2band-1D-mockup} (a).
In the phase transitions at \(m_I=\pm t\) the band gap closes at every \(k\) simultaneously. 
This full-dimensional nodal structure is a result of tuning \(m_R\) to precisely zero, for generic values the system will instead show nodal points moving through the parameter space.

A winding number can be calculated in this system following Eq.~\eqref{eq_2BandWinding}.
It protects the band gap topologically in the regime $|m_I|<t$, while the protection is lifted in the regime $|m_I|>t$.
The two regimes correspond to sectors of different spontaneous symmetry breaking and are separated by gap closing at all $k$.
To see this, we show the transitions between different winding numbers in these two sectors in Fig.~\ref{fig:2band-1D-mockup}.
In the \PT-preserving sector, the winding number can only be changed with the accompany of band gap closings. The implication of this winding number is that operators \(H(k)\) with a non-zero winding number in the  \PT-preserving sector cannot be contracted to a constant \(H(k_0)\) without closing the band gap, since operators without dependence on \(k\) automatically carry zero winding number.
Meanwhile in the \PT-breaking sector, although the winding number can be calculated as well, it may change without closing the band gap, as seen from the lowest loops in Fig.~\ref{fig:2band-1D-mockup} (a).

The results of this model tell that the winding number still serves as a topological invariant protecting the gap, but the protection depends on $\pi_0$ given in Eq.~\eqref{eq_2bdpi0}.
Its protection extends from the pure Hermitian model to the entire \PT-symmetry preserving sector. So the Hermitian topology has a certain robustness to non-Hermiticity.
Only after an exceptional transition that converts the whole system to the \PT-symmetry breaking sector, this protection is lifted. In return, the  \PT-symmetry breaking sector is enriched with another topological index, $\nu\in\mathbb Z_2$ as in the zero-dimensional model $H^{\text{0D}}$.

The classification of nodal structures is best showcased in a two-dimensional model, which generically features nodal lines. We can obtain one such model by dimensional extension: parameterizing the mass terms in Eq.~\eqref{eq:2band-1D} as \(m_R\to m_R+t_y\sin(k_y), m_I\to m_I+t_x\cos(k_y)\) results in
\begin{align}
    \notag
    H^{\text{2D}}(\bk) =& 
        t_x\cos(k_x)\sigma_x +
        i \left[t_y\cos(k_y)+m_I\right] \sigma_y 
        \\&
        + \left[t_x\sin(k_x)+t_y\sin(k_y)+m_R\right]\sigma_z,
        \label{eq:2bandmodel-2d}
\end{align}
describing a tight-binding model on a square lattice, consisting of chains described by \(H^{(1D)}\) in \(x\)-direction together with inter-cell hopping of strength \( t_y\) in \(y\)-direction.  
This means that the one-dimensional model can be thought of as a slice of constant \(k_y\) of this two-dimensional model;
\(H^{\text{1D}}(k) = H^{\text{2D}}(k=k_x,k_y=0)\), allowing for a shift in the mass terms.
To allow for a gapped phase with winding number \(1\) in \(k_x\)-direction, we must choose the hopping strength \(t_y<t_x\).

\begin{figure}
    \centering
    \def\svgwidth{\columnwidth}
    %--
    
    \begingroup%
  \makeatletter%
  \providecommand\color[2][]{%
    \errmessage{(Inkscape) Color is used for the text in Inkscape, but the package 'color.sty' is not loaded}%
    \renewcommand\color[2][]{}%
  }%
  \providecommand\transparent[1]{%
    \errmessage{(Inkscape) Transparency is used (non-zero) for the text in Inkscape, but the package 'transparent.sty' is not loaded}%
    \renewcommand\transparent[1]{}%
  }%
  \providecommand\rotatebox[2]{#2}%
  \newcommand*\fsize{\dimexpr\f@size pt\relax}%
  \newcommand*\lineheight[1]{\fontsize{\fsize}{#1\fsize}\selectfont}%
  \ifx\svgwidth\undefined%
    \setlength{\unitlength}{246bp}%
    \ifx\svgscale\undefined%
      \relax%
    \else%
      \setlength{\unitlength}{\unitlength * \real{\svgscale}}%
    \fi%
  \else%
    \setlength{\unitlength}{\svgwidth}%
  \fi%
  \global\let\svgwidth\undefined%
  \global\let\svgscale\undefined%
  \makeatother%
  \begin{picture}(1,0.81300813)%
    \lineheight{1}%
    \setlength\tabcolsep{0pt}%
    \put(0,0){\includegraphics[width=\unitlength,page=1]{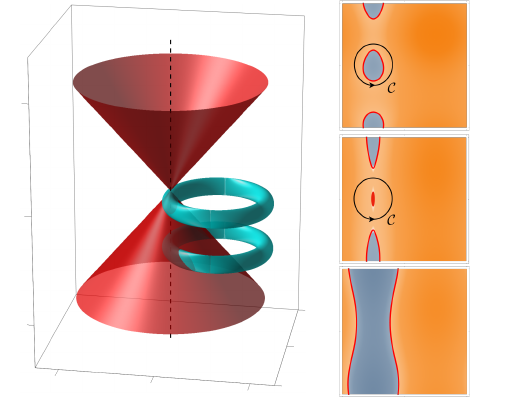}}%
    \put(0.89857665,0.578252){\color[rgb]{0,0,0}\transparent{0.99684298}\makebox(0,0)[rt]{\lineheight{1.25}\smash{\begin{tabular}[t]{r}\(m_I=0\)\end{tabular}}}}%
    \put(0.89857664,0.06045995){\color[rgb]{0,0,0}\transparent{0.99684298}\makebox(0,0)[rt]{\lineheight{1.25}\smash{\begin{tabular}[t]{r}\(m_I=-1\)\end{tabular}}}}%
    \put(0.89857665,0.3221544){\color[rgb]{0,0,0}\transparent{0.99684298}\makebox(0,0)[rt]{\lineheight{1.25}\smash{\begin{tabular}[t]{r}\(m_I=-\frac16\)\end{tabular}}}}%
    \put(0.04781547,0.17082882){\color[rgb]{0,0,0}\transparent{0.99684298}\makebox(0,0)[rt]{\lineheight{1.25}\smash{\begin{tabular}[t]{r}\(-2\)\end{tabular}}}}%
    \put(0.03914874,0.59971406){\color[rgb]{0,0,0}\transparent{0.99684298}\makebox(0,0)[rt]{\lineheight{1.25}\smash{\begin{tabular}[t]{r}\(2\)\end{tabular}}}}%
    \put(0.04290541,0.38186732){\color[rgb]{0,0,0}\transparent{0.99684298}\makebox(0,0)[rt]{\lineheight{1.25}\smash{\begin{tabular}[t]{r}\(0\)\end{tabular}}}}%
    \put(0.1032907,0.05129949){\color[rgb]{0,0,0}\transparent{0.99684298}\makebox(0,0)[t]{\lineheight{1.25}\smash{\begin{tabular}[t]{c}\(-2\)\end{tabular}}}}%
    \put(0.48354864,0.01999582){\color[rgb]{0,0,0}\transparent{0.99684298}\makebox(0,0)[t]{\lineheight{1.25}\smash{\begin{tabular}[t]{c}\(2\)\end{tabular}}}}%
    \put(0.59410476,0.19212872){\color[rgb]{0,0,0}\transparent{0.99684298}\makebox(0,0)[lt]{\lineheight{1.25}\smash{\begin{tabular}[t]{l}\(2\)\end{tabular}}}}%
    \put(0.29162922,0.03583504){\color[rgb]{0,0,0}\transparent{0.99684298}\makebox(0,0)[t]{\lineheight{1.25}\smash{\begin{tabular}[t]{c}\(0\)\end{tabular}}}}%
    \put(0.66745649,0.57170676){\color[rgb]{0,0,0}\transparent{0.99684298}\makebox(0,0)[rt]{\lineheight{1.25}\smash{\begin{tabular}[t]{r}\(-\pi\)\end{tabular}}}}%
    \put(0.66745648,0.67406482){\color[rgb]{0,0,0}\transparent{0.99684298}\makebox(0,0)[rt]{\lineheight{1.25}\smash{\begin{tabular}[t]{r}\(k_y\)\end{tabular}}}}%
    \put(0.66745649,0.77642288){\color[rgb]{0,0,0}\transparent{0.99684298}\makebox(0,0)[rt]{\lineheight{1.25}\smash{\begin{tabular}[t]{r}\(\pi\)\end{tabular}}}}%
    \put(0.66745649,0.30951161){\color[rgb]{0,0,0}\transparent{0.99684298}\makebox(0,0)[rt]{\lineheight{1.25}\smash{\begin{tabular}[t]{r}\(-\pi\)\end{tabular}}}}%
    \put(0.66745648,0.41186966){\color[rgb]{0,0,0}\transparent{0.99684298}\makebox(0,0)[rt]{\lineheight{1.25}\smash{\begin{tabular}[t]{r}\(k_y\)\end{tabular}}}}%
    \put(0.66745649,0.51422773){\color[rgb]{0,0,0}\transparent{0.99684298}\makebox(0,0)[rt]{\lineheight{1.25}\smash{\begin{tabular}[t]{r}\(\pi\)\end{tabular}}}}%
    \put(0.66745649,0.05036515){\color[rgb]{0,0,0}\transparent{0.99684298}\makebox(0,0)[rt]{\lineheight{1.25}\smash{\begin{tabular}[t]{r}\(-\pi\)\end{tabular}}}}%
    \put(0.66745648,0.15272329){\color[rgb]{0,0,0}\transparent{0.99684298}\makebox(0,0)[rt]{\lineheight{1.25}\smash{\begin{tabular}[t]{r}\(k_y\)\end{tabular}}}}%
    \put(0.66922622,0.0061808){\color[rgb]{0,0,0}\transparent{0.99684298}\makebox(0,0)[t]{\lineheight{1.25}\smash{\begin{tabular}[t]{c}\(-\pi\)\end{tabular}}}}%
    \put(0.81432781,0.0061808){\color[rgb]{0,0,0}\transparent{0.99684298}\makebox(0,0)[t]{\lineheight{1.25}\smash{\begin{tabular}[t]{c}\(k_x\)\end{tabular}}}}%
    \put(0.91091902,0.0061808){\color[rgb]{0,0,0}\transparent{0.99684298}\makebox(0,0)[t]{\lineheight{1.25}\smash{\begin{tabular}[t]{c}\(\pi\)\end{tabular}}}}%
    \put(0.66745649,0.25508142){\color[rgb]{0,0,0}\transparent{0.99684298}\makebox(0,0)[rt]{\lineheight{1.25}\smash{\begin{tabular}[t]{r}\(\pi\)\end{tabular}}}}%
    \put(0.54464796,0.04860917){\color[rgb]{0,0,0}\transparent{0.99684298}\makebox(0,0)[lt]{\lineheight{1.25}\smash{\begin{tabular}[t]{l}\(-2\)\end{tabular}}}}%
    \put(0.57450492,0.12505768){\color[rgb]{0,0,0}\transparent{0.99684298}\makebox(0,0)[lt]{\lineheight{1.25}\smash{\begin{tabular}[t]{l}\(0\)\end{tabular}}}}%
    \put(0.0014591,0.78213435){\color[rgb]{0,0,0}\transparent{0.99684298}\makebox(0,0)[lt]{\lineheight{1.25}\smash{\begin{tabular}[t]{l}(a)\end{tabular}}}}%
    \put(0.562652,0.15839878){\color[rgb]{0,0,0}\transparent{0.99684298}\makebox(0,0)[rt]{\lineheight{1.25}\smash{\begin{tabular}[t]{r}\(d_x\)\end{tabular}}}}%
    \put(0.31434582,0.08688651){\color[rgb]{0,0,0}\transparent{0.99684298}\makebox(0,0)[lt]{\lineheight{1.25}\smash{\begin{tabular}[t]{l}\(d_z\)\end{tabular}}}}%
    \put(0.05849914,0.50184923){\color[rgb]{0,0,0}\transparent{0.99684298}\makebox(0,0)[rt]{\lineheight{1.25}\smash{\begin{tabular}[t]{r}\(d_y\)\end{tabular}}}}%
    \put(0,0){\includegraphics[width=\unitlength,page=2]{fig7.pdf}}%
    \put(0.95598303,0.77917537){\color[rgb]{0,0,0}\transparent{0.99684298}\makebox(0,0)[lt]{\lineheight{1.25}\smash{\begin{tabular}[t]{l}\(15\)\end{tabular}}}}%
    \put(0.95598303,0.41386858){\color[rgb]{0,0,0}\transparent{0.99684298}\makebox(0,0)[lt]{\lineheight{1.25}\smash{\begin{tabular}[t]{l}\(0\)\end{tabular}}}}%
    \put(0.95598303,0.04796811){\color[rgb]{0,0,0}\transparent{0.99684298}\makebox(0,0)[lt]{\lineheight{1.25}\smash{\begin{tabular}[t]{l}\(-15\)\end{tabular}}}}%
    \put(0.95598303,0.62700968){\color[rgb]{0,0,0}\transparent{0.99684298}\makebox(0,0)[lt]{\lineheight{1.25}\smash{\begin{tabular}[t]{l}\(E^2\)\end{tabular}}}}%
    \put(0.64176524,0.77874217){\color[rgb]{0,0,0}\transparent{0.99684298}\makebox(0,0)[rt]{\lineheight{1.25}\smash{\begin{tabular}[t]{r}(b)\end{tabular}}}}%
    \put(0.64176524,0.51581065){\color[rgb]{0,0,0}\transparent{0.99684298}\makebox(0,0)[rt]{\lineheight{1.25}\smash{\begin{tabular}[t]{r}(c)\end{tabular}}}}%
    \put(0.64176524,0.25828225){\color[rgb]{0,0,0}\transparent{0.99684298}\makebox(0,0)[rt]{\lineheight{1.25}\smash{\begin{tabular}[t]{r}(d)\end{tabular}}}}%
    \put(0,0){\includegraphics[width=\unitlength,page=3]{fig7.pdf}}%
  \end{picture}%
\endgroup%

%--
\caption{
    The model in Eq.~\eqref{eq:2bandmodel-2d} illustrates the $\pi_1$-protection of nodal loops in two-dimensional systems. 
    (a) 
    For two-dimensional systems, $\mathbf{d}(\bk)$ embeds a torus instead of a circle. 
    Plots of $\mathbf{d}(\bk)$ for $m_R=1$, $t_x=1$, $t_y=\frac{1}{6}$ and $m_I=-1 (-\frac{1}{6})$ for the lower (upper) torus. 
    The band gap closes where the torus intersects the double cone. 
    This intersection changes with increasing $m_I$.
    When parts of the torus extend around the cone, there is a contour with winding number, which protects the existence of a contained nodal structure.
    (b)-(d) 
    Illustrations of the squared eigenvalues $E^2$ of the model in Eq.~\eqref{eq:2bandmodel-2d}, with nodal loops displayed in red. 
    (b) A contour \(\mathcal C\) around a nodal loop carries winding number \(W=1\).
    (c) Under deformation of the model in (b), the nodal loop contracts to a point, but cannot vanish, since it is protected by a winding number along \(\mathcal C\).
    (c) and (d) comprise the spectrum of the upper and lower torus in (a), respectively. 
    }
    \label{fig:2band-2D}
\end{figure}

As for the one-dimensional model, \(m_I\) interpolates between the three components of \(X_2\) for zero real mass, and \(m_R\) interpolates between winding number zero and one along the \(k_x \in[0,2\pi]\)-loop.
Similarly, the winding number occurs as an identifier of distinct gapped phases.
Meanwhile, in two dimensions it also protects localized nodal structures of codimension \(1\).
Setting, for example, \(m_R=t_x=1\) and sweeping \(m_I\) through \([-2,2]\) a pair of nodal loops occur, which we show in Fig.~\ref{fig:2band-2D}.
These structures carry \(\pi_0\) charge, meaning they are interfaces between \PT-preserving and \PT-breaking phases.
Furthermore, they carry non-trivial \(\pi_1\)-charge, since they can be encircled by a loop with a non-trivial winding number. 
They are therefore protected and robust in two dimensions which becomes apparent when, by increasing the imaginary mass, they shrink to nodal points but do not vanish, due to the non-zero winding number.
We illustrate this process in Fig.~\ref{fig:2band-2D}.
This model showcases that \(\pi_1\)-invariants are relevant to nodal structures of codimension \(1\) if they are local in the parameter space.
This means that in a two-dimensional periodic parameter space, there exist different kinds of stable nodal rings, namely isolated loops local in parameter space (which must be protected by \(\pi_1\)), as well as extended loops (which in this case carry only \(\pi_0\) charges). 

As an Abelian invariant, the total winding numbers over a torus must sum to zero. Thus, the non-zero winding numbers calculated on localized loops satisfy a no-go sum rule in the vein of a Nielsen-Ninomiya theorem \cite{NIELSEN198120}.

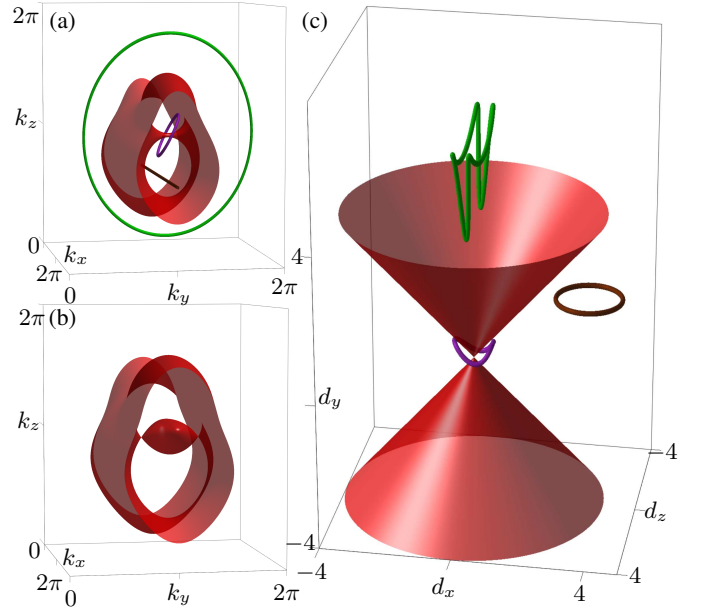
\begin{figure}
    \centering
    \def\svgwidth{\columnwidth}
    %--
    
    \begingroup%
  \makeatletter%
  \providecommand\color[2][]{%
    \errmessage{(Inkscape) Color is used for the text in Inkscape, but the package 'color.sty' is not loaded}%
    \renewcommand\color[2][]{}%
  }%
  \providecommand\transparent[1]{%
    \errmessage{(Inkscape) Transparency is used (non-zero) for the text in Inkscape, but the package 'transparent.sty' is not loaded}%
    \renewcommand\transparent[1]{}%
  }%
  \providecommand\rotatebox[2]{#2}%
  \newcommand*\fsize{\dimexpr\f@size pt\relax}%
  \newcommand*\lineheight[1]{\fontsize{\fsize}{#1\fsize}\selectfont}%
  \ifx\svgwidth\undefined%
    \setlength{\unitlength}{246bp}%
    \ifx\svgscale\undefined%
      \relax%
    \else%
      \setlength{\unitlength}{\unitlength * \real{\svgscale}}%
    \fi%
  \else%
    \setlength{\unitlength}{\svgwidth}%
  \fi%
  \global\let\svgwidth\undefined%
  \global\let\svgscale\undefined%
  \makeatother%
  \begin{picture}(1,0.93495929)%
    \lineheight{1}%
    \setlength\tabcolsep{0pt}%
    \put(0,0){\includegraphics[width=\unitlength,page=1]{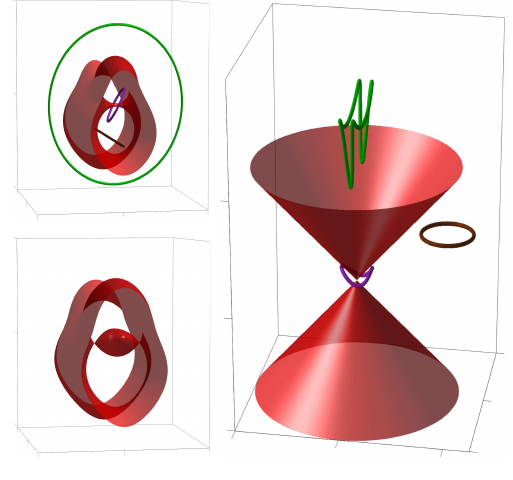}}%
    \put(0.0637516,0.89569889){\color[rgb]{0,0,0}\transparent{0.9940688}\makebox(0,0)[t]{\lineheight{1.25}\smash{\begin{tabular}[t]{c}(a)\end{tabular}}}}%
    \put(0.0637516,0.43830904){\color[rgb]{0,0,0}\transparent{0.9940688}\makebox(0,0)[t]{\lineheight{1.25}\smash{\begin{tabular}[t]{c}(b)\end{tabular}}}}%
    \put(0.45171033,0.89569889){\color[rgb]{0,0,0}\transparent{0.9940688}\makebox(0,0)[t]{\lineheight{1.25}\smash{\begin{tabular}[t]{c}(c)\end{tabular}}}}%
    \put(0.02823859,0.90206411){\color[rgb]{0,0,0}\transparent{0.9940688}\makebox(0,0)[rt]{\lineheight{1.25}\smash{\begin{tabular}[t]{r}\(2\pi\)\end{tabular}}}}%
    \put(0.0291318,0.54882684){\color[rgb]{0,0,0}\transparent{0.9940688}\makebox(0,0)[rt]{\lineheight{1.25}\smash{\begin{tabular}[t]{r}\(0\)\end{tabular}}}}%
    \put(0.07543044,0.47207731){\color[rgb]{0,0,0}\transparent{0.9940688}\makebox(0,0)[t]{\lineheight{1.25}\smash{\begin{tabular}[t]{c}\(0\)\end{tabular}}}}%
    \put(0.40538635,0.4849503){\color[rgb]{0,0,0}\transparent{0.9940688}\makebox(0,0)[t]{\lineheight{1.25}\smash{\begin{tabular}[t]{c}\(2\pi\)\end{tabular}}}}%
    \put(0.06416791,0.50504909){\color[rgb]{0,0,0}\transparent{0.9940688}\makebox(0,0)[rt]{\lineheight{1.25}\smash{\begin{tabular}[t]{r}\(2\pi\)\end{tabular}}}}%
    \put(0.03784519,0.43800812){\color[rgb]{0,0,0}\transparent{0.9940688}\makebox(0,0)[rt]{\lineheight{1.25}\smash{\begin{tabular}[t]{r}\(2\pi\)\end{tabular}}}}%
    \put(0.02973896,0.07973842){\color[rgb]{0,0,0}\transparent{0.9940688}\makebox(0,0)[rt]{\lineheight{1.25}\smash{\begin{tabular}[t]{r}\(0\)\end{tabular}}}}%
    \put(0.40836278,0.01426423){\color[rgb]{0,0,0}\transparent{0.9940688}\makebox(0,0)[t]{\lineheight{1.25}\smash{\begin{tabular}[t]{c}\(2\pi\)\end{tabular}}}}%
    \put(0.06470052,0.03092843){\color[rgb]{0,0,0}\transparent{0.9940688}\makebox(0,0)[rt]{\lineheight{1.25}\smash{\begin{tabular}[t]{r}\(2\pi\)\end{tabular}}}}%
    \put(0.06286449,0.06934462){\color[rgb]{0,0,0}\transparent{0.9940688}\makebox(0,0)[lt]{\lineheight{1.25}\smash{\begin{tabular}[t]{l}\(k_x\)\end{tabular}}}}%
    \put(0.24292545,0.01553924){\color[rgb]{0,0,0}\transparent{0.9940688}\makebox(0,0)[t]{\lineheight{1.25}\smash{\begin{tabular}[t]{c}\(k_y\)\end{tabular}}}}%
    \put(0.03066356,0.28004057){\color[rgb]{0,0,0}\transparent{0.9940688}\makebox(0,0)[rt]{\lineheight{1.25}\smash{\begin{tabular}[t]{r}\(k_z\)\end{tabular}}}}%
    \put(0.43376875,0.53180316){\color[rgb]{0,0,0}\transparent{0.9940688}\makebox(0,0)[rt]{\lineheight{1.25}\smash{\begin{tabular}[t]{r}\(4\)\end{tabular}}}}%
    \put(0.45302212,0.09150075){\color[rgb]{0,0,0}\transparent{0.9940688}\makebox(0,0)[rt]{\lineheight{1.25}\smash{\begin{tabular}[t]{r}\(-4\)\end{tabular}}}}%
    \put(0.44620239,0.04651183){\color[rgb]{0,0,0}\transparent{0.9940688}\makebox(0,0)[t]{\lineheight{1.25}\smash{\begin{tabular}[t]{c}\(-4\)\end{tabular}}}}%
    \put(0.98525899,0.23104439){\color[rgb]{0,0,0}\transparent{0.9940688}\makebox(0,0)[t]{\lineheight{1.25}\smash{\begin{tabular}[t]{c}\(-4\)\end{tabular}}}}%
    \put(0.93459711,0.0408307){\color[rgb]{0,0,0}\transparent{0.9940688}\makebox(0,0)[lt]{\lineheight{1.25}\smash{\begin{tabular}[t]{l}\(4\)\end{tabular}}}}%
    \put(0.85884447,0.01325596){\color[rgb]{0,0,0}\transparent{0.9940688}\makebox(0,0)[t]{\lineheight{1.25}\smash{\begin{tabular}[t]{c}\(4\)\end{tabular}}}}%
    \put(0.95788987,0.14036024){\color[rgb]{0,0,0}\transparent{0.9940688}\makebox(0,0)[lt]{\lineheight{1.25}\smash{\begin{tabular}[t]{l}\(d_z\)\end{tabular}}}}%
    \put(0.4566757,0.31697099){\color[rgb]{0,0,0}\transparent{0.9940688}\makebox(0,0)[lt]{\lineheight{1.25}\smash{\begin{tabular}[t]{l}\(d_y\)\end{tabular}}}}%
    \put(0.06524941,0.53523894){\color[rgb]{0,0,0}\transparent{0.9940688}\makebox(0,0)[lt]{\lineheight{1.25}\smash{\begin{tabular}[t]{l}\(k_x\)\end{tabular}}}}%
    \put(0.24264729,0.47693377){\color[rgb]{0,0,0}\transparent{0.9940688}\makebox(0,0)[t]{\lineheight{1.25}\smash{\begin{tabular}[t]{c}\(k_y\)\end{tabular}}}}%
    \put(0.02695091,0.745935){\color[rgb]{0,0,0}\transparent{0.9940688}\makebox(0,0)[rt]{\lineheight{1.25}\smash{\begin{tabular}[t]{r}\(k_z\)\end{tabular}}}}%
    \put(0.65028437,0.02337397){\color[rgb]{0,0,0}\transparent{0.9940688}\makebox(0,0)[t]{\lineheight{1.25}\smash{\begin{tabular}[t]{c}\(d_x\)\end{tabular}}}}%
    \put(0.07664254,0.00484206){\color[rgb]{0,0,0}\transparent{0.9940688}\makebox(0,0)[t]{\lineheight{1.25}\smash{\begin{tabular}[t]{c}\(0\)\end{tabular}}}}%
  \end{picture}%
\endgroup%

    %--
    \caption{
    Nodal structure of the three-dimensional model given in Eq.~\eqref{eq:2bandmodel-3d}.
    (a) For parameter values \(t_x=1, t_y=2, t_z=3, m_R=1 \text{ and } m_I=5.1\) the model hosts a nodal torus (red) with two handles in its Brillouin zone.
    One handle extends through the BZ, the other is local.
    The green, purple, and brown parametric loops correspond to topological invariants \((m=1,\nu=+1),~(m=0,W=1),~(m=0,W=0)\), respectively. 
    The non-trivial winding invariant of the purple loop protects the handle of the nodal torus.
    (b)
    Changing the imaginary mass parameter to \(m_I=4\) contracts the nodal torus's local handle. 
    As it is protected by a winding invariant, it contracts to points but cannot vanish; the genus of the torus remains invariant.  
    (c) 
    Embedding of the loops shown in (a) into \(\mathbf{d}(\mathbf{k}(\mathcal C))\in\R^3\).
    }
    \label{fig:2band-3D}
\end{figure}

To showcase the features of higher-dimensional nodal structures, we extend our model to a three-dimensional parameter space, by shifting \(m_I\to m_I+ t_z \cos(k_z)\) in Eq.~\eqref{eq:2bandmodel-2d}, i.e.,
\begin{align}
     \notag
    H^{\text{3D}}(\bk) =& 
    t_x\cos(k_x)\sigma_x +\left[t_x\sin(k_x)+t_y\sin(k_y)+m_I\right]\sigma_z
    \\&
    + i \left[t_y\cos(k_y) + t_z\cos(k_z) + m_R\right] \sigma_y.
    \label{eq:2bandmodel-3d}
\end{align}
This corresponds to a simple cubic lattice, in which stacked \(xy\)-planes described by \(H^{(2D)}\) are coupled by an imaginary hopping parameter \(t_z\).
For a range of parameters, this model shows nodal structures that form tori of higher genera. In Fig.~\ref{fig:2band-3D}, we show an exemplary 2-torus, which has one handle that extends through parameter space, and one local handle that is protected by a winding number. We make this apparent by decreasing \(m_I\), during which the handle contracts but cannot vanish.

In this section we have shown how topology manifests in two band models, starting from few dimensions to the higher dimensional case. The dimensional extension procedure we used is completely generic; our models can be extended to arbitrary dimensions or numbers of control parameters.
As noted before, the models in this section also represent chiral symmetric Hermitian models, perturbed by a non-Hermitian \(i \sigma_z\) term. In particular, the one-dimensional model can be thought of as the SSH model perturbed by an imaginary staggered potential $i\sigma_z$, via a transformation $\exp(i\pi\sigma_x/2)$ of orbitals. 
To compare the models directly, we only need to swap $d_y\leftrightarrow d_z$. 
The results on the zeroth homotopy set and the fundamental group, i.e., the orientation \(\nu\) and winding \(W\),
carry over directly to, e.g., the corresponding non-Hermitian SSH model. This was also observed in Ref.~\cite{PhysRevB.100.075403}, where the existence of a $\pi_1$ topological invariant has been used to diagnose phases in chiral-symmetric models. The $\pi_1$ topological invariant considered in Ref.~\cite{PhysRevB.100.075403} around the singularity ring is exactly the winding number around the double cone in our analysis.

\section{Topological properties of nodal and band-gapped phases}\label{sc_ggp}

In this section, we will study the homotopy properties for general $N\times N$ non-degenerate matrices, corresponding to the classification of nodal structures and band-gapped $N$-band systems. 
We will show that the first two non-trivial results are $\pi_0$ and $\pi_1$. 
For the topological classification procedure shown in Fig.~\ref{fig_hptiv}, these will be the most common and decisive.
To arrive at the results, we utilize the homotopy groups of the real and the complex general linear groups, listed in Appendix~\ref{ap_hpgl}. 
Our main mathematical tool is the long exact sequence of homotopy groups for a fibration, a relation between a series of group homomorphisms.
A detailed introduction can be found in Refs.~\onlinecite{bott1982differential,hatcherAlgebraicTopology2002}. 
The generic results derived here contain the two-band results derived in Sec.~\ref{sec_2bd} as a special case at \(N=2\).

In general, a real square matrix $H\in \operatorname{Mat}(N,\mathbb R)$, even if non-degenerate, could have complex eigenvalues. Hence it may not be diagonalized with the similarity transformation $GL_N(\mathbb R)$ by real matrices.
However, as long as $H$ is non-degenerate, it can be transformed into a canonical form.
More precisely, if the eigenvalues of $H$ are $a_1,a_2,\dots,a_n$ (mutually different real numbers), and $a_{n+j}\pm ib_j$ ($1\leq j \leq m$, $b_j>0$) (mutually different complex numbers, that must arise in complex conjugate pairs), we can write
\begin{align}\label{eq_hdcp}
    H &= VH_DV^{-1},
\end{align}
where $V\in GL_{n+2m}(\mathbb R)$ and 
\begin{align}
    H_D &= \operatorname{diag}\\
    &\left(a_1, \ldots, a_n,
        \begin{pmatrix}a_{n+1} & b_1\\-b_1 & a_{n+1}\end{pmatrix}, \ldots, 
    \begin{pmatrix}a_{n+m} & b_m\\-b_m & a_{n+m}\end{pmatrix}\right).\nonumber
\end{align}
All these block diagonal $H_D$ are captured by a mathematical concept called configuration space. A configuration space $\mathrm{Conf}_n(X)$ is made up of $n$-tuples of pairwise distinct points in the space $X$ \cite{PhysRevB.103.155129,PhysRevB.101.205417,kassel2008braid} (also see introduction in Appendix~\ref{ap_hpgl}). From this definition, the $n$ distinct real eigenvalues of $H_D$ are characterized by the space $\mathrm{Conf}_n(\mathbb R)$ and the $m$ distinct pairs of complex eigenvalues correspond to $ \mathrm{Conf}_m(\mathbb C^+)$, where $\mathbb C^+$ is the upper half complex plane. 
All possible block diagonal matrices $H_D$ are given by the space $\mathrm{Conf}_n(\mathbb R)\times \mathrm{Conf}_m(\mathbb C^+)$.

Note however that there are redundancies in the canonical-form parameterization Eq.~\eqref{eq_hdcp}, which should be quotiented out. 
The first redundancy consists of matrices $V$ that commute with $H_D$. They are the stabilizers of $H_D$ under the action of $V$. By direct calculation,
it turns out that the stabilizers are given by
\begin{align}
    &V_s= \operatorname{diag}\\
    &\left(\alpha_1, \ldots, \alpha_n,
        \begin{pmatrix}\alpha_{n+1} & \beta_1\\-\beta_1 & \alpha_{n+1}\end{pmatrix}, \ldots, 
    \begin{pmatrix}\alpha_{n+m} & \beta_m\\-\beta_m & \alpha_{n+m}\end{pmatrix}\right)\nonumber,
\end{align}
where $\alpha_j\ne 0~(j\le n), \alpha_{n+j}+i\beta_j\ne 0$. Here all $\alpha_j$ and $\beta_j$ are real numbers. 
From this parameterization, the above form of $V_s$ may be identified with the group $(\mathbb R^{\times})^n\times (\mathbb C^{\times})^m$, where $\mathbb R^{\times}$ is the multiplicative group of nonzero real numbers and $\mathbb C^{\times}$ is the multiplicative group of nonzero complex numbers (see Appendix.~\ref{ap_hpgl}). 
They correspond to multiplying the real eigenvectors with real numbers or the complex eigenvectors with complex numbers; a gauge transformation of the state.
The second redundancy is the invariance of the Hamiltonian \(H\) when one permutes the blocks in $H_D$ and the column vectors in $V$ simultaneously. This action is the product of two permutation (symmetric) group elements in $S_n\times S_m$, one acting on the real, and one on the complex eigenvalues. 

To summarize, the total space $X^{(m,n)}$ of real gapped matrices with $m$ pairs of complex eigenvalues and $n$ real eigenvalues is given by block diagonal \(H_D\), a change of basis \(V\), and a subsequent removal of the gauge degrees of freedom; leading to the space
\begin{equation}
    X^{(m,n)}=\frac{\Big(\mathrm{Conf}_n(\mathbb R)\times \mathrm{Conf}_m(\mathbb C^+)\Big)\times \frac{GL_{n+2m}(\mathbb R)}{(\mathbb R^{\times})^n\times (\mathbb C^{\times})^m}}{S_n\times S_m}.
\end{equation}
We will call the three constituting parts the eigenvalue part $\mathcal E=\mathrm{Conf}_n(\mathbb R)\times \mathrm{Conf}_m(\mathbb C^+)$, the eigenstate part $M^{(m,n)}=GL_{n+2m}(\mathbb R)\allowbreak/[(\mathbb R^{\times})^n\times (\mathbb C^{\times})^m]$, and the simultaneous permutation \(S_n\times S_m\). 

The above formula can be further simplified, if we notice we can use the symmetric group $S_n$ to uniquely sort all real eigenvalues. 
In other words, we can assume $a_1<a_2<\dots<a_n$. 
Using the map $(a_1,a_2\dots a_m) \to (a_1,a_2-a_1,a_3-a_2,\dots)$,
such lists of real eigenvalues can be identified with $\mathbb R\times(\mathbb R^+)^{n-1}$, which is homeomorphic to $\mathbb R^n$ itself. 
Combined with the homeomorphism between the upper half complex plane $\mathcal{\mathbb C^+}$ and the complex plane itself, we have the following identification
\begin{equation}
     X^{(m,n)}=\frac{\mathbb R^n\times \mathrm{Conf}_m(\mathbb C)\times \frac{GL_{n+2m}(\mathbb R)}{(\mathbb R^{\times})^n\times (\mathbb C^{\times})^m}}{ S_m}.\label{eq_rdndsp}
\end{equation}

The total space $X_N$ of all \(N\)-band gapped Hamiltonians is the union of all these $X^{(m,n)}$ with \(N\) bands,
\begin{equation}
    X_N=\underset{n+2m=N}{\bigcup_{m,n}} X^{(m,n)}.
\end{equation}
This form generalizes the scheme of Sec.~\ref{sec_2bd}, where we decomposed the two-band case as \(X_2 =  X^{(1,0)} \cup  X^{(0,2)}\). 

In the following subsections, we apply homotopy theory to study topological invariants associated with band gap preserving deformations, providing a direct generalization of the studies already done for two-band models. 

\subsection{Zeroth homotopy set}\label{sec-PTband0}

As we have already seen in the two-band situation, exceptional degeneracies of codimension $1$ in systems with \PT{} symmetry are associated with the zeroth homotopy set of $X_N$, i.e., the set of path-connected components of $X_N$. We now work this out for a general $N$-band system.

First of all, the two spaces $X^{(m,n)},X^{(m',n')}$ with $m\ne m'$ or $ n\ne n'$ are disjoint by definition. 
Each $X^{(m,n)}$ is open in $X_N$ due to the gap (non-degeneracy),
so two matrices with different $(m,n)$ must lie in different connected components of $X_N$. 
Physically, if we want to continuously change $m$ or $n$, at least two real eigenvalues must merge and form a twofold degeneracy, then split into pairs of complex conjugate non-real eigenvalues (or vice-versa).

Furthermore, even fixing $(m,n)$, the space $X^{(m,n)}$ may still not be connected.
Let us first look at the ``numerator" in Eq.~\eqref{eq_rdndsp}.
Since both $\mathbb R^n$ and $\mathrm{Conf}_m(\mathbb C^+)$ are connected, 
its $\pi_0$ is determined by $\frac{GL_{n+2m}(\mathbb R)}{(\mathbb R^{\times})^n\times (\mathbb C^{\times})^m}$.
The real general linear group has two disjoint connected components, distinguished by the sign of the determinant (Appendix~\ref{ap_hpgl}). 
If $n\geq 1$, under the action of $\mathbb R^\times$, pairs of points from these two disjoint components are identified, which means that the quotient space is connected and so do $X^{(m,n)}$. However,
if $n=0$, then $\frac{GL_{2m}(\mathbb R)}{(\mathbb C^{\times})^m}$ still has two connected components. Moreover, they are not identified by the $S_m$, since the $S_m$ action permutes pairs of eigenvalues together.
So $X^{(m,0)}$  has two components for $n=0$, which we denote as $X^{(N/2,0)}_+$ and $X^{(N/2,0)}_-$.

To conclude, the zeroth homotopy set is given by
\begin{align}
    \pi_0(X_N)=&\{X^{(0,N)},X^{(1,N-2)},\dots, X^{(N/2,0)}_+,X^{(N/2,0)}_-\},\nonumber\\ &\textrm{$N$ even};\nonumber\\
    \pi_0(X_N)=&\{X^{(0,N)},X^{(1,N-2)},\dots, X^{((N-1)/2,1)}\},\, \textrm{$N$ odd}.\label{eq_zhmtpy}
\end{align}
When $N=2$, we reproduce the three elements of $\pi_0(X_2)$ in Sec.~\ref{sec_2bd}: $X^{(0,2)},X^{(1,0)}_+$ and $X^{(1,0)}_-$, with the first corresponding to the out-cone region while the latter twos correspond to the upper and the lower in-cone regions. A generic $X^{(m,n)}$ is a generalization of such regions, separated from each other by level crossings. Here we see that more bands give rise to more sectors of band-gapped phases, distinguished by how many eigenvalues/eigenvectors are spontaneously breaking the symmetry. Notice that the frame orientation topology, the $\pm$ subsript in $X^{(N/2,0)}_\pm$ only appears for even $N$.

Similar to the argument for two-band models, the zeroth homotopy set here gives a topological reason for the abundance of nodal structures in \PT-symmetric systems. 
When only Hermitian terms are allowed, the space $X_N$ is connected and there is only one element in $\pi_0(X)$. 
Therefore a given parameter space is not in general separated into disjoint parts by the gapless regions, since the gapless regions often have codimension larger than 1. 
However, when non-Hermitian terms are allowed, the space of gapped Hamiltonians is not connected; it has boundaries where a gap closing happens. 
To break a $d$-dimensional parameter space into disconnected regions, $d-1$ dimensional nodal structures must occur. 
This shows why one needs fewer parameters to tune a system to show specific nodal structures.

\subsection{Fundamental groups: \PT-preserving phases}\label{sc_fdPTbk}

\renewcommand{\arraystretch}{1.5}
\begin{table}[t]
    \centering
    \begin{tabular}{c|c|c|c}
        \parbox[t]{2cm}{
            \(\pi_1\left(X^{(m,n)}\right)\) 
            \\ 
            $\textrm{[Topology]}$ 
        } & 
        \multirow{2}{*}{\(m=0\)} &
        \multirow{2}{*}{\(m=1\)} &
        \multirow{2}{*}{\(m>1\)} 
    \\
        \hline
        \(n=0\) & \multicolumn{2}{c|}{\multirow{2}{*}{\(\{1\}\) [trivial]}} & \multirow{2}{*}{\(\mathbb{B}_m\)[braid]}
    \\
        \cline{1-1}
        \(n=1\) & \multicolumn{2}{c|}{}  & 
    \\
        \hline
        \(n=2\) & \(\mathbb{Z}\) [winding] &
        \multicolumn{2}{c}{\multirow{2}{*}{
        }} 
    \\
        \cline{1-2}
        \(n>2\) & \(Q(n)\)[frame$_{N}$] &
        \multicolumn{2}{c}{\multirow{-2}{*}{
            \parbox[t]{2.5cm}{
                \(\mathbb{B}_m\times \mathbb{Z}_2^{n-1}\)
                \\ 
                \([\)braid \(\times\) frame\(_A]\)
            }}}
    \\
    \end{tabular}
    \caption{Homotopy groups of \PT-symmetric gapped Hamiltonians with \(n\) real and \(m\) complex-conjugate pairs of non-real eigenvalues. They correspond to the topology along a loop in the BZ. The term ``winding'' is the rotation angle of the two real eigenvectors, divided by $\pi$. The phrase ``frame$_N$'' corresponds to non-Abelian frame charges and ``frame$_A$'' is its Abelianization, a generalized Zak phase.}
    \label{tab:homotopy-groups}
\end{table}

Having calculated the zeroth homotopy groups $\pi_0$, we now move to the fundamental groups $\pi_1$. 
We start from the situation when there is no spontaneous symmetry breaking, $m=0$. 

In this case, the space of non-degenerate matrices given in Eq.~\eqref{eq_rdndsp} simplifies to
\begin{equation}
     X^{(0,n)}=\mathbb R^n\times \frac{GL_n(\mathbb R)}{(\mathbb R^{\times})^n}.
\end{equation}
The space $\mathbb R^n$ representing eigenvalues is a contractible space, meaning that its homotopy property is the same as a single point and does not contribute any nontrivial topology. The eigenvector parts can be simplified via the homotopy equivalences   $GL_n(\mathbb R)\simeq O(n)$, $\mathbb R^{\times}\simeq \mathbb Z_2$ (see Appendix~\ref{ap_hpgl}). So the homotopy property of $X^{(0,n)}$ is determined by $M^{(0,n)}=O(n)/(\mathbb Z_2)^n$. As we have shown, the $\mathbb Z_2$ is to flip the sign of an eigenvector, an action $x_j\to -x_j$ in $\mathbb R^n$. We can employ one action $x_1\to-x_1$ to reduce $O(n)$ to $SO(n)$,
the final result is further simplified to 
\begin{equation}
    X^{(0,n)}\simeq M^{(0,n)}=SO(n)/(\mathbb Z_2)^{n-1}.\label{eq_spfbgs}
\end{equation}
Here, each $\mathbb Z_2$ acts on $SO(n)$ by simultaneously reflecting two column vectors. 
As a convention, we assume the $k^{th}$ $\mathbb Z_2$ in Eq.~\eqref{eq_spfbgs} acts as $\pi$ rotation in the plane $(x_1,x_{k+1})~(2\leq k\leq n)$. 

This space Eq.~\eqref{eq_spfbgs} is the real flag manifold, which has been extensively studied in Refs.~\onlinecite{wu2019non,bouhon2020non,PhysRevB.102.115135}, although within a different physical context. 
The topology entirely comes from the eigenvectors. 
These linear-independent eigenvectors span a frame and this frame rotates non-trivially along a loop.
For $n=2$, we simply have 
\begin{equation}
\pi_1(M^{(0,2)})\cong\mathbb Z,
\end{equation}
reproducing the two-band result, the winding number of the $d_x,d_z$ vectors outside the double cone. It is worth noticing that this $\mathbb Z$-winding number comes from a double covering of the fundamental group $\pi_1(\big(SO(2)\big)=\mathbb Z$. However, the frame topology is non-Abelian for $n\geq 3$, as we have more quotients in Eq.~\eqref{eq_spfbgs} and the fundamental groups of $SO(n)$ is stablized to $\mathbb Z_2$ for $n\ge 3$. Below we briefly present how the non-Abelian structure arises and review the results of Refs.~\onlinecite{wu2019non,bouhon2020non}.

A standard way to study homotopy properties of such spaces is to use the long exact sequence of homotopy groups associated with a fibration:
\begin{align}
    \cdots\to& \pi_{d}\big(\mathbb Z_2^{n-1}\big) \to \pi_d\big(SO(n)\big)\to \pi_d(M^{(0,n)})
    \nonumber\\
    \to&\pi_{d-1}\big(\mathbb Z_2^{n-1}\big) \to \pi_{d-1}\big(SO(n)\big)\to\cdots.
\end{align}
Taking $d=1$ gives the following exact sequence (for $n\ge 3$):
\begin{equation}
    0\to \mathbb Z_2\to \pi_1(M^{(0,n)})\to (\mathbb Z_2)^{n-1}\to 0.\label{eq_sqpts}
\end{equation}
The above sequence tells us that $\pi_1(M^{(0,n)})$ is an extension of $ (\mathbb Z_2)^{n-1}$ by an $\mathbb Z_2$ element from $\pi_1\big(SO(n)\big)$.
However, the above sequence does not necessarily split and we do not know the group structure for these generators a priori.

The way to circumvent this problem is to represent $SO(n)/(\mathbb Z_2)^{n-1}$ as another quotient space.
We know $SO(n)$ has a simply-connected double cover $\textrm{Spin}(n)$.
We also know those $\pi$ rotations should be lifted to some $\pm e\in\textrm{Spin}(n)$ such that $e^2=-1$ and that anticommute with each other.
Therefore, we arrive at $X^{(0,n)}\simeq \text{Spin}(n)/Q(n)$ and
\begin{equation}
    \pi_1(M^{(0,n)})\cong Q(n),~~(n\geq 3),
\end{equation}
where $ Q(n)$ is one of the Salingaros vee groups \cite{wu2019non}:
\begin{align}
    Q(n)&=\{\pm e_1^{k_1}e_2^{k_2}\dots e_{n-1}^{k_{n-1}}|k_i\in\{0,1\}\},\nonumber\\
    e_ie_j&=-e_je_i ~(i\ne j),\quad e^2_j=-1.\label{eq_dfsvg}
\end{align}
From these expressions, the non-Abelian structure comes from the group element $-1$. This the lift of the fundamental group $\pi_1\big(SO(n)\big)=\mathbb Z_2$, the first $\mathbb Z_2$ in the sequence Eq.~\eqref{eq_sqpts}. Its physical meaning is a $2\pi$ rotation of the eigenvector frame.

Let us consider the case $n=3$ as a minimal example, to illustrate the meaning of this group. The corresponding $ Q(3)$ is the quaternion group and contains eight elements that can easily be represented using Pauli matrices \(\sigma_j\), $ Q(3)=\{\pm 1,\pm i\sigma_x,\pm i\sigma_y,\pm i\sigma_z\}$. 
One may verify directly that the prefactor $i$ in front of the Pauli matrices enables this set to be a group. 
We can use $i\sigma_x$ to represent $\pi$-rotation of frames. This element has order $4$: $(i\sigma_x)^4=1$, instead of order $2$, since a $2\pi$ rotation of frames is non-trivial in $GL_{n>2}(\mathbb R)$. It rather gives the element $(i\sigma_x)^2=-1$ in the quaternion group. To write the quaternion group in the form of Eq.~\eqref{eq_dfsvg}, we can put $e_1=-i\sigma_x$ and $e_2=-i\sigma_y$. The minus sign in $e_i$ is only to keep the same conventions as previous literature. 

As the fundamental group is non-Abelian, there are several cautions when using it to characterize degeneracy and gapped phases \cite{RevModPhys.51.591,wu2019non}. When characterizing a degeneracy, we need to fix a base point in the Brillouin zone. Different choices of base points correspond to permutations of the topological characters. A fixed base point enables us to compare different nodal structures on the same ground. The loops employed to enclose defects must start and end at the base point. After fixing the base point, each degeneracy is uniquely described by the $\pi_1$ element along the loop enclosing it. The composition laws of degeneracy follow from the group product. 

To characterize one-dimensional gapped phases with a non-Abelian fundamental group, we need to further work out the conjugacy classes in the group. The reason is that a change of base point induces a conjugate action $gGg^{-1}$ on the group itself. But choosing another base point does not induce another gapped phase---this is only a different way of labeling. When the group is Abelian, the conjugate action is trivial and we do not need to worry about this. But the conjugate action on a non-Abelian group is often nontrivial. For the quaternion group, there are five conjugacy classes \cite{RevModPhys.51.591}:
\begin{equation}
    \left\{\{1\},\{-1\},\{\pm i\sigma_x\},\{\pm i\sigma_y\},\{\pm i\sigma_z\}\right\}.
\end{equation}
So the gapped phase corresponding to $i\sigma_j$ is the same gapped phase as that corresponding to $-i\sigma_j$. For a non-Abelian fundamental group, we often have fewer distinct one-dimensional gapped phases than the order of the fundamental group. Nevertheless, all information about the phase is still contained in the fundamental group.

\subsection{Fundamental groups: spontaneously \PT-breaking phases}\label{sec-PTband1SSB}

More interesting phases emerge when the \PT{} symmetry is spontaneously broken (namely, $m\geq 1$ in Eq.~\eqref{eq_rdndsp}). 
In this subsection, we work out all the homotopy groups of $X^{(m,n)}$.

We still denote the eigenvalue part $\mathbb R^n\times \mathrm{Conf}_m(\mathbb C)$ as $\mathcal E$.
The space $\mathbb R^n$ representing real eigenvalues is contractible, and thus we have $\pi_d(\mathbb R^n)=0$. The complex eigenvalues, in comparison, have nontrivial topology. The configuration space of complex numbers is the Eilenberg–MacLane space $K(P\mathbb B_m,1)$ \cite{kassel2008braid,PhysRevB.103.155129,PhysRevB.101.205417}, where $P\mathbb B_m$ is the pure braid group, the braid group that does not permute any elements. It has a nontrivial fundamental group $\pi_1\big(\mathrm{Conf}_m(\mathbb C)\big)=P\mathbb B_m$ and vanishing (trivial) higher homotopy groups.

Then let us look at the eigenvector part, denoted as $M^{(m,n)}=GL_{n+2m}(\mathbb R)/[(\mathbb R^{\times})^n\times (\mathbb C^{\times})^m]$. 
It fits in a fibration $GL_{n+2m}(\mathbb R)\to M^{(m,n)}$ with the fiber $(\mathbb R^{\times})^n\times (\mathbb C^{\times})^m$
and we have the following long exact sequence:
\begin{align}
    \cdots\to \pi_d(GL_{n+2m}(\mathbb R))\to \pi_d(M^{(m,n)})\nonumber\\
    \to\pi_{d-1}((\mathbb R^{\times})^n\times (\mathbb C^{\times})^m)\to\cdots.\label{eq_sqfg}
\end{align}

We first consider the case where $n>0$ so that $M^{(m,n)}$ is path-connected.
Since we are considering the \PT-symmetry breaking phase, $m>0$, and hence $\pi_1(GL_{n+2m}(\mathbb R))\cong\mathbb Z_2$.
Plugging in these relations, we have 
\begin{equation}\label{eq-pi1PTbreaking}
    \mathbb Z^m\to \mathbb Z_2\to \pi_1(M^{(m,n)})\to (\mathbb Z_2)^n\to\mathbb Z_2\to 0.
\end{equation}
As discussed in Appendix~\ref{ap_hpgl}, the first map $\mathbb Z^m\to \mathbb Z_2$ is induced by several inclusion maps $SO(2)\hookrightarrow SO(n+2m)$, which sends an integer number to its mod $2$ value. 
It is surjective as long as $m\geq 1$, and the exactness implies that the map $\mathbb Z_2\to \pi_1(M^{(m,n)})$ is trivial. The fourth map $(\mathbb Z_2)^n\to\mathbb Z_2$ is the product of the signs of $\mathbb R^\times$. Its kernel is $(\mathbb Z_2)^{n-1}$. As a result, we know that the fundamental group of the eigenvector part is 
\begin{equation}\label{eq-pi1mn}
    \pi_1(M^{(m,n)})=(\mathbb Z_2)^{n-1}~~ (m>0,n>0).
\end{equation}
Note that we need $m\geq 1$ for this calculation, which is why the \PT-breaking phase has a different topology compared with the \PT-preserving phase.

Now we combine the eigenvalue and the eigenvector part. 
Classifying loops in $X^{(m,n)}$ amounts to classifying paths in $\mathcal E\times M^{(m,n)}$ from $(e,v)$ to $(ge,gv)$ for all possible $g\in  S^m$. Here $e\in\mathcal E$ and $v\in M^{(m,n)}$ are base points.
For any such path, projecting it to $\mathcal E$ then to $\mathcal E/S_m$ gives a loop in $\mathcal E/S_m$, characterized by the braid group $\mathbb B_m$, which also in turn determines $g$ by forgetting the braiding details. 
So the rest is to classify paths from $v$ to $gv$,  which is given by the fundamental group $\pi_1(M^{(m,n)})$ as we can compare two paths by joining them into a loop.
In Appendix~\ref{ap_pdtetv}, we will rigorously prove that:
\begin{equation}\label{eq-pi1isproduct}
    \pi_1(X^{(m,n)})=\mathbb B_m \times (\mathbb Z_2)^{n-1}~~ (m>0,n>0).
\end{equation}
This is quite intuitive since $\mathbb B_m$ comes from the motion of complex eigenvalues while $(\mathbb Z_2)^{n-1}$ comes from the motion of real eigenvectors (recall the derivation towards Eq.~\eqref{eq-pi1mn}) and they should not interfere with each other.

Finally, we go back to the special situation $n=0$. 
In this situation, $M^{(m,0)}=GL_{2m}(\mathbb R)/(\mathbb C^{\times})^m$ has two path-connected components which are homeomorphic to each other. We only need to look at the homotopy groups for one of them. The sequence for the fundamental group then looks as
\begin{equation}
    \mathbb Z^m\to \mathbb Z_2\to \pi_1\big(M^{(m,0)}\big)\to 0.
\end{equation}
As before, the first map is surjective, so the fundamental group $\pi_1\big(M^{(m,0)}\big)$ is trivial. Then the fundamental group of non-degenerate matrices is given entirely by the eigenvalue part:
\begin{equation}
    \pi_1(X^{(m,0)})=\mathbb B_m,\, m>0,n=0.
\end{equation}
Again, if we choose $m=1$, this gives the two-band result in the symmetry-breaking regime. The braid group of one strand is trivial $\mathbb B_1=\{1\}$, since a single strand alone cannot braid. Thus, this is again consistent with the conclusions for symmetry-breaking regimes of two-band models in Sec.~\ref{sec_2bd}.

This completes the homotopy groups of arbitrary numbers of bands. The results are summarized in Table~\ref{tab:homotopy-groups}. Their exact interpretation will be presented in Sec.~\ref{sc_expti}. We give a systematic method of constructing braided models in Appendix~\ref{appendix:braiding-models}, which also includes the model we present in Fig.~\ref{fig_nphs} (c). In Appendix~\ref{ap_2hmtpbg}, we give the second homotopy groups. They represent topological invariants that appear on a two-dimensional surface (c.f. Fig.~\ref{fig_hptiv}). When $\pi_0$ and $\pi_1$ are nontrivial, the homotopy groups $\pi_2$ are usually the second or the third topological index to appear for a gapped phase or nodal structure.

\section{Topology of separation-gapped phases}\label{sc_spphs}

We turn to the topological properties related to separation gaps in our system. A $\mathcal{PT}$-symmetric system generically allows for the appearance of both real and complex eigenvalues. A typical separation is the separation between these real and complex eigenvalues: each band is either entirely on the real axis, which we abbreviate as a \emph{real band}, or entirely on one half complex plane, abbreviated as a \emph{complex band}. These bands carry different dynamic properties as naturally distinguished by the imaginary parts of their spectra. We show that a single-gap description is forbidden for this separation gap due to symmetry constraints in Sec.~\ref{sec:specflattfail}. The classification turns out to be beyond $K$-theory. Therefore they cannot be described through the reference point/line classification. The phases exhibit no topology along a loop, i.e., trivial $\pi_1$. The first nontrivial topology beyond $\pi_0$ comes from $\pi_2$. From this, we will show that there are new Chern-Euler and Chern-Stiefel-Whitney invariants protecting the separation gap.

An important tool that will be used in this section is spectral flattening, which has played an important role in Hermitian single-gap classifications \cite{kitaevperio}. Although diagonalizable non-Hermitian matrices can be flattened by deforming their eigenvalues, similarly to Hermitian matrices, this native approach does not apply to non-diagonalizable matrices, i.e., matrices corresponding to EPs. In Ref.~\onlinecite{PhysRevX.9.041015}, the authors argue that EPs can be pairwise annihilated and neglect them in spectral flattening. However, a recent study shows that this pair-annihilation may not always be true \cite{konig2022braid}. 
In Sec.~\ref{sc_mspfl}, we will show how to generalize the naive spectral flattening to non-Hermitian matrices. Based on the analytic properties of band projections, both diagonalizable and non-diagonalizable matrices can be reduced to a unified canonical form for a large class of separation gaps.

\subsection{Failure of single-gap descriptions} \label{sec:specflattfail}

We demonstrate that the separation gaps between the real and the complex bands in \PT-symmetric and pseudo-Hermitian systems are correlated. These gaps have to be opened and closed simultaneously. They cannot be reduced to a single gap.

In \PT-symmetric and pseudo-Hermitian systems, the eigenvalues can be classified into three types: 1. those on the real axis; 2. those in the upper complex plane, and; 3. those in the lower complex plane. If there is no band carrying both real and complex eigenvalues, we can partition our band indices into three sets corresponding to these three types: $ J_0,J_\uparrow$ and $J_\downarrow$. Following this, we have two separation gaps. One is the separation gap between the bands on the upper complex plane and those on the real axis. The other is the separation gap between the bands on the lower complex plane and those on the real axis. 

\begin{figure}
    \centering
    \def\svgwidth{\columnwidth}
%--
    
    \begingroup%
  \makeatletter%
  \providecommand\color[2][]{%
    \errmessage{(Inkscape) Color is used for the text in Inkscape, but the package 'color.sty' is not loaded}%
    \renewcommand\color[2][]{}%
  }%
  \providecommand\transparent[1]{%
    \errmessage{(Inkscape) Transparency is used (non-zero) for the text in Inkscape, but the package 'transparent.sty' is not loaded}%
    \renewcommand\transparent[1]{}%
  }%
  \providecommand\rotatebox[2]{#2}%
  \newcommand*\fsize{\dimexpr\f@size pt\relax}%
  \newcommand*\lineheight[1]{\fontsize{\fsize}{#1\fsize}\selectfont}%
  \ifx\svgwidth\undefined%
    \setlength{\unitlength}{246bp}%
    \ifx\svgscale\undefined%
      \relax%
    \else%
      \setlength{\unitlength}{\unitlength * \real{\svgscale}}%
    \fi%
  \else%
    \setlength{\unitlength}{\svgwidth}%
  \fi%
  \global\let\svgwidth\undefined%
  \global\let\svgscale\undefined%
  \makeatother%
  \begin{picture}(1,0.69105691)%
    \lineheight{1}%
    \setlength\tabcolsep{0pt}%
    \put(0,0){\includegraphics[width=\unitlength,page=1]{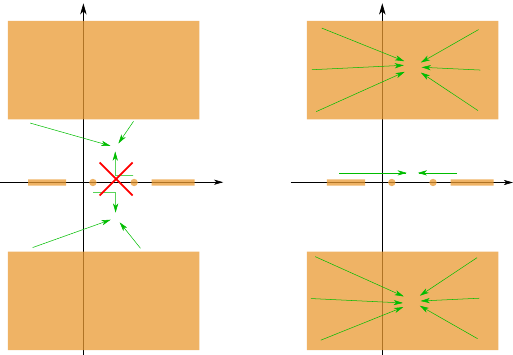}}%
    \put(0.16908056,0.65836471){\color[rgb]{0,0,0}\makebox(0,0)[lt]{\lineheight{1.25}\smash{\begin{tabular}[t]{l}Im\(E\)\end{tabular}}}}%
    \put(0.75495932,0.65836471){\color[rgb]{0,0,0}\makebox(0,0)[lt]{\lineheight{1.25}\smash{\begin{tabular}[t]{l}Im\(E\)\end{tabular}}}}%
    \put(0.43189162,0.29788968){\color[rgb]{0,0,0}\makebox(0,0)[rt]{\lineheight{1.25}\smash{\begin{tabular}[t]{r}Re\(E\)\end{tabular}}}}%
    \put(0.99741826,0.29788968){\color[rgb]{0,0,0}\makebox(0,0)[rt]{\lineheight{1.25}\smash{\begin{tabular}[t]{r}Re\(E\)\end{tabular}}}}%
    \put(0,0){\includegraphics[width=\unitlength,page=2]{fig9.pdf}}%
    \put(0.36708354,0.4190387){\color[rgb]{0,0,0}\makebox(0,0)[t]{\lineheight{1.25}\smash{\begin{tabular}[t]{c}Forbidden\end{tabular}}}}%
    \put(0.91068823,0.4190387){\color[rgb]{0,0,0}\makebox(0,0)[t]{\lineheight{1.25}\smash{\begin{tabular}[t]{c}Allowed\end{tabular}}}}%
  \end{picture}%
\endgroup%

    %--
    \caption{
    Multiple-gap descriptions for separation gaps between real bands and complex bands. (a) One can not perform a single-gap description with respect to the separation gaps. The real bands always split and move towards opposite directions into the complex plane, it cannot be grouped with either the upper complex spectrum or the lower complex plane. (b) Instead, the separation gap between complex and real eigenvalues can be simplified to a two-gap problem. This is done by flattening all eigenvalues on the upper complex plane to $i$, all eigenvalues on the real axis to $0$, and all eigenvalues on the lower complex plane to $-i$.}
    \label{fig_spfl}
\end{figure}

If we want to perform a single-gap description for such a system, there are two choices. The first is to group the bands on the lower complex plane and the real axis together and study their separation with the bands on the upper complex plane. The second choice is to group the bands on the upper complex plane and the real axis together and study their separation with the bands on the lower complex plane. We show that neither of these two ways of grouping are legal. The reason is that for a separation gap, the bands belonging to the same partition set are free to touch. However, due to the symmetry, such touch will inevitably close their separation gap with the other partition set if we do these two groupings. Therefore a single-gap description is not proper for classification. 

Let us look at the first choice and similar arguments apply to the second choice. If we want to allow the spectra below the upper complex plane free to touch, we have to either lift the real eigenvalues to the lower half complex plane, or push the eigenvalues on the lower half complex plane towards the real axis. However, neither of these two processes are allowed, as depicted in Fig.~\ref{fig_spfl} (a):
\begin{enumerate}
    \item If we lift the real eigenvalues towards the lower complex plane, due to the eigenvalue symmetry, this can only happen after two real eigenvalues collide on the real axis. But after this collision, the two real eigenvalues split away from the real axis in opposite directions. One of them goes to the upper complex plane and the other goes to the lower complex plane. The one going to the upper half plane destroys the separation.
    \item If we push the complex eigenvalues on the lower complex plane towards the real axis, according to the symmetry, the complex eigenvalues on the upper complex plane will also be pushed towards the real axis. Therefore, the complex eigenvalues on the upper complex plane will touch the real axis during this spectral flattening. This procedure closes the separation gap between the bands on the upper complex plane and the bands on the real axis, hence it is not allowed.
\end{enumerate}

From the above argument, we see that the failure of a single-gap description exactly comes from the symmetry constraints on the motion of eigenvalues. 
Due to the $\mathcal{PT}$ symmetry, we have to consider at least two gaps if we want to separate eigenvalues with different imaginary parts.
As a result, the classification cannot be captured by K-theory or a single reference line.

\subsection{Multiple-gap spectral-flattening}\label{sc_mspfl}

The natural partition  $ J_0,J_\uparrow$ and $J_\downarrow$ leads us to consider a multiple-separation-gap classification. This reduces the universal separation gaps in \PT-symmetric systems and pseudo-Hermitian systems to a two-gap problem. We first show how the spectral flattening around the two separation gaps can be done for diagonalizable non-Hermitian Hamiltonians.

A spectral flattening for separation gaps is to gradually deform the spectra in each partition set into constants, such that $E_j(\mathbf k)=\lambda_\alpha,\forall\, j\in J_\alpha$  without closing the separation gaps \cite{kitaevperio}. Here $\lambda_j$ are distinct complex constants. If spectral flattening is permitted in a system, the classification can be reduced to that of eigenvectors as the eigenvalues are constant.

The spectral-flattening for diagonalizable matrices,  matrices containing no EPs, is similar to Hermitian situations and is fairly straightforward: 
we simply diagonalize the matrix and replace all the eigenvalues on the real axis (and the upper/lower half complex plane) by $0$ (and $\pm i$, respectively), without changing the eigenvectors [see Fig.~\ref{fig_spfl} (b)].
More precisely, we decompose a matrix $H$ as
\begin{equation}
    H = VH_{D}V^{-1},
\end{equation}
with $H_D$ the diagonal matrix representing the eigenvalues. Then the spectral-flatten of $H$ is defined as
\begin{equation}\label{eq-sfdiag}
    H_{F} = VH_{DF}V^{-1},
\end{equation}
where
\begin{equation}
    H_{DF} = \operatorname{diag}    \left(0, \ldots, 0,
        i, \ldots, i,-i, \ldots, -i\right).
\end{equation}
Here, $``D"$ stands for ``diagonalized" and $``F"$ stands for ``spectral-flattened".
It can be checked that $H_{DF}$ is unambiguously determined by $H_D$ in the sense that it does not depend on the choice of $V$. The eigenvectors of a matrix are continuous when away from EPs \cite{kato2013perturbation}, it is conceivable (and will be proved later) that $H_{DF}$ depends continuously on $H_D$.

For non-diagonalizable matrices, the above procedure does not apply immediately.
For example, consider one eigenvalue $\lambda$ of $H$ forming a non-trivial Jordan structure:
\begin{equation}
    H=V\begin{pmatrix}
 \ddots & 0 & 0 & \dots &0& 0 \\
0 & \lambda & 1 & \dots &0  & 0\\ 
0 & 0 & \lambda & \dots &0 & 0\\
\vdots & \vdots & \vdots & \ddots &1& 0\\
0 & 0 & 0 & 0 & \lambda& 0 \\ 
0 & 0 & 0 & 0 & 0& \ddots  \\ 
\end{pmatrix}V^{-1}.\label{eq_hsgjd}
\end{equation}
Since the spectral decomposition process could be singular near those EPs \cite{kato2013perturbation,PhysRevB.107.144304,PhysRevLett.128.010402}, one may wonder whether a proper generalization of spectral flattening exists or not. 
If it exists, should we spectral-flatten such a structure to a scalar matrix $\lambda_\alpha I$ ($\lambda_\alpha=0,\pm i$ in our $\mathcal{PT}$ case), or should we keep the off-diagonal $1$'s as it is in Eq.~\ref{eq_hsgjd}?

Here, we claim that the correct spectral-flattening is the scalar matrix $\lambda_\alpha I$. 
More precisely, 
we define the spectral-flattening function $sf()$ as follows: 
we simply Jordan decompose a matrix and replace each Jordan block with eigenvalue $\lambda$ by $i\times sgn(\Im{\lambda})\times$ identity matrix.
Equivalently, let us denote $P_\lambda$ as the projection (not necessarily orthogonal) onto the generalized-eigenspace corresponding to the eigenvalue $\lambda$.
Then we define
\begin{align}
    sf(H)=&iP_+(H)-iP_-(H) \nonumber\\
    =&i\sum_{\lambda\in\text{spec}_+(H)}P_\lambda-i\sum_{\lambda\in\text{spec}_-(H)}P_\lambda,
\end{align}
where $\text{spec}_\pm(H)$ denotes the set of eigenvalues on the upper (and the lower) complex plane.
A useful representation of the projection $P_\pm(H)$, valid even for non-diagonalizable matrices, is as follows \cite{kato2013perturbation}:
\begin{equation}\label{eq-complexint}
   P_\pm(H)=\frac{1}{2\pi i}\int_\mathcal{C_{\pm}}(z-H)^{-1}dz,
\end{equation}
where $\mathcal{C_\pm}$ is an arbitrary complex contour on the upper-half-plane or the lower-half-plane that surrounds the corresponding spectrum $\text{spec}_\pm(H)$.  

In Appendix~\ref{ap_sft}, using Eq.~\eqref{eq-complexint}, we proved the following lemma:
\begin{lemma*}\label{th_fl}
    Fixing the number of real eigenvalues, $sf()$ is continuous in the input matrix. 
\end{lemma*}
Note that $sf()$ matches the flattened form Eq.~\eqref{eq-sfdiag} for diagonalizable matrices.
Since diagonalizable matrices are dense in all $N\times N$ matrices, $sf()$ must be the \emph{unique} generalization of the spectral-flattening for general matrices.

For later reference, here we give some comments on the above-defined spectral-flattening function.
\begin{itemize}
    \item $sf()$ is continuous only after fixing the number of real (and complex) eigenvalues.
    Otherwise, consider $H_\epsilon=\mathbf1_2+i\epsilon\sigma_y$, then $sf()$ is not continuous near $\epsilon=0$.
    \item $sf(H)$ is always diagonalizable even if $H$ is not.
    \item The spectral-flattening can also be viewed as a continuous procedure. 
    We can define a deformation retraction (which gives a homotopy equivalence) from the space of separation-gapped matrices to a particular subspace with assigned eigenvalues:
\begin{equation}
    H_t=(1-t)H+t\cdot sf(H).\label{eq_sfl}
\end{equation}
According to lemma~\ref{th_fl}, $H_t$ is continuous as a multi-variable function in $H$ and $t$.
\item The spectral-flattening respects the spectral gaps.
If $H$ is separation gapped in the sense of Fig.~\ref{fig_spfl}, so are $sf(H)$ and $H_t$ for $\forall t\in[0,1]$.
This is because the eigenvalues of $H_t$ have the form of $t\lambda+i(1-t)sgn(\lambda)$, which are clearly also separation gapped.
\item The spectral-flattening respects the symmetries of $H$. 
For example, if $H$ is real, so are $sf(H)$ and $H_t$ for $\forall t\in[0,1]$.
This can be seen from Eq.~\eqref{eq-complexint}, since it implies
\begin{equation}
    P_\pm(H)^\ast=P_\mp( H^\ast).
\end{equation}
\item Although we present the spectral-flattening for matrices with two separation gaps in $\mathcal{PT}$-symmetric systems, the procedure can be similarly performed for other systems with a single gap, or even more separation gaps. We remark on its applicability in Appendix~\ref{ap_sft}.
\end{itemize}

\subsection{Homotopy groups}\label{sc_hpsgp}

In this subsection, we consider the classification of \PT-symmetric Hamiltonians which have separation gaps between the bands on the real axis and the bands on the upper/lower complex planes.

Let us denote $X_S^{(m,n)}$ to be the space of real matrices with $m$ pairs of complex eigenvalues and $n$ real eigenvalues ($2m+n=N$).
Due to symmetry, a jump of $m$ (and hence $n$) cannot happen without a gap closing. 
Therefore, $X_S^{(m,n)}$ with different $(m,n)$ are mutually disconnect.
 
Now fixing $(m,n)$, let us spectral-flatten all matrices in $X_S^{(m,n)}$ according to Eq.~\eqref{eq_sfl}.
Such a procedure is a homotopy equivalence and does not change the topological information.
Hence we only need to study the space of real matrices that are diagonalizable and whose eigenvalues are $0$ (multiplicity=$n$) and $\pm i$ (multiplicity=$m$), denoted as $M^{(m,n)}_S$.

Similar to Eq.~\eqref{eq_hdcp}, any such spectral-flattened matrix can be represented as:
\begin{equation}
    H_F=V
    \tilde H_{DF}
    V^{-1},
\end{equation}
where $V\in GL_{n+2m}(\mathbb R)$ and
\begin{equation}
    \tilde H_{DF}=\begin{pmatrix}
        &\mathbf1_m&\\
        -\mathbf1_m&&\\
        &&0_n
    \end{pmatrix}.
\end{equation}
For such a block-diagonal matrix, all $V$ that commutes with $\tilde H_{DS}$ must have a form of $\big(\begin{smallmatrix}
        A_{2m}&\\
        &B_n
    \end{smallmatrix}\big)$ where $A\in GL_{2m}(\mathbb R)$ commutes with $\big(\begin{smallmatrix}
        0&\mathbf1_m\\
        -\mathbf1_m&0
    \end{smallmatrix}\big)$ and $B\in GL_n(\mathbb R)$.
Equivalently, $A_{2m}$ can be identified as a $GL_m(\mathbb C)$ matrix (intuitively, $GL_m(\mathbb C)$ is a ``gauge'' transformation for the $m$ complex eigenvectors corresponding to $\lambda=i$; since eigenvectors corresponding to $\lambda=-i$ must be conjugate to those of $\lambda=i$, there is no more ``gauge'' freedom corresponding to $\lambda=-i$; see more in Appendix~\ref{ap_hpgl}).
Therefore, after quotienting out those $V$ commuting with $\tilde H_{DS}$, we have
\begin{equation}
    X_S^{(m,n)}\simeq M^{(m,n)}_S=\frac{GL_{n+2m}(\mathbb R)}{GL_m(\mathbb C)\times GL_n(\mathbb R)}.\label{eq-thequotient}
\end{equation}

To get a feeling for how the topology is different in non-Hermitian systems, we compare it to the single-gap Hermitian insulator with \PT{} symmetry: $O(N_1+N_2)/[O(N_1)\times O(N_2)]$ \cite{wu2019non,bouhon2020non}. The groups appearing in the quotient Eq.~\eqref{eq-thequotient} are not homogeneous, one being the complex general linear group while the other being the real general linear group. The action of $GL_m(\mathbb{C})$ on $GL_{n+2m}(\mathbb{R})$ only rotates pairs of complex eigenvectors, instead of an individual eigenvector like what $O_{N_1}$ does. An important observation is that this space cannot be captured by those symmetric spaces in K-theory \cite{kitaevperio} except for $n=0$ (which can be seen using the homotopy equivalence $GL_n(\mathbb R)\simeq O_n,GL_m(\mathbb C)\simeq U_m,$). Therefore it is necessary to go back to homotopy theory to obtain the topological properties.

For the zeroth homotopy set, the result is similar to the band-gapped case:
\begin{equation}\label{eq-PTseppi0}
    \pi_0\left(M_S^{(m,n\ne 0)}\right)=\{1\},\, \pi_0\left(M_S^{(m,0)}\right)=\mathbb Z_2.
\end{equation}

Higher homotopy groups can be worked out using the long exact sequence
\begin{align} \label{eq:seqEV}
    \cdots
    \to \pi_d\big( GL_{n+2m}(\mathbb R)\big)
    \to \pi_d\left(M_S^{(m,n)}\right)\nonumber\\
    \to\pi_{d-1}\big(GL_m(\mathbb C)\times GL_n(\mathbb R)\big)
    \to\cdots.
\end{align}
Since $\pi_{0}\big(GL_m(\mathbb C)\times GL_n(\mathbb R)\big)\to \pi_0\big( GL_{n+2m}(\mathbb R)\big)$ is always injective, we have
\begin{equation}
\begin{aligned}
    &\,\pi_1(M_S^{(m,n)})\\
\cong&\text{coker}[\pi_{1}(GL_m(\mathbb C)\times GL_n(\mathbb R))\to \pi_1( GL_{n+2m}(\mathbb R))].
\end{aligned}
\end{equation}
Similarly, since $\pi_{2}(GL_{n+2m}(\mathbb R))=0$, we know
\begin{equation}\label{eq-MSpi2}
\begin{aligned}
    &\,\pi_2(M_S^{(m,n)})\\
    \cong&\text{ker}[\pi_{1}(GL_m(\mathbb C)\times GL_n(\mathbb R))\to \pi_1( GL_{n+2m}(\mathbb R))].
\end{aligned}
\end{equation}

The map $\pi_{1}\big(GL_m(\mathbb C)\times GL_n(\mathbb R)\big)\to \pi_1\big( GL_{n+2m}(\mathbb R)\big)$ can be analyzed with information listed in Appendix~\ref{ap_hpgl}.
There are several cases:
\begin{itemize}
    \item If $m=0$, or if $m=1$ and $n=0$, then the map is an isomorphism, hence both the kernel and cokernel are trivial.

\item If $m>0,n>2$, then the map concerned is $\mathbb Z\times\mathbb Z_2\to \mathbb Z_2$. Denoting the generators of $\mathbb Z$ and $\mathbb Z_2$ as $w_1$ and $w_2$, then the map is $(aw_1,bw_2)\mapsto a+b ~(mod\, 2)$.
It is surjective.
The kernel is a free Abelian group generated by $w_1+w_2$ alone:
\begin{equation}
    \pi_2\left(M^{(m>0,n>2)}_S\right)\cong\mathbb Z.
\end{equation}

\item If $m>0,n=2$, we need to consider the map $\mathbb Z\times\mathbb Z\to \mathbb Z_2$. 
Denoting their generators as $w'_1$ and $w'_2$, the map is $(aw'_1+bw'_2)\mapsto a+b ~(mod \,2)$.
It is also surjective.
The kernel is generated by $w'_1-w'_2$ and $w'_1+w'_2$. There is no further relation between these two elements, so we have:
\begin{equation}
    \pi_2\left(M_S^{(m>0,n=2)}\right)=\mathbb Z\times \mathbb Z.
\end{equation}

\item If $m>0,n=1$, or if $m>1,n=0$, the map is $\mathbb Z\to \mathbb Z_2$, which is simply the mod 2 map.
It is still surjective.
Its kernel gives 
\begin{equation}
    \pi_2\left(M^{(m>0,n=1)}\right)\cong \pi_2\left(M^{(m>1,n=0)}\right) \cong\mathbb Z. 
\end{equation}
\end{itemize}
To summarize, in all cases, we have
\begin{equation}
    \pi_1(M_S^{(m,n)})=0.
\end{equation}
And we have the second homotopy groups as listed in Table~\ref{tb_fdgsg}.

With this information, we can extract all the topological properties of \PT-symmetric $N$-band systems with gaps separating real and complex eigenvalues. The total space of such separation-gapped matrices is, up to homotopy equivalence, given by
\begin{equation}
    X_{N,S}=\underset{n+2m=N}{\bigcup_{m,n}} X^{(m,n)}_S\simeq\underset{n+2m=N}{\bigcup_{m,n}} M^{(m,n)}_S.
\end{equation}

 \renewcommand{\arraystretch}{1.5}
 \begin{table}[t]
     \centering
     \begin{tabular}{c|c|c|c}
        
             $\pi_2\left(X_S^{(m,n)}\right)$
          & 
        \(m=0\)&
         \(m=1\) &
        \(m>1\)
    \\
         \hline
       \(n=0\) & \multicolumn{2}{c|}{\(\{1\}\)} & $\mathbb Z$ $\left[\frac{C}{2}\right]$
    \\
         \hline
        \(n=1\) &\multirow{3}{*}{$\{1\}$} & \multicolumn{2}{c} {$\mathbb Z$ $\left[\frac{C}{2}\right]$} 
    \\
         \cline{1-1}\cline{3-4}
         \(n=2\) & &\multicolumn{2}{c}{
           $\mathbb Z\times \mathbb Z$ $\left[\frac{C+\chi}{2},\frac{C-\chi}{2}\right]$   }
    \\
         \cline{1-1}\cline{3-4}
         \(n>2\) & &\multicolumn{2}{c}{
           $\mathbb Z$ $[C]$      }
    \\
     \end{tabular}
     \caption{The second homotopy groups corresponding to separation-gapped \PT-symmetric Hamiltonians with $n$ real eigenvalues and $m$ pairs of conjugate complex eigenvalues. 
     They play the dominant role, since the corresponding fundamental groups are trivial $ \pi_1\left(X_S^{(m,n)}\right)=0$. 
     They are often the first topological characters \cite{PhysRevLett.51.51} after fixing the number $n$ of real eigenvalues and the number $m$ of pairs of complex eigenvalues. The numbers $C$ and $\chi$ are the Chern number and the Euler number, see Sec.~\ref{sc_cheu}. Note that only the first line $n=0$ is captured by the previous reference line gap approach.
     }
     \label{tb_fdgsg}
 \end{table}

A slight difference here from band-gapped situations is that these $X^{(m,n)}_S$ do not have a natural correspondence to $\pi_0$ of the total space $X_{N,S}$. The reason is that not all of these spaces are open.
In fact, the total space $X_{N,S}$ itself is connected; it is exactly the space of $N\times N$ real matrices $\mathrm{Mat}(N,\mathbb R)$.
Nevertheless, the sets $X^{(m,n)}_S$ are still disjoint and each separation-gapped phase can only belong to one of them:
\begin{align}
     &\{M_S^{(0,N)},M_S^{(1,N-2)},\dots, M^{(N/2,0)}_{S+},M^{(N/2,0)}_{S-}\},\,\textrm{$N$ even};\nonumber\\
     &\{M_S^{(0,N)},M_S^{(1,N-2)},\dots, M_S^{((N-1)/2,1)}\},\,\textrm{$N$ odd}.\label{eq_zhmtlgp}
\end{align}

The higher homotopy groups depend on which element [Eq.~\eqref{eq_zhmtlgp}] the system belongs to. After fixing $X^{(m,n)}_S$, the fundamental group is always trivial, meaning there is no topology along a loop in the Brillouin zone. An isolated nodal structure that can be enclosed by a loop is not protected in general. However, there can be topological invariants in the bulk of a two-dimensional Brillouin zone, described by the groups in Table~\ref{tb_fdgsg}. An isolated degeneracy that can be enclosed by a two-dimensional sphere between complex and real spectra may also be protected by these topological invariants.

\section{Expressions for topological invariants}\label{sc_expti}

To connect the rather abstract discussions in the previous section to physics, we devote this section to interpreting the general topological invariants obtained from the homotopy classification. This is done by explicitly expressing the homotopy invariants in terms of operators and quantities familiar in physics. We will employ some knowledge from differential geometry. A good reference introducing the relation between geometry and topological invariants is Ref.~\onlinecite{tu2017differential}.

\subsection{Nodal structures and band-gapped phases}

We begin by calculating the invariants of band-gapped phases, which also classify nodal structures. We provide formulae for the zeroth homotopy set \(\pi_0\) and continue to the first homotopy group \(\pi_1\).

\subsubsection{Connected components}
We start with the zeroth homotopy set, which comprises the most fundamental example. 
For $N$-band band-gapped systems, as discussed in Sec.~\ref{sec-PTband0}, the zeroth homotopy set is first determined by the number $n$ of real and the number $m$ of pairs of complex eigenvalues. 
Each pair of $(m,n\neq0)$ satisfying $n+2m=N$ corresponds to one different element of the zeroth homotopy set. 

Furthermore, when $n=0$ (which can only happen when $N=2m$ is even), there is another topological invariant $\nu$ giving the two elements $X^{(N/2,0)}_{\nu=\pm}\in \pi_0(X_N)$. 
As described in Sec.~\ref{sec-PTband0}, this \(\nu\) is given by the determinant of a $GL_N(\mathbb R)$ matrices formed by the eigenvectors. 
As in the two-band case, we label all right eigenvectors corresponding to eigenvalues in the upper half complex plane as $\ket{u_j}=|u_{j,R}\rangle+i|u_{j,I}\rangle,\, 1\le j\le m$. We can form a $N\times N$ invertible matrix $\{|u_{1,R}\rangle,|u_{1,I}\rangle,\dots,|u_{N/2,R}\rangle,|u_{N/2,I}\rangle\}$ from these eigenvectors. 
The sign of
$X^{(N/2,0)}_\nu$ is determined by
\begin{equation}\label{eq-invPTnu}
    \nu=\textrm{sgn}\det \{|u_{1,R}\rangle,|u_{1,I}\rangle,\dots,|u_{N/2,R}\rangle,|u_{N/2,I}\rangle\}.
\end{equation}

\subsubsection{Fundamental groups}
We then turn to the fundamental groups. These are topological invariants along a loop traversing the Brillouin zone or encircling a degeneracy. 
We focus on the \PT-symmetry breaking case since the symmetry-preserving case is already contained in Ref.~\cite{wu2019non} and, for the case of two bands, explicitly discussed in Sec.~\ref{sec_2bd}.
As discussed in Sec.~\ref{sec-PTband1SSB}, the fundamental group is
\begin{equation}
    \pi_1(X^{(m,n)})=\mathbb B_m \times (\mathbb Z_2)^{n-1}~~ (m>0,n>0).
\end{equation}
It includes two parts, the eigenvalue part, and the eigenvector part.

Since it is complicated even to represent elements of the braid group \footnote{People usually represent elements in the braid group by generators and relations, and sometimes reduce them to normal forms.}, the most direct way to obtain the braid structure of complex eigenvalues is to plot the spectrum as in Fig.~\ref{fig_nphs} (c).
The braid Abelianization (an integer in $\mathbb Z$) is often a useful proxy quantity to determine whether a given braid is non-trivial. In group theory, the Abelianization of a group $G$ is the quotient $G/[G,G]$.
The braid Abelianization is expressed as the eigenvalue winding carried by the braid: 
\begin{align}
    \mathbb B_m\to \mathbb Z=\sum_{i< j}\frac{1}{\pi}\int \partial_{\mathbf k}\textrm{arg }\left[E_i(\mathbf k)-E_j(\mathbf k)\right]d\mathbf k,
\end{align}
where the sum is taken over all distinct pairs $i,j$. 
If the Abelianization is non-zero, the braid must be non-trivial, but not vice versa if $m\geq 3$. 
The case $m=2$ is an exception since the braid group $\mathbb B_2$ is isomorphic to $\mathbb Z$. 
The braid Abelianization corresponds to the mutual energy winding number (not to be confused with the reference-point winding, see Appendix~\ref{ap_dfgs}), or vorticity, studied earlier in Refs.~\onlinecite{PhysRevLett.120.146402, PhysRevLett.126.086401}.

To extract/interpret the topology corresponding to the eigenvector part, which are the $(\mathbb Z_2)^{n-1}$ invariants in Table~\ref{tab:homotopy-groups}, we can apply the idea of Wilson loops \cite{wu2019non,PhysRevB.100.195135}. From the sequence in Eq.~\eqref{eq_sqfg}, the $(\mathbb Z_2)^{n-1}$ invariants come from the identification of real eigenvectors different by a sign, $\mathbb R^\times\to -\mathbb R^\times$. This is a non-trivial rotation of the frame formed by the $n$ real eigenvectors when traveling around a loop. 
Assuming that the left and right real eigenvectors are given by continuous $|u^R_j(s)\rangle$ and $\langle u^L_j(s)|$ at each point $s\in[0,1]$ of a path, then the total rotation is given by an $n\times n$ matrix
\begin{equation}\label{eq-nbynWilson}
    \langle u^L_i(1)|u^R_j(0)\rangle. 
\end{equation}

To compute it, we can insert a series of complete basis (eigenvectors) along the path, 
\begin{align}
    &\sum_{j_1,j_2\dots j_W}\langle u^L_i(1)|u^R_{j_W}(s_W)\rangle\langle u^L_{j_W}(s_W) |u^R_{j_{W-1}}(s_{W-1})\rangle\nonumber\\
    &\cdots\langle u^L_{j_1}(s_{1})|u^R_j(0)\rangle,\quad 1>s_W>s_{W-1}>\cdots>s_1>0,
\end{align}
where each summation index $j_1,j_2,\cdots,j_W$ should be taken from $1$ to $2m+n$. The vectors corresponding to complex eigenvalues can also be understood as the pairs of real vectors appearing in $V$ of Eq.~\eqref{eq_hdcp}. 
By taking $W\to\infty$, this is equivalent to the non-Abelian Wilson loop:
\begin{equation} \label{eq:NABC}
    \left(\mathcal Se^{i\int\mathcal A(s)ds}\right)_{ij},
\end{equation}
where the $(n+2m)\times (n+2m)$ Berry connection is $\mathcal A_{ij}(s)=i\langle u^L_i(s)|\frac{d}{ds}u^R_j(s)\rangle$ and $\mathcal S$ means that the integral is path-ordered.

Eq.~\eqref{eq:NABC} is conceptually nice but computationally challenging due to its involvement of eigenvectors corresponding to complex eigenvalues (although we only care about rotations of real eigenvectors) and the necessity of knowing the continuous path of eigenvectors before the calculation to evaluate the Berry connection. 
In practice, once we have found the continuous path of eigenvectors, we can simply use Eq.~\eqref{eq-nbynWilson} to calculate the $\mathbb Z_2^{n-1}$ invariant.
Such continuous paths can be defined by the following procedure, which only involves the real eigenvectors.
We observe that the projectors to the real eigenspaces, $\ket{u^R_j(s)}\bra{u^L_j(s)}=P_j(s)$, $1\le j\le n$, are continuous and periodic for $0\le s\le 1$ (this can be understood by comparing to Chern insulators, where the eigenvectors have a singularity in the BZ but the projection is smooth/periodic). 
In fact, the projector is entirely determined by the operator according to Eq.~\eqref{eq-complexint}:
\begin{equation}
    P_j(s)=\frac{1}{2\pi i}\oint_{\mathcal C\textrm{ around }E_j}\frac{dt}{t-H(s)},
\end{equation}
where the loop is taken around $E_j$. Choosing a well-defined loop $\mathcal C$ is always possible as all energy modes are distinct. The projection satisfies $P^2_j=P_j$, although it may not be orthogonal. With this, we can construct a set of continuous right eigenvectors along a loop. Start with some fixed $|u^R_j(0)\rangle$, we define $|u^R_j(s)\rangle$ as:
\begin{equation}
\ket{u_j^R(s)}\propto\lim_{W\to\infty}P_j(s)P_j\left(\frac{W-1}{W}s\right)\cdots P_j\left(\frac{1}{W}s\right)\ket{u_j^R(0)}.\label{eq_wlev}
\end{equation}
(Note: in fact, if we use  $\ket{u^R}\bra{u^R}$ instead of $\ket{u^R}\bra{u^L}$ in Eq.~\eqref{eq_wlev}, the above-defined $\ket{u^R(s)}$ is normalized automatically in the limit, roughly due to $\lim_{W\to\infty}(1+\frac{1}{W^2})^W=1$.)
The functions constructed above may not be periodic, i.e., $|u^R_j(0)\rangle\ne |u^R_j(1)\rangle$. Assuming that it is non-vanishing, the total rotation along the path is then given by the sign of $\langle u^L_j(0)|u^R_j(1)\rangle$. 
In other words, we can consider
\begin{equation}
    \lim_{W\to\infty}\langle u_j^L(0)|P_j(s)P_j\left(\frac{W-1}{W}s\right)\cdots P_j\left(\frac{1}{W}s\right)\ket{u_j^R(0)}\label{eq_sbbc},
\end{equation}
with different $j$ chosen from $1$ to $n$. 
Their signs tell us how eigenvectors have been flipped during the loop, represented by a list of $n$ signs $(\pm,\pm,\dots,\pm)$. The total number of sign flips must be even in order to preserve the eigenvector frame orientation. So this list of signs can be generated by $(n-1)$ $\mathbb Z_2$ generators, each flipping a pair of eigenvectors: 
\begin{align}
    &(-,-,+,\dots,+),\, (-,+,-,+,\dots,+),\, \dots,\nonumber\\
    &(-,+,\dots,+,-)
\end{align}
The $(\mathbb Z_2)^{n-1}$ invariant is given by the sign flips obtained from the Wilson loop procedure, written in terms of these generators.

Another way of writing the Wilson loop is to allow $u^R_j$ to be complex vectors.
With such freedom, we can actually choose $u^R_j(s)$ to be periodic and continuous simultaneously.
For example, the path $u^R_j(s)\to u^R_j(s)\exp(-i\pi s)$ is periodic if there is a sign flip. 
In this periodic but complex gauge, we can write Eq.~\eqref{eq_sbbc} as $\mathcal P\exp(i\int\mathcal A_{jj}(s)ds)$.
Note that this is a single number (not to be confused with the $jj$ matrix element of a non-Abelian Wilson loop). 
Namely, the integral of Berry connection (holonomy) now gives the sign flip. 
This can be proved by first noticing that for real $\ket{u_j^R(s)}$, 
$\mathcal S\exp(i\int\mathcal A_{jj}(s)ds)$ must be positive. 
When transformed to the complex gauge, $\mathcal S\exp(i\int\mathcal A_{jj}(s)ds)$ then must give the correct sign, due to the covariance of the Wilson loop.
In this point of view, the $(\mathbb Z_2)^{n-1}$ is a generalized Zak phase, existing in $(n-1)$ pairs of bands.

\subsection{Separation-gapped phases}\label{sc_cheu}
\subsubsection{Connected components}
As discussed in Sec.~\ref{sc_hpsgp}, we first need to specify the number $n$ of real and the number $m$ of pairs of complex eigenvalues. 
Each pair of $(m,n)$ satisfying $n+2m=N$ corresponds to different separation-gapped phases. 

Furthermore, according to Eq.~\eqref{eq-PTseppi0}, if $N$ is even and $n=0$, there is another topological invariant $\nu\in\mathbb Z_2$.
The formula for it is exactly the same as Eq.~\eqref{eq-invPTnu}, since they both come from $\pi_0(GL_{2m}(\mathbb R))=\mathbb Z_2$.
Note that this formula works even with degeneracies or exceptional points, as long as $\{\ket{u_{j,R}}+i\ket{u_{j,I}}\}$ $(1\leq j\leq m)$ spans the direct sum of generalized eigenspaces for eigenvalues in the upper half plane.

\subsubsection{Second homotopy group}
The first nontrivial topology beyond $\pi_0$ comes from the second homotopy group $\pi_2\left(M^{(m,n)}_S\right)$. 
This kind of topology appears when the BZ, or the surface we study in the space of control parameters, is a two-dimensional sphere (it also applies if BZ is a two-dimension torus, since $\pi_1(M^{(m,n)}_S)=0$ \cite{PhysRevLett.51.51}). 
To see its physical meaning, we will use the language of vector bundles \cite{tu2017differential} and their characteristic classes, which are familiar in physical literature via 
Berry connections and Berry curvatures
\cite{PhysRevLett.49.405,berry1984quantal,PhysRevLett.52.2111}. 

To start with, the Bloch states form an $N$-dimensional Hilbert space at each $\mathbf k$. By putting all these Hilbert spaces together, we obtain a total space $\textrm{BZ}\times \mathbb R^N$ \cite{PhysRevLett.51.51},
the trivial vector bundle of rank $N$ over the BZ/sphere. 
Next, the bands of the system give a decomposition of this bundle into two subbundles, which are topological, as follows. 
At each point $\mathbf k$, the corresponding eigenvectors in $M^{(m,n)}_S$ give a decomposition of the $N=2m+n$ dimensional real vector space into an $2m$-dimensional real vector subspace and an $n$-dimensional real vector subspace.
The former has a complex structure $\big(\begin{smallmatrix}
    &\mathbf1_m\\-\mathbf1_m&
\end{smallmatrix}\big)$ on it, and can be identified as a $m$-dimensional complex vector space.
It is canonically isomorphic to the complex eigenspace corresponding to eigenvalues with positive imaginary part. The latter $n$-dimensional real vector subspace is the eigenspace corresponding to real eigenvalues.
By joining together all such vector subspaces at every $\mathbf k$, 
we see that we can split the trivial bundle over the BZ  as a direct sum of two vector bundles:
\begin{equation}
    \textrm{BZ}\times \mathbb R^N
    = \tilde F_{\mathbb R} \oplus F \label{eq_vbdcp}
\end{equation}
Here, $\tilde F$ is the $m$-dimensional complex vector bundle, $\tilde F_{\mathbb R}$ is the same as $\tilde F$ as a topological space, but we view it as a $2m$-dimensional real vector bundle; $F$ is the $n$-dimensional real vector bundle. In plainer language, the above equation tells us that the (spontaneous) symmetry-breaking bands of the system give rise to a complex vector bundle $\tilde F$, and the symmetry-preserving bands form a real vector bundle $F$.
These bundles $\tilde F$ and $F$ are often called Bloch bundles in physical literature.
With this decomposition, the topology $\pi_2\left(M^{(m,n)}_S\right)$ is converted to the topology of these vector bundles. 
It is natural to use characteristic classes  \cite{milnor1974characteristic} to characterize the vector bundles.

For the complex vector bundle $\tilde F$, we have Chern classes; while for the real vector bundle $F$, we may have Stiefel-Whitney, Euler, and Pontryagin classes. The Pontryagin class starts to appear from $d=4$ dimension, so it does not enter into our questions. The integrals of the first Chern class and the Euler class over the BZ are often known as the Chern number $C$ and the Euler number $\chi$.
Due to Eq.~\eqref{eq_vbdcp}, these characteristic classes/numbers are not independent. 
Using the Whitney product formula, we have
\begin{equation}
    w(\tilde F_{\mathbb R})\smile  w(F)=1,
\end{equation}
where $w()$ is the total Stiefel–Whitney class and $\smile$ is the cup product \cite{hatcherAlgebraicTopology2002}.
Also note that $w(\tilde F_{\mathbb R})=c(\tilde F) ~(\text{mod}~2)$, where $c()$ is the total Chern class.
The equation tells us the following relation between the first Chern class of the complex bands the second Stiefel-Whitney class of the real bands
\begin{equation}
    c_1(\tilde F) ~(\text{mod}~2)=w_2(F).\label{eq_cswrl}
\end{equation}
Moreover, being a complement of $\tilde F_{\mathbb R}$, $F$ must be orientable and hence has a Euler class $e(F)$, which gives rise to the top Stiefel-Whitney class by mod 2.
Hence,
\begin{equation}
        c_1(\tilde F) \equiv e(F) ~(\text{mod} ~2),~~\text{if $n=2$}.
\end{equation}
The same equation also applies to the corresponding characteristic numbers $C$ and $\chi$.

These equations can be viewed as a generalization of the Chern number summation rule (Chern numbers of all bands must sum to zero) for Hermitian systems. 
More precisely,
\begin{itemize}
    \item When $m>0$ and $n=2$, the system is characterized by $c_1(\tilde F)$ and $e(F)$, and accordingly by the first Chern number $C$ and the Euler number $\chi$, with the constraint that the sum of them must be even.
    \item When $m>0$ and $n>2$, the system is characterized by $c_1(\tilde F)$ and $w_2(F)$, but Eq.~\eqref{eq_cswrl} tells us that $w_2(F)$ is completely determined by the Chern number of the complex bands. So the system can be characterized by a single Chern number $C$.
    \item  When $m>0, n=1$ or $m>1, n=0$, $w_2(F)=0$ due to dimensional constraint, and Eq.~\eqref{eq_cswrl} tells us that $c_1(\tilde F)$, the Chern number, must be even, so the system can be characterized a single integer $C/2$.
    \item When $m=1$ and $n=0$, $\tilde F_{\mathbb R}$ itself is trivial, so $c_1(\tilde F)=e(\tilde F_{\mathbb R})=0$.
\end{itemize}
The above rules also match the elements in Table~\ref{tb_fdgsg}.
This is natural: inspecting the derivation of $\pi_2(M_S^{(m,n)})$ there, we see that on a sphere $S^2$, the Chern number comes from $\pi_1(GL_m(\mathbb C))$, the Euler/Stiefel-Whitney number comes from $\pi_1(GL_n(\mathbb R))$, and Eq.~\eqref{eq-MSpi2} is essentially a restatement of Eq.~\eqref{eq_vbdcp}.

Some characteristic classes can be expressed through Berry curvatures on the vector bundle. In the following we briefly the formalism.
We will also explain our choice in the left and right eigenvectors for connection and curvature matrices.

A connection on a vector bundle is a rule for how to take a derivative within that bundle \cite{tu2017differential}. This follows from the idea of defining a parallel transport.
Let's first consider the complex vector bundle $\tilde F$.
As discussed above, we can think of each fiber as spanned by $m$ right complex eigenvectors corresponding to the complex eigenvalues with positive imaginary parts. 
As a subbundle of the trivial bundle, there is a naturally defined induced connection $\nabla_a$ on it:
\begin{equation}
   \nabla^{\tilde F}_a|u^R_j(\mathbf k)\rangle\equiv
   P_{\tilde F}\partial_a|u^R_j(\mathbf k)\rangle=-i\sum_{j'} \mathcal A_{j'j,a}(\mathbf k)|u^R_{j'}(\mathbf k)\rangle.\label{eq_dfci}
\end{equation}
Here the second equation is the expansion of $P_{\tilde F}\partial_a|u^R_j(\mathbf k)\rangle$ in terms of the basis $|u^R_{j'}(\mathbf k)\rangle$. 
The operator $P_{\tilde F}$ is a projection (does not need to be orthogonal) onto the $m$-dimensional subspace we want to consider;
$j$ and $j'$ can only take $m$ values also.
A natural choice of $P_{\tilde F}$ is given by the eigen-decomposition of $H$.
Introducing the left eigenvectors and using the biorthogonal condition $\langle u^L_i(\mathbf k)|u^R_j(\mathbf k)\rangle=\delta_{ij}$, $P_{\tilde F}$ can be represented as $P_{\tilde F}=\sum_{ij} \ket{u^R_i}\bra{u^L_j}$, and the connection matrices $\mathcal A$ are:
\begin{equation}
    \mathcal A_{ij,a}(\mathbf k)= i\langle u^L_i(\mathbf k)|\partial_au^R_j(\mathbf k)\rangle.\label{eq_bcm}
\end{equation}
This expression gives back the Berry connection \cite{PhysRevLett.49.405,berry1984quantal,PhysRevLett.52.2111} in Hermitian cases. The corresponding Berry curvature $\Omega$ is a matrix-valued two-form, given by the Berry connection through
\begin{equation}
     \Omega_{ij,xy}=\partial_x\mathcal A_{ij,y}-\partial_y\mathcal A_{ij,x}-i[\mathcal A_x,\mathcal A_y]_{ij}.\label{eq_bc}
\end{equation}

The basis $\{|u^R_j(\mathbf k)\rangle\}$ does not have to be orthonormal, yet Eq.~\eqref{eq_dfci} is still a well-defined connection since $P$ and $\partial_a$ are well-defined globally.
The well-definedness can also be verified by observing that under a linear transformation $V(\mathbf k)$ on the eigenvectors, the connection matrix transforms as
\begin{equation}
    \mathcal A_{a}(\mathbf k)\to V^{-1}\mathcal A_{a}(\mathbf k)V+iV^{-1}(\mathbf k)\partial_a V(\mathbf k).\label{eq_bctr}
\end{equation}
This transformation will leave the Berry curvature Eq.~\eqref{eq_bc} covariant in the sense that $\Omega\to V^{-1}\Omega V$. 
This is to say, $\Omega$ is a $(1,1)$-tensor well-defined globally on the BZ.
The first Chern number is then expressed in terms of the non-Abelian Berry curvature through \cite{tu2017differential,ryu2010topological}:
\begin{equation}\label{eq-Berryintegral}
    C=\int\frac{1}{2\pi} \textrm{tr }_{\uparrow} \Omega_{ij,xy} (\mathbf k)d^2 k=\int\frac{1}{2\pi}\sum_{\textrm{Im }E_j>0} \Omega_{jj,xy} (\mathbf k)d^2 k,
\end{equation}
Again $\textrm{tr }_{\uparrow}$ is over the bands with energies in the upper half complex plane. The bands from the lower half complex plane are their complex conjugate and thus not independent. In fact, those bands have Chern number $-C$. This can be seen by computing the corresponding Berry curvature and noticing that the last commutator in the Eq.~\eqref{eq_bc} actually does not contribute to the Chern number.

Note that we have used the ``LR" convention here: replace the bra and ket states in standard definition in physical literature \cite{berry1984quantal,PhysRevLett.52.2111} by the left and right eigenvectors respectively. 
From the above discussions, it is a natural choice for calculating the Berry connection and Berry curvature in non-Hermitian systems.
In comparison, if we choose right-right or left-left eigenvectors in Eq.~\eqref{eq_bcm}, the connection matrix does not transform according to Eq.~\eqref{eq_bctr}. Additional care is needed to let it be well-defined for the entire BZ (details in Appendix~\ref{ap_cnbc}). After fixing this issue, one can prove that different choices of left and right eigenvectors for the Berry connection give the same Chern number (see Ref.~\cite{PhysRevLett.120.146402} and Appendix~\ref{ap_cnbc}). The biorthogonal connection that we use serves as the most covariant form.

Now we look at the topological invariants for the bands of real energy. 
As we have discussed, the real vector bundle $F$ corresponding to real eigenvectors is orientable, it can have a nontrivial Euler class when $n=2$. We denote the corresponding eigenvectors as $|u_1^R(\mathbf k)\rangle, |u_2^R(\mathbf k)\rangle$.
Similar to the Chern class, the Euler class is an intrinsic topological invariant that does not depend on the details of the geometry. But in order to write the Euler class through geometric quantities, there are some restrictions \cite{tu2017differential,milnor1974characteristic}:
we need to pick up a connection that is compatible with a Riemannian metric on the vector bundle and write down the curvature matrix under an orthonormal basis (more elaborate explanation in Appendix~\ref{ap_eulcl}). In Hermitian systems, these two conditions are automatically satisfied due to the orthogonality of eigenvectors. 
In contrast, we lose this convenience in non-Hermitian systems. The right eigenvectors $|u_1^R(\mathbf k)\rangle, |u_2^R(\mathbf k)\rangle$ of a non-Hermitian matrix usually have a non-vanishing overlap. 
To amend this, we need to orthogonalize the real eigenvectors by hand and obtain an orthonormal basis $\{|\tilde u_1(\mathbf k)\rangle,|\tilde u_2(\mathbf k)\rangle\}$ that is compatible with the orientation and spans the eigenspace of real bands. 
The new connection matrix is to be evaluated by:
\begin{equation}     
\tilde{\mathcal A}_{jj',a}(\mathbf k)=i\langle \tilde u_j(\mathbf k)|\partial_a \tilde u_{j'}(\mathbf k)\rangle.
\end{equation}
The new curvature $\widetilde\Omega_{ij,xy}$ is obtained by inserting this connection into Eq.~\eqref{eq_bc}. Its detailed expression is in Appendix~\ref{ap_eulcl}. 
This will be a skew-symmetric matrix and its off-diagonal part is its Pfaffian:
\begin{equation}
     \textrm{Pf }\widetilde\Omega_{ij,xy}\big\vert_{i,j\in\{1,2\}}=\widetilde\Omega_{12,xy}=-\widetilde\Omega_{21,xy},
\end{equation}
The Euler number is then given by the integral of this component \cite{PhysRevX.9.021013,bouhon2020non,tu2017differential}:
\begin{equation}
    \chi=\int \frac{1}{2\pi i} \widetilde\Omega_{12,xy}d^2k.
\end{equation}
The additional factor of $i$ in the denominator comes from the fact that the Berry curvature matrix is different from the usual definition of curvature matrix in mathematical literature by a factor of $-i$ [see Eq.~\eqref{eq_dfci}]. 

At the end of this section, we note that although the Stiefel-Whitney numbers usually cannot be expressed through invariant polynomials of the curvature, it is not a problem here. They can always be derived from the Chern number by the summing rule in our situation [c.f. discussion after Eq.~\eqref{eq_cswrl}]. Another observation is that Euler, Chern and Stiefel-Whitney numbers  intrinsically come from characteristic classes. After identifying them with the homotopy invariants, the results of $\pi_2$ apply to any two-dimensional surfaces due to the universal validity of characteristic classes. We do not need to worry about whether the invariants are living on a sphere or a torus.

\section{Fragile and stable topology}\label{sc_fgstbto}

We now turn to a discussion about the stability of the topological invariants derived in the previous sections. 
We will demonstrate that some stable Hermitian topology can be trivialized by adding 
trivial non-Hermitian bands. This leads to more interesting phenomena that do not have counterparts in Hermitian systems, especially a different kind of stability and fragility of the $2\pi$ frame topology. This frame topology has been demonstrated to possess edge states in recent experiments \cite{guo2021experimental}.

Let us first introduce the notion of fragile and stable topology. A topological gapped phase of an $N$-band system cannot be reduced to a trivial phase unless the gap closes at some point. This is to say, the $N\times N$-matrix-valued function $H_N(\mathbf k)$ can not be deformed into a constant matrix-valued function while keeping the eigenvalues non-degenerate. 
Now let us add trivial bands and convert the system to $H_N(\mathbf k)\oplus H_{\textrm{trivial}}$, with  $H_{\textrm{trivial}}$ a constant non-degenerate matrix function. 
This new matrix function $H_N(\mathbf k)\oplus H_{\textrm{trivial}}$ acts in an enlarged Hilbert space. 
The topological nature of $H_N(\mathbf k)\oplus H_{\textrm{trivial}}$ may be different from $H_N(\mathbf k)$. 
The $N$-band topological phase (character) is said to be \emph{fragile}, if there exists $H_{\textrm{trivial}}$ such that $H_N(\mathbf k)\oplus H_{\textrm{trivial}}$ can be reduced to a constant matrix function during some gap-preserving deformation.  Topological phases that are robust to such inclusion of trivial bands are said to be \emph{stable}.

\subsection{Band-gapped phases and nodal structures}

Whether a topological phase (character) is stable or not can be found out by checking Table~\ref{tab:homotopy-groups}. Adding trivial bands can be considered as increasing the number $m$ and $n$ by $\delta m$ and $\delta n$ respectively. By doing so, we have a homomorphism $\pi_1(X^{(m,n)})\to \pi_1(X^{(m+\delta m,n+\delta n)})$, induced by the inclusion map. 
If the kernel of this homomorphism is nontrivial, 
The topological phase in $\pi_1(X^{(m,n)})$ living in the kernel (if nontrivial) of this map is mapped to a trivial phase in $\pi_1(X^{(m+\delta m,n+\delta n)})$ and are thus fragile.

\begin{figure}
    \centering
    \def\svgwidth{\columnwidth}
    %--

    \begingroup%
  \makeatletter%
  \providecommand\color[2][]{%
    \errmessage{(Inkscape) Color is used for the text in Inkscape, but the package 'color.sty' is not loaded}%
    \renewcommand\color[2][]{}%
  }%
  \providecommand\transparent[1]{%
    \errmessage{(Inkscape) Transparency is used (non-zero) for the text in Inkscape, but the package 'transparent.sty' is not loaded}%
    \renewcommand\transparent[1]{}%
  }%
  \providecommand\rotatebox[2]{#2}%
  \newcommand*\fsize{\dimexpr\f@size pt\relax}%
  \newcommand*\lineheight[1]{\fontsize{\fsize}{#1\fsize}\selectfont}%
  \ifx\svgwidth\undefined%
    \setlength{\unitlength}{246bp}%
    \ifx\svgscale\undefined%
      \relax%
    \else%
      \setlength{\unitlength}{\unitlength * \real{\svgscale}}%
    \fi%
  \else%
    \setlength{\unitlength}{\svgwidth}%
  \fi%
  \global\let\svgwidth\undefined%
  \global\let\svgscale\undefined%
  \makeatother%
  \begin{picture}(1,0.85721973)%
    \lineheight{1}%
    \setlength\tabcolsep{0pt}%
    \put(0,0){\includegraphics[width=\unitlength,page=1]{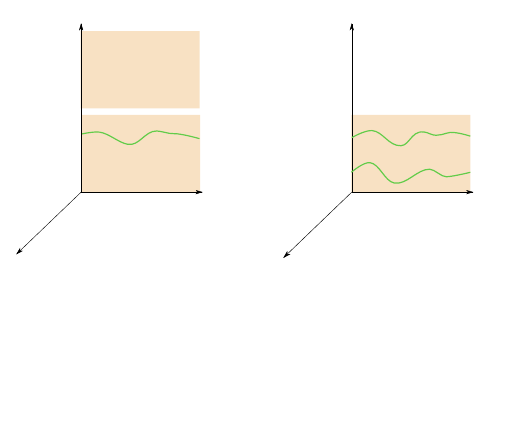}}%
    \put(0.37577258,0.4486832){\color[rgb]{0,0,0}\makebox(0,0)[lt]{\lineheight{1.25}\smash{\begin{tabular}[t]{l}\(k\)\end{tabular}}}}%
    \put(0,0){\includegraphics[width=\unitlength,page=2]{fig10.pdf}}%
    \put(0.15447166,0.79109692){\color[rgb]{0,0,0}\makebox(0,0)[rt]{\lineheight{1.25}\smash{\begin{tabular}[t]{r}Re\(E\)\end{tabular}}}}%
    \put(0.6830884,0.79109692){\color[rgb]{0,0,0}\makebox(0,0)[rt]{\lineheight{1.25}\smash{\begin{tabular}[t]{r}Re\(E\)\end{tabular}}}}%
    \put(0.03874563,0.38998262){\color[rgb]{0,0,0}\makebox(0,0)[t]{\lineheight{1.25}\smash{\begin{tabular}[t]{c}Im\(E\)\end{tabular}}}}%
    \put(0.55176763,0.38388506){\color[rgb]{0,0,0}\makebox(0,0)[t]{\lineheight{1.25}\smash{\begin{tabular}[t]{c}Im\(E\)\end{tabular}}}}%
    \put(0.55176763,0.63465887){\color[rgb]{0,0,0}\makebox(0,0)[t]{\lineheight{1.25}\smash{\begin{tabular}[t]{c}Im\(E\)\end{tabular}}}}%
    \put(0.02202948,0.82476818){\color[rgb]{0,0,0}\makebox(0,0)[t]{\lineheight{1.25}\smash{\begin{tabular}[t]{c}(a)\end{tabular}}}}%
    \put(0.52311674,0.82735405){\color[rgb]{0,0,0}\makebox(0,0)[t]{\lineheight{1.25}\smash{\begin{tabular}[t]{c}(b)\end{tabular}}}}%
    \put(0.12344254,0.52329242){\color[rgb]{0,0,0}\makebox(0,0)[rt]{\lineheight{1.25}\smash{\begin{tabular}[t]{r}\(\psi_1\)\end{tabular}}}}%
    \put(0.66654314,0.52329242){\color[rgb]{0,0,0}\makebox(0,0)[rt]{\lineheight{1.25}\smash{\begin{tabular}[t]{r}\(\psi_1\)\end{tabular}}}}%
    \put(0.13300678,0.58587833){\color[rgb]{0,0,0}\makebox(0,0)[rt]{\lineheight{1.25}\smash{\begin{tabular}[t]{r}\(\psi_2\)\end{tabular}}}}%
    \put(0.66959196,0.58587833){\color[rgb]{0,0,0}\makebox(0,0)[rt]{\lineheight{1.25}\smash{\begin{tabular}[t]{r}\(\psi_2\)\end{tabular}}}}%
    \put(0.91867273,0.51390535){\color[rgb]{0,0,0}\makebox(0,0)[lt]{\lineheight{1.25}\smash{\begin{tabular}[t]{l}\(E_1(k)\)\end{tabular}}}}%
    \put(0.91867273,0.58065185){\color[rgb]{0,0,0}\makebox(0,0)[lt]{\lineheight{1.25}\smash{\begin{tabular}[t]{l}\(E_2(k)\)\end{tabular}}}}%
    \put(0.39101671,0.58065185){\color[rgb]{0,0,0}\makebox(0,0)[lt]{\lineheight{1.25}\smash{\begin{tabular}[t]{l}\(E_2(k)\)\end{tabular}}}}%
    \put(0.39101671,0.51390535){\color[rgb]{0,0,0}\makebox(0,0)[lt]{\lineheight{1.25}\smash{\begin{tabular}[t]{l}\(E_1(k)\)\end{tabular}}}}%
    \put(0.89929175,0.4486832){\color[rgb]{0,0,0}\makebox(0,0)[lt]{\lineheight{1.25}\smash{\begin{tabular}[t]{l}\(k\)\end{tabular}}}}%
    \put(0.10773071,0.40814371){\color[rgb]{0,0,0}\makebox(0,0)[lt]{\lineheight{1.25}\smash{\begin{tabular}[t]{l}stable \(2\pi\)-frame topology\end{tabular}}}}%
    \put(0.64804763,0.40814371){\color[rgb]{0,0,0}\makebox(0,0)[lt]{\lineheight{1.25}\smash{\begin{tabular}[t]{l}fragile \(2\pi\)-frame topology\end{tabular}}}}%
    \put(0.16503284,0.76880501){\color[rgb]{0,0,0}\makebox(0,0)[lt]{\lineheight{1.25}\smash{\begin{tabular}[t]{l}add bands in \(\mathbb{R}\)\end{tabular}}}}%
    \put(0.690772,0.76880501){\color[rgb]{0,0,0}\makebox(0,0)[lt]{\lineheight{1.25}\smash{\begin{tabular}[t]{l}add bands in \(\mathbb{C}\setminus\mathbb{R}\)\end{tabular}}}}%
    \put(0.48046777,0.17059787){\color[rgb]{0,0,0}\makebox(0,0)[lt]{\lineheight{1.25}\smash{\begin{tabular}[t]{l}\(\mathcolor{red}{\Z_2}\)\end{tabular}}}}%
    \put(0.48046777,0.06453692){\color[rgb]{0,0,0}\makebox(0,0)[lt]{\lineheight{1.25}\smash{\begin{tabular}[t]{l}\(\mathcolor{red}{ ( \Z_2 )^{n-1} }\)\end{tabular}}}}%
    \put(0.68491023,0.17889511){\color[rgb]{0,0,0}\makebox(0,0)[lt]{\lineheight{1.25}\smash{\begin{tabular}[t]{l}\(\mathcolor{blue}{\B_m}\)\end{tabular}}}}%
    \put(0.26593474,0.01453764){\color[rgb]{0,0,0}\makebox(0,0)[rt]{\lineheight{1.25}\smash{\begin{tabular}[t]{r}\(n\)\end{tabular}}}}%
    \put(0.84793419,0.29206195){\color[rgb]{0,0,0}\makebox(0,0)[lt]{\lineheight{1.25}\smash{\begin{tabular}[t]{l}\(m\)\end{tabular}}}}%
    \put(0.35103623,0.29206195){\color[rgb]{0,0,0}\makebox(0,0)[t]{\lineheight{1.25}\smash{\begin{tabular}[t]{c}\(m=0\)\end{tabular}}}}%
    \put(0.67879843,0.29206195){\color[rgb]{0,0,0}\makebox(0,0)[t]{\lineheight{1.25}\smash{\begin{tabular}[t]{c}\(m>0\)\end{tabular}}}}%
    \put(0.16788653,0.17276853){\color[rgb]{0,0,0}\makebox(0,0)[lt]{\lineheight{1.25}\smash{\begin{tabular}[t]{l}\(n=2\)\end{tabular}}}}%
    \put(0.16788653,0.06453683){\color[rgb]{0,0,0}\makebox(0,0)[lt]{\lineheight{1.25}\smash{\begin{tabular}[t]{l}\(n>2\)\end{tabular}}}}%
    \put(0.11300853,0.23221974){\color[rgb]{0,0,0}\makebox(0,0)[lt]{\lineheight{1.25}\smash{\begin{tabular}[t]{l}\(n\in\{0,1\}\)\end{tabular}}}}%
    \put(0.35103623,0.17276853){\color[rgb]{0,0,0}\makebox(0,0)[t]{\lineheight{1.25}\smash{\begin{tabular}[t]{c}\(\Z\)\end{tabular}}}}%
    \put(0.67879842,0.23913749){\color[rgb]{0,0,0}\makebox(0,0)[t]{\lineheight{1.25}\smash{\begin{tabular}[t]{c}\(\B_m\)\end{tabular}}}}%
    \put(0,0){\includegraphics[width=\unitlength,page=3]{fig10.pdf}}%
    \put(0.67879842,0.11865272){\color[rgb]{0,0,0}\makebox(0,0)[t]{\lineheight{1.25}\smash{\begin{tabular}[t]{c}\((\Z_2)^{n-1} \times \B_m\)\end{tabular}}}}%
    \put(0.35476849,0.11865272){\color[rgb]{0,0,0}\makebox(0,0)[lt]{\lineheight{1.25}\smash{\begin{tabular}[t]{l}\(\mathcolor{red}{\Z_4}\)\end{tabular}}}}%
    \put(0.35103623,0.06453683){\color[rgb]{0,0,0}\makebox(0,0)[t]{\lineheight{1.25}\smash{\begin{tabular}[t]{c}\(Q(n)\)\end{tabular}}}}%
    \put(0,0){\includegraphics[width=\unitlength,page=4]{fig10.pdf}}%
    \put(0.02412228,0.30520919){\color[rgb]{0,0,0}\makebox(0,0)[t]{\lineheight{1.25}\smash{\begin{tabular}[t]{c}(c)\end{tabular}}}}%
  \end{picture}%
\endgroup%

    %--
    \caption{Illustration of fragile and stable topology. (a) The minus one frame charge (represented by a $2\pi$ rotation of eigenvectors along a loop) leading to non-Abelian frame topology is stable in Hermitian situation or \PT{}-symmetry preserved phases. (b) The $2\pi$ topology can be trivialized by adding two bands spontaneously breaking \PT{} symmetry. (c) To determine whether a topological character is stable or fragile, we do an excursion in Table~\ref{tab:homotopy-groups} by increasing $m$ and $n$. The system with smaller $m$ and $n$ can be included in the system with larger $m$ and $n$. The inclusion map is indicated beside the colored arrows. If the inclusion map is injective (blue), the topology is stable. If the inclusion map has a non-vanishing kernel (red), we have loss and the topology can be fragile.}
    \label{fig_fragile}
\end{figure}

We first look at the Hermitian situation or the \PT-preserving situation, the element in Table~\ref{tab:homotopy-groups} corresponding to $m=0$, $n=2$. 
We claim that some of the topological phases, described by the winding number of $(d_x,d_z)$, is fragile: by increasing from $n=2$ to $n>2$, the inclusion $\pi_1(X^{(0,2)})\to \pi_1(X^{(0,n)})$ maps $\mathbb Z$ to a $\mathbb Z_4$ subgroup in $Q(n)$. 
Indeed, a generator of $\pi_1(X^{(0,2)})$ is a $2\pi$ rotation of $(d_x,d_z)$, causing a $\pi$ rotation of $(\ket{u_1},\ket{u_2})$:
$|u_1\rangle\to-|u_1\rangle,|u_2\rangle\to -|u_2\rangle$. 
This is lifted to $e_1\in Q(n)$ as discussed in Sec.~\ref{sc_fdPTbk}.
From the group relations Eq.~\eqref{eq_dfsvg}, we see that $e_1^4=1$, meaning that the kernel is non-trivial. All $2$-band phases with a winding number of multiples of $4$ are fragile upon adding trivial bands with real eigenvalues. In contrast, the topology in $ Q(n)$ is stable as long as there is no spontaneous $\mathcal{PT}$ symmetry breaking, reflected in Fig.~\ref{fig_fragile} (a). By adding more gapped trivial bands with real eigenvalues, we just increase the number of generators $e_j$. There is no loss of topology in the many-real-band limit.

The fragile topology of winding numbers has long been known in Hermitian systems. However, adding trivial bands with complex energy to our systems will bring very different phenomena. This corresponds to moving right in Table~\ref{tab:homotopy-groups}.

From Table~\ref{tab:homotopy-groups}, adding complex bands reduce the non-Abelian group $Q(n)$ to the Abelian group $(\mathbb Z_2)^{n-1}$. This is a notable feature of spontaneous $\mathcal{PT}$ symmetry breaking: Abelianization of the frame charges. When all eigenvalues are real, the systems are featured by frame charges that ``anticommute'' with each other $e_ie_j=-e_je_i$ (note that the right-hand side is the product of three group elements, $-1,e_j$ and $e_i$). The key of this non-Abelian topology is the $-1$ frame charge, appearing in the commutators between $e_i$ and $e_j$. This charge comes from the fundamental group $\pi_1\left[GL_N(\mathbb R)\right]=\mathbb Z_2$. When the \PT{} symmetry is spontaneously broken, this $-1$ charge is killed by the complex eigenvalues, albeit the complex bands are trivially gapped. As a consequence, the frame charges are Abelianized, as shown in Fig.~\ref{fig_fragile} (b). 
A hint can be found by comparing Eq.~\eqref{eq_sqpts} and Eq.~\eqref{eq-pi1PTbreaking} and notice the difference between $m=0$ and $m>0$. 
Below we give a simple explicit example of how this $-1$ frame charge is annihilated by the complex eigenvalues. 

The charge $-1$ in $\pi_1\left[GL_N(\mathbb R)\right]=\mathbb Z_2$ can be understood as a $2\pi$ rotation of a pair of real eigenvectors, via the natural map $\pi_1\left[GL_2(\mathbb R)\right]\to \pi_1\left[GL_N(\mathbb R)\right]$.
Without loss of generality, we take the real eigenvalues fixed to be $1,-1$.
Now let us add a pair of complex eigenvalues at $i,-i$.
The topological phase with charge $-1$ can be understood as the following loop of matrices:
\begin{equation}
    H_0(\theta)=V_0(\theta) D_0 V_0^{-1}(\theta), ~~\theta\in [0,2\pi].
\end{equation}
Here
\begin{equation}
\begin{aligned}
    &D_0=\begin{pmatrix}
        \sigma_z&\\
        &i\sigma_y
    \end{pmatrix},\\
    &V_0(\theta)=\begin{pmatrix}
    R_{\theta}&\\
    & \mathbf1_{2}
    \end{pmatrix},~~
        R_\theta=\begin{pmatrix}
    \cos\theta  &\sin\theta\\
    -\sin\theta&\cos\theta 
    \end{pmatrix}.
\end{aligned}
\end{equation}
where $ \mathbf1_{2}$ is the $2\times 2$ identity matrix. When $\theta$ goes from $0$ to $2\pi$, the first two real eigenvectors perform a $2\pi$ rotation in the plane they span. The Hamiltonian goes back to itself after the rotation $H_0(0)=H_0(2\pi)=D_0$ (this is why we call it a loop of matrices). Notice, if we replace $\theta$ by $k_x$ and treat $H_0(k_x)$ as a Bloch Hamiltonian in one dimension, then the top-left blocks in $D_0$ and $V_0$ carry the same topology as the model studied in \cite{guo2021experimental}.

Now we want to show that this loop of matrices $H_0(\theta)$ can be continuously deformed via a family of loops $H_t(\theta)$ into the constant loop 
\begin{equation}\label{eq-constloop}
    H_1(\theta)\equiv D_0
\end{equation}
without gap closing.
To do so, we demonstrate that we can gradually deform the loop $V_0(\theta)$ via a family of loops $V_t(\theta)$ (continuous in both $t$ and $\theta$, $V_t(\theta)\in GL_4(\mathbb R)$) to a new loop $V_1(\theta)$:
\begin{equation}
    V_1(\theta)=\begin{pmatrix}
       \mathbf1_{2} &\\
        &  R_{\theta}
    \end{pmatrix}. 
\end{equation}
Defining the family of Hamiltonian loops via:
\begin{equation}\label{eq-Httheta}
        H_t(\theta)=V_t(\theta) D_0 V_t^{-1}(\theta),
\end{equation}
then Eq.~\eqref{eq-constloop} will be satisfied, since $R_\theta$ is a linear combination of $ \mathbf1_{2}$ and $\sigma_y$. 
As $V_t(\theta)$ does not change eigenvalues, the system is gapped everywhere during the deformation. This proves that the $-1$ charge is lifted in the presence of complex eigenvalues.
Simply speaking, we have ``traded" the $2\pi$ rotation of two real eigenvectors into a $e^{2\pi i}$ rotation of the complex eigenvector, the latter is trivial loop since the phase of a complex eigenvector does not affect the matrix $H$.

The existence of $V_t(\theta)$ is proven as follows. We observe that $V_0(\theta)$ can be transformed to $V_1(\theta)$ by the following action
\begin{equation}
    V_1(\theta)=\begin{pmatrix}
        & \mathbf1_{2}\\
        \mathbf1_{2} &
    \end{pmatrix}V_0(\theta)\begin{pmatrix}
        & \mathbf1_{2}\\
        \mathbf1_{2} &
    \end{pmatrix}.
\end{equation}
The determinant of $\begin{pmatrix}& \mathbf1_{2}\\ \mathbf1_{2} &\end{pmatrix}$ is $1$, as this is a general linear transformation switching two pairs of coordinates. Since $GL_4(\mathbb R)$ only has two path-connected components, distinguished by the sign of the determinant, all matrices with the same determinant can be continuously deformed to each other. Hence, we can find a continuous function of matrix $K_t\in GL^+_4(\mathbb R)$ interpolating the identity matrix and $\begin{pmatrix}& \mathbf1_{2}\\ \mathbf1_{2} &\end{pmatrix}$:
\begin{equation}
    K_0=\mathbf1_{4},\, K_t\in GL^+_4(\mathbb R),\,  K_1=\begin{pmatrix}& \mathbf1_{2}\\ \mathbf1_{2} &\end{pmatrix}.
\end{equation}
One such realization of \(K_t\) is 
\begin{widetext}
\begin{equation}
   K_t =  \mqty(
     \cos \left(\frac{t\pi}{2}\right) & -\sin \left(\frac{t\pi}{2}\right)
   (\cos \left(\frac{t\pi}{2}\right)) & \sin ^2\left(\frac{t\pi}{2}\right) & 0 \\
 \sin \left(\frac{t\pi}{2}\right) \cos \left(\frac{t\pi}{2}\right) & \cos
   \left(\frac{t\pi}{2}\right) \cos (t\pi ) & -2
   \sin \left(\frac{t\pi}{2}\right) \cos ^2\left(\frac{t\pi}{2}\right) &
   \sin ^2\left(\frac{t\pi}{2}\right) \\
 \sin ^2\left(\frac{t\pi}{2}\right) & 2 \sin \left(\frac{t\pi}{2}\right)
   \cos ^2\left(\frac{t\pi}{2}\right) & \cos \left(\frac{t\pi}{2}\right)
   \cos (t\pi ) & -\sin \left(\frac{t\pi}{2}\right)
   (\cos \left(\frac{t\pi}{2}\right)) \\
 0 & \sin ^2\left(\frac{t\pi}{2}\right) & \sin\left(\frac{t\pi}{2}\right) \cos \left(\frac{t\pi}{2}\right) & \cos \left(\frac{t\pi}{2}\right)
   ).
\end{equation}
\end{widetext}
With the help of $K_t$, we can build $V_t(\theta)$ as
\begin{equation}\label{eq-Vttheta}
    V_t(\theta)=K_tV_0(\theta)K^{-1}_t.
\end{equation}
According to previous discussions, $H_t(\theta)$ defined by Eq.~\eqref{eq-Httheta} and Eq.~\eqref{eq-Vttheta} will be a continuous interpolation between $H_0(\theta)$, the loop of matrices carrying $-1$ frame charge, to $H_1(\theta)$, the constant matrix loop. 
This proves that the complex eigenspace identifies the frame charge $-1$ with the trivial charge $1$. All frame charges now commute with each other and give Abelian eigenvector topology.

The braid $\mathbb B_m$ topology and $(\mathbb Z_2)^{n-1}$ topology in Table~\ref{tab:homotopy-groups} are always stable. The braid comes from the nontrivial mutual winding of the complex energy. This cannot be destroyed if more trivial gapped bands are included. From Sec.~\ref{sc_expti}, the remaining $(\mathbb Z_2)^{n-1}$ topology is the sign change of the real eigenvectors following a loop. This manifests the discrete gauge transformation of real eigenvectors. It does not change upon adding trivial gapped bands. The results are summarized in Fig.~\ref{fig_fragile}.

We also remark in the end that the frame orientation $\nu$ for an even number of bands $N=2m$ is a fragile topology. It is removed when one more band of real energy is added to the system.

\subsection{Separation gap}

As we have shown in Table~\ref{tb_fdgsg}, the Euler number is not stable. 
It only appears in models with two bands of real eigenvalues. 
When more such real bands are added, the Euler class of the original two bands is reduced to a $\mathbb Z_2$ invariant in the second Stiefel-Whitney class, with the latter being stable. This reproduces the fragile topology of the Euler class in Hermitian models \cite{PhysRevX.9.021013}. 
The Chern numbers on the other hand are stable and always protect the separation gaps.

\section{Pseudo-Hermitian systems}\label{sc_psedh}

The \PT-symmetric system is closely related to pseudo-Hermitian systems. An operator $H$ is said to be $\eta$-pseudo-Hermitian if there exists an invertible operator $\eta$ such that \cite{MostafazadehPSH}
\begin{equation}\label{eq-etarelation}
   \eta H^\dagger\eta^{-1}=H.  
\end{equation}
In our work, we further demand $\eta$ to be unitary \cite{Bernard2002}, such that it does not change the norms of physical states. The pseudo-Hermiticity generalizes the usual notion of Hermiticity, which can be recovered for $\eta=1$. 
It is not hard to show each \PT-symmetric Hamiltonian belongs to some $\eta$-pseudo-Hermitian class \cite{MostafazadehPSH}.
The classification of $\eta$-pseudo-Hermitian Hamiltonians requires precise knowledge of $\eta$, which depends on the details of systems.
In fact, the signatures of $\eta$ have close connections to the nature of a degeneracy  \cite{PhysRevResearch.5.023035,PhysRevA.108.022206}.
We will give a thorough classification for all different classes of $\eta$.

\subsection{Two-band model}
As a simple example, let us consider the two-band case where $\eta=\text{diag}\{1,-1\}$.
We parameterize our matrix $H$ as
\begin{equation}
    H=d_x\sigma_x+d_y\sigma_y+d_z\sigma_z,
\end{equation}
where $d_x, d_y, d_z\in\mathbb{C}$ and we have assumed that $H$ is traceless.
A straightforward calculation shows that, the $\eta$-pseudo-Hermitian condition is equivalent to:
\begin{equation}
    d_x^*=-d_x,~~d_y^*=-d_y,~~d_z^*=d_z.
\end{equation}
Hence we can now parameterize $H$ with real parameters as
\begin{equation}
    H=i\tilde d_x\sigma_x+i\tilde d_y\sigma_y+d_z\sigma_z,
\end{equation}
where now $\tilde d_x, \tilde d_y, d_z\in\mathbb{R}$.
The condition of non-degeneracy becomes:
\begin{equation}
    d_z^2\neq \tilde{d}_x^2+ \tilde d_y^2.
\end{equation}

It is evident that this is the same situation as in Sec.~\ref{sec-PTtwoband} (with $y$ and $z$ switched, and an overall prefactor of $i$), see also Fig.~\ref{fig_2bspace}. 
There are three connected components, two of which are contractible, and the remaining one is homotopy equivalent to $S^1$ (note that different from Sec.~\ref{sec-PTtwoband}, this region now corresponds to $H$ with no real eigenvalues).

\subsection{General results}
In this section, we work out the classification of $\eta$-pseudo-Hermitian Hamiltonian for a general number of bands and unitary $\eta$. 

\subsubsection{Reduction to special cases}
Here we show that the problem can be reduced to the case where $\eta$ is a diagonal matrix with diagonal elements $\pm 1$:
\begin{equation}\label{eq-etamn}
    \eta=\text{diag}(\mathbf1_m,-\mathbf1_n).
\end{equation}

If we take a conjugate of $H$, $H\to UHU^\dagger$, then the $\eta$-pseudo-Hermitian relation Eq.~\eqref{eq-etarelation} becomes:
\begin{equation}
    (UHU^\dagger)^\dagger=(U\eta U^\dagger)(UHU^\dagger)(U\eta U^\dagger)^\dagger.
\end{equation}
Namely, $UHU^\dagger$ is $U\eta U^\dagger$-pseudo-Hermitian.
Note that a unitary matrix $\eta$ can always be diagonalized by a unitary matrix $U$,  
we can assume without of generality that $\eta$ is diagonal: $\eta_{ij}=\eta_i\delta_{ij}$.
The $\eta$-pseudo-Hermitian relation becomes:
\begin{equation}
    H_{ji}^*=\eta_i \eta^*_j H_{ij},
\end{equation}
which implies (upon switching $i,j$) that
\begin{equation}
    {H}^*_{ji}=\eta_i^2 \eta_j^{*2}{H}^*_{ji}.
\end{equation}
Hence $H_{ij}=0$ as long as $\eta_i\neq\pm\eta_j$.
Therefore, we can organize the eigenvalues of $\eta$ so that $H$ is block diagonal.
Moreover, multiplying $\eta$ by a global phase does not change the $\eta$-pseudo-Hermitian relation, hence we can assume without loss of generality that $\eta$ has the form of Eq.~\eqref{eq-etamn}. 
We can also choose the larger block to be positive, such that $m\geq n$.

\subsubsection{Parameterization}

We claim that a non-degenerate matrix $H$ is $\eta$-pseudo-Hermitian if and only if can be diagonalized as $H=VH_DV^{-1}$ such that $\eta$ satisfies: 
\begin{align}
        V^{\dagger}\eta V&=\text{diag}\nonumber\\
        &\renewcommand{\arraystretch}{1.3}\left(1,\cdots,1,-1,\cdots,-1,\begin{pmatrix} & 1\\1&\end{pmatrix},\cdots,\begin{pmatrix} & 1\\1&\end{pmatrix}\right),\label{eq-etaV}
\end{align}
and the diagonal part is
\begin{equation}\label{eq-etaD}    H_D=\text{diag}\left(\lambda_i,\cdots,\lambda_k,\lambda_{k+1},\lambda^*_{k+1},\cdots,\lambda_{k+l},\lambda^*_{k+l}\right),
\end{equation}
where $\lambda_i\in\mathbb R~(1\leq i\leq k)$ and $\lambda_{k+j}$ is imaginary $(1\leq j\leq l)$. So we have $k$ real eigenvalues and $l$ pairs of complex conjugate eigenvalues.

We explain how these canonical forms are obtained. First, being non-degenerate, $H$ can be diagonalized: $H=VDV^{-1}$ where $D$ is diagonal and the columns of $V$ correspond to the eigenvectors of $H$.
Second, being pseudo-Hermitian, the eigenvalues of $H$ must be real or come in conjugation pairs (since the characteristic polynomial of $H$ is real), hence $D$ has a form of Eq.~\eqref{eq-etaD}.
Third, two eigenvectors of $H$ corresponding to non-conjugate eigenvalues must be $\eta$-orthogonal 
($Hv_i=\lambda_i v_i$ and $Hv_j=\lambda_j v_j$  imply that $\lambda_j v_i^\dagger \eta v_j=v_i^\dagger\eta H v_j=v_i^\dagger H^\dagger \eta v_j=\lambda^*_i v_i^\dagger \eta v_j$, which implies $v_i^\dagger \eta v_j=0$ as long as $\lambda_i\neq\lambda^*_j$).
This implies that
\begin{equation}
     V^{\dagger}\eta V=\text{diag}\renewcommand{\arraystretch}{1.3}\left(v_1^\dagger \eta v_1,\cdots,v_k^\dagger \eta v_k,\begin{pmatrix} & *\\ *&\end{pmatrix},\cdots,\begin{pmatrix} & *\\ *&\end{pmatrix}\right).
\end{equation}
Then Eq.~\eqref{eq-etaV} is achieved using the scalar degree of freedom of eigenvectors. 

We note that the signs of the diagonal elements in Eq.~\eqref{eq-etaV} are unchanged under scalar multiplication of eigenvectors. 
Instead, due to Sylvester's law of inertia, the number of $1$ minus the number of $-1$ must equal $m-n$.
Hence the number of $1$s equals $p=m-l$ and the number of $(-1)$s equals $q=n-l$.
As a byproduct, it also implies that $l\leq n=\min\{m,n\}$, namely, the number of non-real eigenvalues are upper bounded sign structure of $\eta$.

Let us find the space of $V$ such that Eq.~\eqref{eq-etaV} holds.
To do so, we diagonalize the right-hand side of Eq.~\eqref{eq-etaV} by  right multiplying $V$ with  
\begin{equation}
    W=\text{diag}\renewcommand{\arraystretch}{1.3}\left(\mathbf1_k,\frac{1}{\sqrt{2}}\begin{pmatrix} 1&1\\ 1&-1\end{pmatrix},\cdots,\frac{1}{\sqrt{2}}\begin{pmatrix} 1&1\\ 1&-1\end{pmatrix}\right).
\end{equation}
After this action,
we have
\begin{equation}
    \begin{aligned}
    &(VW)^\dagger\eta VW\\
    =&\text{diag}
    (1,\cdots,1,-1\cdots,-1,1,-1,\cdots,1,-1).
\end{aligned}
\end{equation}
Again, the total number of $1$s and $(-1)$s equal $m$ and $n$ due to Sylvester's law of inertia, so the diagonal elements are $p$ $1$s followed by $q$ $(-1)$s followed by $l$ pairs of $(1,-1)$.
We can further make it equal $\eta$ by a suitable permutation matrix $S$ ($S$ is uniquely determine by $m,n,l$):
\begin{equation}
    (VWS)^\dagger\eta VWS=\eta.
\end{equation}
Therefore, the space of $VWS$ is exactly $U(m,n)$, the generalized unitary group with respect to $\eta$.
The space of $V$ is related to $U(m,n)$ by an invertible linear transformation, hence homeomorphic to each other.

We also note that the parameterization $H=VH_DV^{-1}$  has some redundancies. 
First, we can permute eigenvalues together with their eigenvectors among those diagonal $1$s. There are $p$ of them, giving rise to a $S_p$ redundancy.
The same applies to those diagonal (-1)s ($q$ of them) and those $l$ conjugation pairs.
Second, each $v_i~(i\leq k)$ has a $U(1)$ phase degree of freedom. Each pair of eigenvectors corresponding to a conjugate pair of eigenvalues also comes with a $U(1)$ phase degree of freedom.

Combining the above observations, we can represent the space of non-degenerate $\eta$-pseudo-Hermitian Hamiltonians as
\begin{equation}
    X=\bigcup_{l=0}^n X^{(l)},
\end{equation}
where
\begin{equation}\label{eq-etaX}
X^{(l)}=\frac{\text{Conf}_k(\mathbb{R})\times\text{Conf}_l(\mathbb{C}^+)\times\frac{U(m,n)}{U(1)^{k+l}}}
{S_p\times S_q \times S_l}
\end{equation}
Here again $p=m-l$ and $q=n-l$; $k=p+q$.
Each $U(1)$ in $U(1)^{p}$ (and $U(1)^{q}$) acts on $U(m,n)$ by multiplying a ``positive" (and ``negative") column with a phase;  
Each $U(1)$ in $U(1)^{l}$ acts on $U(m,n)$ by jointly multiplying a ``positive" column and a ``negative" column with the same phase.

\subsubsection{Topology}
Let us first calculate the zeroth homotopy set.
Both $\text{Conf}_l(\mathbb{C}^+)$ and $U(m,n)$ are connected,
hence non-connectedness comes from $\frac{\text{Conf}_k(\mathbb{R})}
{S_p\times S_q}$:
\begin{equation}
    |\pi_0(X^{(l)})|=\binom{k}{q}.
\end{equation}
It can be understood as follows. 
We say a real eigenvalue to be of type $A$ (or $B$) if the corresponding eigenvector $\nu$ satisfies $\nu^\dagger \eta\nu=1$ (or -1).
The binomial number is just the number of ways to arrange $p$ type $A$ points (indistinguishable) and $q$ type $B$ points (indistinguishable) in order.
Also note that $q$ runs over 0 to $n$,  
the total number of connected components is therefore:
\begin{equation}\label{eq-etapi0}
    |\pi_0(X)|=\sum_{q=0}^{n} \binom{m-n+2q}{q}.
\end{equation}
The cardinality of this zeroth homotopy set is much larger than the situations in $\mathcal{PT}$-symmetric systems for multiple band systems. This means it is more common to have nodal structures in pseudo-Hermitian systems; they appear with codimension one. The comparison is listed in Table~\ref{tb_cpptph}.

\renewcommand{\arraystretch}{2.5}
\begin{table}[t]

\begin{tabular}{ |P{2.5cm}|P{3.5cm}|P{1.5cm}|  }
 \hline
  \centering Symmetry & $X\sim$& $|\pi_0(X)|$\\
 \hline
 \centering \PT/CP & $\frac{ \mathrm{Conf}_m(\mathbb C)\times \frac{GL_{n+2m}(\mathbb R)}{(\mathbb R^{\times})^n\times (\mathbb C^{\times})^m}}{S_m}$& $\sim N$
 \\
\hline
 \centering Pseudo-Hermitian/\newline pseudo-chiral & $\frac{\text{Conf}_k(\mathbb{R})\times\text{Conf}_l(\mathbb{C}^+)\times\frac{U(m,n)}{U(1)^{k+l}}}
{S_p\times S_q \times S_l}$
 &
 $\sim \exp N$
 \\
\hline
\end{tabular}
\caption{ A comparison between  $\mathcal{PT}$-symmetric and pseudo-Hermitian systems.
 We present the most general forms of their corresponding band-gapped matrices. The cardinality of
the zeroth homotopy set for  $\mathcal{PT}$-symmetric systems scales linearly with the total number $N$ of bands, while it can scale exponentially in pseudo-Hermitian systems (for $\eta$ containing $\Theta(N)$ numbers of positive and negative signatures). }\label{tb_cpptph}
\end{table}

The higher homotopy groups can be worked out in a straightforward way similar to previous sections. 
To start, we first note that the natural embedding $U(m)\times U(n)\to U(m,n)$ is a homotopy equivalence.
Let us again write a connect component of $X^{(l)}$ as $(\mathcal{E}\times M)/S_l$. 
We have the following exact sequence for $M^{(l)}=U(m,n)/U(1)^{k+l}$:
\begin{equation}
    0\to\pi_2(M^{(l)})\to\mathbb Z^{k+l}\to\mathbb \pi_1(U(m,n))\to\pi_1(M^{(l)})\to 0.
\end{equation}
According to the comments below Eq.~\eqref{eq-etaX}, the map $\mathbb Z^{k+l}\to\mathbb \pi_1(U(m,n))$ is defined by:
\begin{equation}
\begin{aligned}
    &(a_1,\cdots, a_p, b_1, \cdots, b_q, c_1, \cdots, c_l)\\
    \mapsto  &
    (\sum a+\sum c,\sum b+\sum c)
\end{aligned}
\end{equation}
if $n\geq 1$; and
\begin{equation}
        (a_1,\cdots, a_p)\mapsto\sum a
\end{equation}
if $n=0$.
Therefore, the homotopy groups of $M^{(l)}$ are given by:
\begin{equation}\label{eq-etapi1M}
    \pi_1(M^{(l)})\cong\begin{cases}
        \mathbb Z~~(k=0)\\
        0
    \end{cases},
\end{equation}
and 
\begin{equation}
    \pi_2(M^{(l)})\cong\begin{cases}
        \mathbb Z^{k+l-1}~~(n=0~\text{or}~k=0)\\
        \mathbb Z^{k+l-2}
    \end{cases}.
\end{equation}

Since $\text{Conf}_l(\mathbb C^+)$ has trivial $\pi_2$ and $S_m$ is discrete, we have $\pi_2(X)\cong\pi_2(M^{(l)})$. 
Regarding $\pi_1$, using a similar proof as that in Appendix \ref{ap_pdtetv}, we have
\begin{equation}\label{eq-etapi1X}
    \pi_1(X^{(l)})\cong \mathbb B_m\times\pi_1(M^{(l)}).
\end{equation}

As a simple check, the two-band model corresponds to $m=n=1$.
Eq.~\eqref{eq-etapi0} predicts that $X$ has 3 connect components $X^{(1)}$, $X^{(0)}_+$, $X^{(0)}_-$.
The component $X^{(1)}$ corresponds to $k=0$;
Eq.~\eqref{eq-etapi1M} and \eqref{eq-etapi1X} then predict that $\pi_1(X^{(1)})\cong\mathbb Z$, consistent with the results in the previous subsection.

\section{Conclusions and outlook}\label{sc_cocloutl}
\subsection{Main results}

In this work, we applied homotopy theory to give a comprehensive description of topology in non-Hermitian systems with \PT{} or pseudo-Hermitian symmetry. 
We formulated the concepts of band-gapped operators and separation-gapped operators for non-Hermitian systems, based on some earlier observations and experimental relevance. 
The former is highly pertinent to single-mode and bosonic systems, determining the cost of single-particle excitations and whether bands cross.
Its classification implies which properties are invariant when the system is deformed without closing the single-mode excitation gap, as well as dictating how nodal points, lines, or other structures deform and merge. 
The concept of separation gaps generalizes the insulating many-body gap in Hermitian systems, and characterizes excitations that cross bands of very different static or dynamical properties.
In \PT-symmetric systems, we employed the natural separation originating from the spontaneous symmetry breaking of the system. This separation gap turns out to be closely related to the long-time behaviors. The classification of separation gaps determines when spectral crossings take place in general non-Hermitian systems, signaling a transition in the dynamical or static properties of the system.  Although here the band concepts are motivated by the relevant \PT{} and pseudo-Hermitian symmetries in experiments, they also apply to systems with other symmetries or without symmetries.

With these definitions, we tackled the key question of what happens to Hermitian eigenvector topology when eigenvalues diverge from the real axis and become topological themselves. This is systematically done within the framework of band gaps. Most remarkably,
our study has unveiled different levels of familiar and novel topology, contingent upon the extent to which the symmetry is spontaneously broken by the eigenvalues and eigenstates. This is entirely distinct from Hermitian examples, where the topology is highly homogeneous.

Homotopy theory first gives a topological origin for the lower codimension of degeneracies in \PT-symmetric and pseudo-Hermitian systems. 
The number of bands spontaneously breaking the symmetry serves as an order parameter of the system. Phases of different such orders must be separated by nodal structures of codimension one, similar to domain walls in ordered media \cite{RevModPhys.51.591}. 
We found that the Hermitian topology from eigenvectors persists to a certain extent, inside the regions where all bands have real eigenvalues.
However, compared to their Hermitian antecedents, such topology can protect nodal structures of exceptional degeneracy.
In the presence of complex eigenvalues, the conventional eigenvector topology can be partially reduced. 
This led us to a new kind of fragile topology: the topology stable with respect to addition of bands with real eigenvalues, but fragile to addition of bands with complex eigenvalues. One instance of such fragility is the Abelianization of the Hermitian frame topology.
In turn, novel structure arises for multiple eigenvalues on the complex plane, including frame orientation and eigenmode braid topology. All of these results led to a series of phases from non-Abelian eigenvector topology to non-Abelian eigenvalue topology $\times$ Abelian eigenvector topology. In \PT-symmetric systems, non-Abelian topology turns out to be a ubiquitous phenomenon.

We then studied the separation gaps between these real and complex eigenvalues. The separation directly measures how many bands are \PT{}-symmetry preserving or breaking and the dynamical properties of the system.
We showed that this dynamical gap can be accompanied by an unconventional topological description. 
It hosts topological phases that are protected by Chern numbers for bands of complex eigenvalues and Euler numbers or Stiefel-Whitney classes for bands of real eigenvalues. 
As the Chern number describes very different topological structures from the Euler number and the Stiefel-Whitney class, they do not appear simultaneously in Hermitian physics. 
Non-Hermitian \PT-symmetric systems constitute a unique platform to bridge these distinct types of topology. The coexistence of Chern and Euler numbers, albeit being entirely wave function topology, forbids any reduction to Hermitian band structures. This contradicts the earlier conclusion \cite{PhysRevX.9.041015} that all spectrally separated bands can be adiabatically connected to Hermitian systems. We expect a multitude of exotic non-Hermitian phases to appear in other anti-linear symmetry classes.

We systematically answered a number of technical questions regarding non-Hermitian topology. 
For topological invariants that can be expressed in terms of Wilson loops for Hermitian systems, we showed that the non-Hermitian extension is formulated using biorthogonal eigenvectors. 
The projection operator is directly obtained from the Hamiltonian via a loop integral, and thus gauge invariant. 
We showed how to arrive at the biorthogonal Berry connection from the definition of parallel transport. This generalizes previous studies on a single-Chern band.
In particular, we give the condition when other choices of left/right eigenvectors may give the same Chern number. Meanwhile, the biorthogonal combination turns out to possess a consistent $GL_N$-gauge structure and always a globally well-defined connection over the entire Brillouin zone. 
In contrast, the Euler number requires modifications to accommodate for non-orthogonal eigenvectors. 
We provided a systematic explanation on the necessary adaptations to obtain these topological numbers.
We show how to perform spectral flattening even in the presence of exceptional points. 
Although near exceptional points, eigenvectors coalesce and individual band property becomes singular \cite{kato2013perturbation,PhysRevLett.128.010402}, the combination of all coalescing bands still corresponds to a continuous projection and can thus be flattened.  

Altogether, this work constitutes a framework for the study of band structures of non-Hermitian operators with symmetries. 
Within homotopy theory, we unified topological results at various levels. 
First, we unified the description of band-gapped and nodal phases. 
Second, we unified Hermitian eigenvector and non-Hermitian eigenvalue topology. 
This classification is based on the energy cost of excitations and differs from the reference line/point aspects. 
The topological invariants of the two classifications carry very different physical interpretations which we have investigated in detail in this work yet leaving ample opportunity for further exploration.

\subsection{Outlook}

This work paves the way for ample further studies on topological non-Hermitian band structures. 
The methods showcased for \PT{} and pseudo-Hermitian symmetry presented here provide a guideline to systematically describe and classify non-Hermitian physics subject to other symmetries. 
In distinction to Hermitian systems, we have shown that the band topology of non-Hermitian systems always has far richer structure beyond K-theory. 
Another factor in classifying non-Hermitian systems is the representation of the symmetry. 
When K-theory is not applicable, the specifics of the symmetry representation also play a role. 
This points to further abundance of new topological structures in non-Hermitian systems. Note that homotopy theory applies both to nodal structures and gapped bands. It provides a starting point to understand how topological phase transitions originate from level crossings.

The interplay of homotopy and symmetry with boundaries and defects provides intriguing avenues for future inquiry. 
An important scenario of our work is non-Hermitian matrices as the Bloch Hamiltonians under periodic boundary conditions. 
It is known that non-Hermitian systems may exhibit qualitatively different behavior under open boundary conditions due to the skin effects. 
In this context, the interplay between spectral winding and more conventional eigenvector topology under different boundaries has been intensely studied in the past few years \cite{PhysRevLett.121.086803,PhysRevLett.121.026808,PhysRevLett.121.136802,Lee2016,PhysRevLett.126.216407,Xiong_2018,lin2023,okuma2023,PhysRevB.99.201103,PhysRevLett.124.056802,PhysRevA.99.052118,PhysRevResearch.2.043046,PhysRevB.103.195157,PhysRevB.102.205118,PhysRevLett.122.076801,PhysRevB.99.081302,PhysRevLett.123.016805,PhysRevResearch.2.013058,PhysRevLett.125.180403}
In comparison, it is not clear how other unique aspects of non-Hermitian band topology, such as energy braids, and the Chern-Euler description presented in this paper, should manifest under different boundary conditions. 
We note that an early study has excluded certain edge states in the \PT-symmetry preserving phase \cite{PhysRevB.84.153101}. 
As we have shown however, the topology of the \PT-symmetry breaking phase is even more interesting. 
Intuitively, when two systems with different band topology are put together, there might exist in-gap structure tunneling the different topological phases. 
The non-Hermitian skin effects usually alter this picture and require systematic study.

We have further shown in this article that some Hermitian stable topology can be made fragile by considering non-Hermiticity. 
This indicates that certain Hermitian topological phases may be obtained from trivial phases by including non-Hermiticity. The fragile topology also modifies the nodal structures permitted in Hermitian systems. An early example is the Alice string \cite{PhysRevResearch.2.023226}, where the Chern number of Weyl nodes is flipped via encircling exceptional lines. 
Recently, Hermitian fragile topological phases have been found to possess unusual bulk-boundary correspondence \cite{song2020twisted} and Wannier obstructions \cite{PhysRevLett.121.126402,PhysRevX.9.021013}. 
It may prove interesting to investigate how this may fit within a general theory of non-Hermitian fragile topology.

The topological phases uncovered in this article manifest in several ways. 
First of all, the large nontrivial zeroth homotopy set leads to an abundance of nodal structures of codimension $1$ in \PT-symmetric and pseudo-Hermitian operators. 
The nodal structures of non-Hermitian systems are accompanied by gap closings in the real parts or imaginary parts of the spectrum. This has various static and dynamic consequences \cite{ozdemir2019parity,PhysRevLett.125.227204,yoshidapeterskawakmi,PhysRevResearch.4.L042025}. 
Non-Hermitian systems with \PT{} symmetry have been realized in electric circuits \cite{PhysRevLett.126.215302}, coupled laser systems \cite{doi:10.1126/sciadv.abm7454}, and single-photon interferometry \cite{PhysRevLett.123.230401,doi:10.1126/sciadv.adi0732}. 
The topological properties of eigenvectors and eigenvalues when encircling nodal structures is a direct application of our work, similar to the symmetryless systems. The frame topology can be achieved in transmission line networks \cite{guo2021experimental} and biaxial photonic crystals \cite{PhysRevX.13.021024}. Introducing non-Hermiticity will Abelianize the frame charge and induce non-Abelian eigenmodes. This will change the flow of the frame topology along nodal structures.

The real and complex bands in \PT-symmetric systems correspond to eigenmodes with different types of evolution. The decaying bands (lower complex plane) and the growing bands (upper complex plane) carry opposite Chern numbers. The oscillating bands (real axis) are characterized by Euler numbers or Stiefel-Whitney classes. It will be interesting to test these conventional topological invariants that emerge during dynamical evolution. More importantly, Chern number is a topological invariant for both non-interacting and interacting models. In the latter case, the Chern number can be expressed through the many-body wave function under twisted boundary conditions \cite{PhysRevB.31.3372}. This can be adapted to our separation gaps as well, with $\mathbf k$ replaced by a twisted boundary condition and the wave function now being the many-body wave functions. The separation gap then becomes the lifetime difference between the two many-body wave functions decaying most slowly. Such formulation might lead to distinct non-Hermitian many-body effects as exemplified previously in one-dimensional models \cite{PhysRevB.105.205403}.

Our results indicate that there remains novel and unforeseen topology awaiting discovery in non-Hermitian systems, from both physical and mathematical points of view. We suggest that highly controllable metamaterials ranging from photonics to electrical circuits and mechanical setups will provide ideal platforms to bring these intriguing abstract mathematical concepts to concrete reality.

\emph{Note added.} After posting our work on arxiv, we became aware of a parallel manuscript \cite{li2023braiding} discovering and simulating the braid structures of $\mathcal{PT}$-symmetric systems in electric circuits.

\acknowledgments We thank Piet Brouwer, Vatsal Dwivedi, Janet Zhong, Jack Harris, Masatoshi Sato and Krystof Kolar for stimulating discussions.  K.Y. is supported by the ANR-DFG project (TWISTGRAPH).
J.L.K.K., K.Y. and E.J.B. were supported by 
the Swedish Research Council (VR, grant 2018-00313), 
the Wallenberg Academy Fellows program (2018.0460) and 
the project Dynamic Quantum Matter (2019.0068) of the Knut and Alice Wallenberg Foundation, 
as well as the  G\"oran Gustafsson Foundation for Research in Natural Sciences and Medicine.  L.R. is supported by the Knut and Alice Wallenberg Foundation under Grant No. 2017.0157.
Z.L. is supported by Perimeter Institute; research at Perimeter Institute is supported in part by the Government of Canada through the Department of Innovation, Science and Economic Development and by the Province of Ontario through the Ministry of Colleges and Universities.

\appendix

\section{General representations of \texorpdfstring{$\mathcal{PT}$}{PT} symmetry}\label{ap_PTgU}
In this section we consider the topological classification of $\mathcal{PT}$-symmetric Hamiltonians for general representations. Namely, we allow $U\neq I$ in the $\mathcal{PT}$-symmetry relation
\begin{equation}
    H=U H^\ast U^{-1}.
\end{equation}

\subsection{Two-band model}
As a simple example, consider the two-band case where $U=\sigma_y$.
We parameterize $H$ as
\begin{equation}
    H=d_x\sigma_x+d_y\sigma_y+d_z\sigma_z,
\end{equation}
where $d_x, d_y, d_z\in\mathbb{C}$ and we have assumed that $H$ is traceless.
A simple calculation shows that, the $\mathcal{PT}$-symmetry condition is equivalent to:
\begin{equation}
     d^\ast_x=-d_x,~~ d^\ast_y=-d_y,~~ d^\ast_z=-d_z.
\end{equation}
So we can now parameterize $H$ as
\begin{equation}
    H=i(\tilde d_x\sigma_x+\tilde d_y\sigma_y+\tilde d_z\sigma_z),
\end{equation}
where now $\tilde d_x, \tilde d_y, \tilde d_z\in\mathbb{R}$.
The condition of non-degeneracy is:
\begin{equation}
     \tilde{d}_x^2+ \tilde d_y^2+\tilde d_z^2\neq 0,
\end{equation}
so we have a space of non-degenerate matrices which is homotopy equivalent to $S^2$.

\subsection{General results}
\subsubsection{Reduction of special cases}

If we take a conjugate of $H$, $H\to VHV^\dagger$, then the $\mathcal{PT}$-symmetry relation becomes:
\begin{equation}
    VHV^{-1}=(VUV^T)( V^\ast  H^\ast V^T)( V^\ast U^{-1}V^{-1}).
\end{equation}
Namely, $VHV^{-1}$ is \PT-symmetric with respect to $U'=VUV^T$.

With such transformation, we can always \cite{weinberg_1995} transform the unitary matrix $U$ to be block-diagonal, such that the diagonal is either the unit ($1\times 1$ matrix), or $2\times 2$ of in the following form:
\begin{equation}
    \begin{pmatrix}
        0&e^{i\theta}\\
        e^{-i\theta}&0
    \end{pmatrix},
\end{equation}
where $\theta\in (0,\pi/2]$ ($\theta=0$ can be transformed to identity matrix).
Accordingly, we can rearrange the matrix $U$ as
\begin{equation}
    U=\text{diag}\{I,\big(\begin{smallmatrix}
        0&e^{i\theta_1}\\
        e^{-i\theta_1}&0
    \end{smallmatrix}\big)\otimes I, \cdots, \big(\begin{smallmatrix}
        0&e^{i\theta_k}\\
        e^{-i\theta_k}&0
    \end{smallmatrix}\big)\otimes I,-\sigma_y\otimes I\}.
\end{equation}
Here $\theta_i\in(0,\frac{\pi}{2})$ and are mutually different.

Furthermore, it is easy to check that $H$ is also block diagonal once we write $U$ as above. 
Therefore, we have reduced to the problem into three cases.
\subsubsection{Case 1}
In this case $U=\mathbf1_{N}$, we recover the situation in the main text. The $\mathcal{PT}$-symmetry simply says $H$ is a real $N\times N$ matrix:
\begin{equation}
    H\in \mathrm{Mat}(N,\mathbb R).
\end{equation}

\subsubsection{Case 2}
In this case $U=\big(\begin{smallmatrix}
        0&e^{i\theta}\\
        e^{-i\theta}&0
    \end{smallmatrix}\big)\otimes \mathbf1_{\frac{N}{2}}$ where $\theta\in(0,\frac{\pi}{2})$, the total number of bands $N$ must be even.
It follows that $H$ must be of the form
\begin{equation}
    H=\begin{pmatrix}
        A&0\\0& A^\ast
    \end{pmatrix},
\end{equation}
where $A$ is a $\frac{N}{2}\times \frac{N}{2}$ complex matrix:
\begin{equation}
    A\in \mathrm{Mat}\left(\frac{N}{2},\mathbb C\right).
\end{equation}

It can be regarded as pairing a system without symmetry with its $\mathcal{PT}$ partner, without any couplings between them. 
Hence, the classification of $H$ reduces to that of $A$.
The latter question has been addressed in Ref.~\cite{PhysRevB.103.155129}.

\subsubsection{case 3}
In this case $U=-\sigma_y\otimes \mathbf1_\frac{N}{2}$, this also requires even number of bands.
It follows that $H$ must be of the form
\begin{equation}
    H=\begin{pmatrix}
        A&- B^\ast\\B& A^\ast
    \end{pmatrix}.
\end{equation}
This $H$ may be identified as an $\frac{N}{2}\times\frac{N}{2}$ quaternion matrix: 
\begin{equation}
    H\in \mathrm{Mat}\left(\frac{N}{2},\mathbb{H}\right),
\end{equation}
where $\mathbb{H}$ stands for the quaternion algebra. 
Compared to case 2, couplings between the two subsystems $A$ and $A^\ast$ are allowed in this case.

Eigenvalues of quaternion matrices come in conjugate pairs and real eigenvalues are doubled. 
So we can classify the system similarly according to the number of real and complex eigenvalues. The study of different gapped conditions can be done similarly as in Sec.~\ref{sc_ggp} and Sec.~\ref{sc_spphs}.

For example, we could consider non-degenerated quaternion matrices, which excludes any real eigenvalues.
Such matrices can be decomposed as \cite{wiegmann_1955, zhang2001jordan}:
\begin{equation}
    H=V\begin{pmatrix}
    D&0\\0&\bar D
\end{pmatrix}V^{-1}
\end{equation}
where $V\in GL_n(\mathbb{H})$ and $D$ is diagonal with mutually different diagonal elements from the upper complex plane.
The parameter space can thus be represented as
\begin{equation}
\frac{\text{Conf}_n(\mathbb{C}^+)\times\frac{GL_n(\mathbb{H})}{U(1)^{n}}}{S_n}.
\end{equation}
The homotopy groups can be found following the procedure of Ref.~\cite{PhysRevB.103.155129}.
As a simple check, the two-band case ($n=1$) gives $Sp(1)/U(1)$, which is exactly an $S^2$ (note that $GL_n(\mathbb{H})$ is homotopic to $Sp(n)$, and $Sp(1)\cong SU(2)$).

\section{Differences between gap classification and reference classification}\label{ap_dfgs}

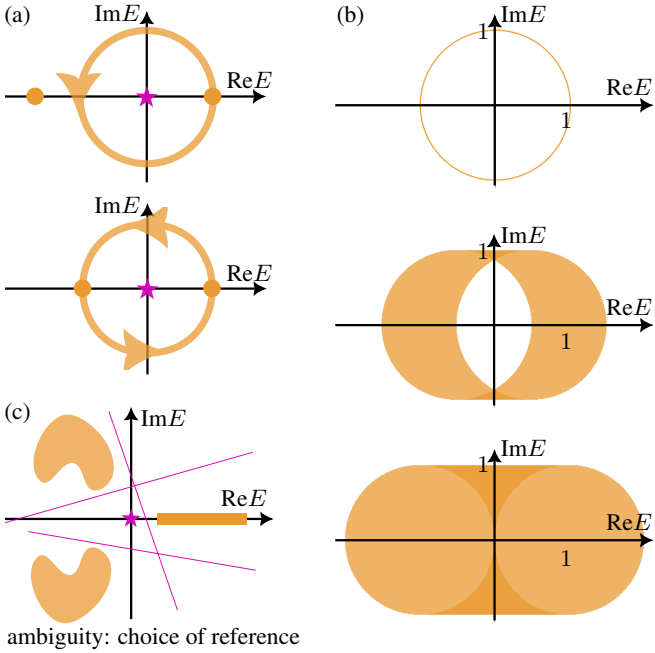
\begin{figure}[htbp]
    \centering
    \def\svgwidth{\columnwidth}
    %--
    
    \begingroup%
  \makeatletter%
  \providecommand\color[2][]{%
    \errmessage{(Inkscape) Color is used for the text in Inkscape, but the package 'color.sty' is not loaded}%
    \renewcommand\color[2][]{}%
  }%
  \providecommand\transparent[1]{%
    \errmessage{(Inkscape) Transparency is used (non-zero) for the text in Inkscape, but the package 'transparent.sty' is not loaded}%
    \renewcommand\transparent[1]{}%
  }%
  \providecommand\rotatebox[2]{#2}%
  \newcommand*\fsize{\dimexpr\f@size pt\relax}%
  \newcommand*\lineheight[1]{\fontsize{\fsize}{#1\fsize}\selectfont}%
  \ifx\svgwidth\undefined%
    \setlength{\unitlength}{246bp}%
    \ifx\svgscale\undefined%
      \relax%
    \else%
      \setlength{\unitlength}{\unitlength * \real{\svgscale}}%
    \fi%
  \else%
    \setlength{\unitlength}{\svgwidth}%
  \fi%
  \global\let\svgwidth\undefined%
  \global\let\svgscale\undefined%
  \makeatother%
  \begin{picture}(1,1)%
    \lineheight{1}%
    \setlength\tabcolsep{0pt}%
    \put(0,0){\includegraphics[width=\unitlength,page=1]{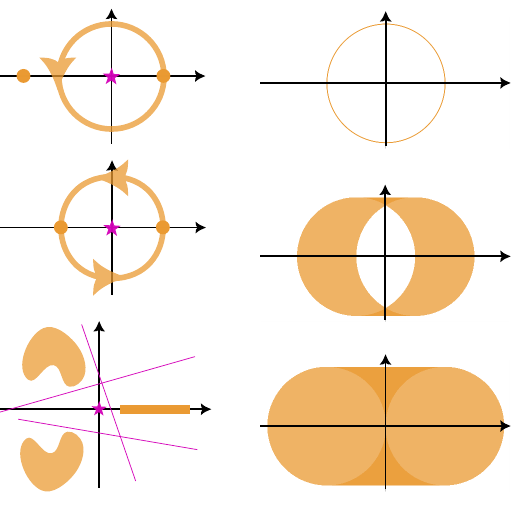}}%
    \put(0.76197097,0.9661721){\color[rgb]{0,0,0}\makebox(0,0)[lt]{\lineheight{1.25}\smash{\begin{tabular}[t]{l}Im\(E\)\end{tabular}}}}%
    \put(0.99493901,0.8537433){\color[rgb]{0,0,0}\makebox(0,0)[rt]{\lineheight{1.25}\smash{\begin{tabular}[t]{r}Re\(E\)\end{tabular}}}}%
    \put(0.74422562,0.94047755){\color[rgb]{0,0,0}\makebox(0,0)[rt]{\lineheight{1.25}\smash{\begin{tabular}[t]{r}\(1\)\end{tabular}}}}%
    \put(0.85338261,0.80316588){\color[rgb]{0,0,0}\makebox(0,0)[lt]{\lineheight{1.25}\smash{\begin{tabular}[t]{l}\(1\)\end{tabular}}}}%
    \put(0.76197097,0.29862702){\color[rgb]{0,0,0}\makebox(0,0)[lt]{\lineheight{1.25}\smash{\begin{tabular}[t]{l}Im\(E\)\end{tabular}}}}%
    \put(0.99493901,0.18448764){\color[rgb]{0,0,0}\makebox(0,0)[rt]{\lineheight{1.25}\smash{\begin{tabular}[t]{r}Re\(E\)\end{tabular}}}}%
    \put(0.74422562,0.27304298){\color[rgb]{0,0,0}\makebox(0,0)[rt]{\lineheight{1.25}\smash{\begin{tabular}[t]{r}\(1\)\end{tabular}}}}%
    \put(0.85338261,0.12963381){\color[rgb]{0,0,0}\makebox(0,0)[lt]{\lineheight{1.25}\smash{\begin{tabular}[t]{l}\(1\)\end{tabular}}}}%
    \put(0.76197097,0.6261586){\color[rgb]{0,0,0}\makebox(0,0)[lt]{\lineheight{1.25}\smash{\begin{tabular}[t]{l}Im\(E\)\end{tabular}}}}%
    \put(0.99493901,0.51506488){\color[rgb]{0,0,0}\makebox(0,0)[rt]{\lineheight{1.25}\smash{\begin{tabular}[t]{r}Re\(E\)\end{tabular}}}}%
    \put(0.74422562,0.60097127){\color[rgb]{0,0,0}\makebox(0,0)[rt]{\lineheight{1.25}\smash{\begin{tabular}[t]{r}\(1\)\end{tabular}}}}%
    \put(0.85338261,0.46353728){\color[rgb]{0,0,0}\makebox(0,0)[lt]{\lineheight{1.25}\smash{\begin{tabular}[t]{l}\(1\)\end{tabular}}}}%
    \put(-0.00137751,0.96668519){\color[rgb]{0,0,0}\makebox(0,0)[lt]{\lineheight{1.25}\smash{\begin{tabular}[t]{l}(a)\end{tabular}}}}%
    \put(0.50963057,0.96668519){\color[rgb]{0,0,0}\makebox(0,0)[lt]{\lineheight{1.25}\smash{\begin{tabular}[t]{l}(b)\end{tabular}}}}%
    \put(-0.00137751,0.35532484){\color[rgb]{0,0,0}\makebox(0,0)[lt]{\lineheight{1.25}\smash{\begin{tabular}[t]{l}(c)\end{tabular}}}}%
    \put(0.20840781,0.34522208){\color[rgb]{0,0,0}\makebox(0,0)[lt]{\lineheight{1.25}\smash{\begin{tabular}[t]{l}Im\(E\)\end{tabular}}}}%
    \put(0.40258086,0.22156044){\color[rgb]{0,0,0}\makebox(0,0)[rt]{\lineheight{1.25}\smash{\begin{tabular}[t]{r}Re\(E\)\end{tabular}}}}%
    \put(0.00293615,0.0094621){\color[rgb]{0,0,0}\makebox(0,0)[lt]{\lineheight{1.25}\smash{\begin{tabular}[t]{l}ambiguity: choice of reference\end{tabular}}}}%
    \put(0.20777141,0.96824462){\color[rgb]{0,0,0}\makebox(0,0)[rt]{\lineheight{1.25}\smash{\begin{tabular}[t]{r}Im\(E\)\end{tabular}}}}%
    \put(0.34048041,0.86409717){\color[rgb]{0,0,0}\makebox(0,0)[lt]{\lineheight{1.25}\smash{\begin{tabular}[t]{l}Re\(E\)\end{tabular}}}}%
    \put(0.20777141,0.67207545){\color[rgb]{0,0,0}\makebox(0,0)[rt]{\lineheight{1.25}\smash{\begin{tabular}[t]{r}Im\(E\)\end{tabular}}}}%
    \put(0.34048041,0.56770709){\color[rgb]{0,0,0}\makebox(0,0)[lt]{\lineheight{1.25}\smash{\begin{tabular}[t]{l}Re\(E\)\end{tabular}}}}%
  \end{picture}%
\endgroup%

    %--
    \caption{
    The difference between excitation gaps and reference gaps. 
    (a) 
    Sketches of two-band spectra.
    Top: A band winds around a reference point. 
    Bottom: Two bands exchange near a reference point. 
    Both spectra are point gapped and possess the same winding number w.r.t. the reference energy shown in purple. 
    However, the bands on the top panel may be continuously deformed into a flat-band spectrum without closing the band gap, while the braiding (eigenvalue exchange) makes this impossible for the bottom panel.
    This distinguishes the reference picture from the band gap / excitation picture. 
    (b) 
    A band-gapped system described by \(H(k_x)+\cos(k_y) \mathbf{1}_2\), where a term proportional to the identity matrix is added to the braiding model \(H\) given in Eq.~\eqref{eq:z-braid-model-nonsym}.
    The three panels show the spectrum for 
    \(
    k_y=\pi/2,
    k_y\in[\frac \pi 3,\frac{2\pi}3],
    k_y\in[0,2\pi ]
    \) 
    respectively.
    The eigenvalue exchange over the course of a braid along \(k_x\) forms the circle seen in the first panel. The shift of the spectrum along \(k_y\) renders the entire spectrum featureless to any references.
    (c)
    In a general multi-band system, there is no unique way to choose a reference. The complicated spectrum geometry makes it hard to judge where the reference should be put. However, if the separation is relevant to a certain physical context, as we discussed in Sec.~\ref{sec:preliminaries}, there is no ambiguity to do the partition. } 
    \label{fig_gapPGLGl}
\end{figure}

In this appendix, we give a more detailed clarification on the difference between gapped Hamiltonian classification presented here and the reference point/line description \cite{PhysRevX.8.031079,PhysRevX.9.041015}. The difference is general and  applies to the situations of all symmetry classes (see also Ref.~\cite{zhong2023eigenenergy}).

In the reference point/line description, a system is said to be gapped if and only if the spectrum of the system has no overlap with the references. 
The reference is often taken to be the origin (point reference), or the real and the imaginary axes (line reference) on the complex plane.
In the classifications of this work (refered as gap classification in the following), the key element is the energy gap between different band.
In the band-gapped systems, we require $E_i(\mathbf k)-E_j(\mathbf k)\ne 0$ for all $i,j$ and $\mathbf k$. There is no requirement on where the locations of $E_j$ are on the complex plane. In the separation gapped systems, the system is featured by $E_i(\mathbf k)-E_j(\mathbf k')\ne 0$ for $i,j$ belonging to different partition sets $J,J'$ and all $\mathbf k,\mathbf k'$.

In the reference description, the key element is the energy gap \emph{relative to a reference}, while in the gap classification, the key point is the energy gap \emph{relative to each other}. 
They are in principle different classification, as shown by the following discussions and explicit examples.

First, the (eigenvalue part of the) gap classification has the feature that it is fully determined by the relative energy of different bands hence is fully invariant under a translation of the spectra on the complex plane. 
Meanwhile, an extra reference has to be introduced in the reference classification. Whether the reference classification is invariant under translation of spectra depends on whether the reference energy also shifts along with the translation, which in turn depends on how the reference is defined (and sometimes this requires physical inputs).
Any fixed references can be crossed by performing such translation of spectra.

Second, two classifications can give different answer on whether two systems are topological equivalent. 
Consider two one-dimensional band-gapped systems, shown in Fig.~\ref{fig_gapPGLGl} (a).
In this first system, one eigenvalue is $k$-independent, while the other winds around the reference point at 0.
In the second system, two bands are swapped as $k$ traverses the BZ (an example is Eq.~\eqref{eq:z-braid-model-nonsym}).
According to our band-gap classification, these two systems are not equivalent and are differed by a distinct \emph{relative} spectral winding number. They cannot be adiabatically joined without band gap closings.
However, if we use a reference point classification, then no matter where the reference is located, two systems have the same reference winding number, i.e., indistinguishable from the reference point of view.

Third, two classification concerns different spaces of Hamiltonians in principle.
In Fig.~\ref{fig_gapPGLGl} (b), we show a nontrivial band-gapped systems in two dimensions where no reference point can exist.
Here, the BZ is a two-dimensional torus and the 2D Hamiltonian is \(H(k_x)+\cos(k_y) \mathbf{1}_2\).
We assume that the function $H(k_x)$ has a braid topology along the meridians of the torus, i.e., along any constant $k_y$ curve in the BZ, so that the system is nontrivial according to our band-gap classification, c.f. Fig.~\ref{fig_hptiv}.
Note that the braid only cares about the relative difference between the two bands, the location of the braid on the complex plane may change for different meridians, namely different $k_y$. 
This is plotted in Fig.~\ref{fig_gapPGLGl} (b), where the braid at different $k_y$ is only a translation of the braid at $k_y=0$. The whole spectrum occupies a simply-connected region on the complex plane. There is no reference line or point that can detect this topological structure. 

Certainly, sometimes one may adopt a generalized notion of reference gap and then make connections with band gaps or separation gaps. 
For example, one may shift a two-band band-gapped system with a $\bf k$-dependent constant so that it becomes point-gapped at $0$. One way to do this is to make $H(\bf k)$ traceless.
However, for a generic multi-band band-gapped system, there is in general no canonical way to perform this reduction.
On the other hand, applying such procedure to the single-band Hatano-Nelson model will make the model trivial, since now spectra always locate at $0$, destroying any nontriviality in the point-gap sense.

We can consider a very general $3$-band systems. In such a system, we can usually have a separation of bands as shown in Fig.~\ref{fig_gapPGLGl} (c). If we want to use a reference to study this system, there are several choices to place the reference according to the spectrum geometry. It becomes a question where the reference should be put. There are several places that one can insert a reference in this geometry. And in many settings, the matrix $H$ is unfixed up to a constant term, representing the adjustable chemical potential in experiments. Such a constant term makes it even harder to place the reference. Especially, the constant term can change the reference winding numbers in \cite{PhysRevX.8.031079,PhysRevX.9.041015}.  In contrast, if we use the separation gap notion, we do not need to worry about the unfixed constant term. The spectral separation cannot be changed by shifting to a new chemical potential. More explicitly, the choice of the partition rule in the separation gap comes from the physical question, as we stressed in its definition, instead of reading out directly from the coordinates of the spectra.

We also note that, although sometimes there is a natural reference line to tell the partition and define the separation gap, there exists situations where there could be no natural reference line, or the reference line has to be inserted in a very ambiguous way. 
For example, as for the Liouvillian gap, if we use the reference line approach, the line has to be placed infinitesimally close to, but not exactly on the real axis.
This brings some conceptual difficulties as infinitesimal is not a number strictly speaking. And for these steady states on the real axis, there can be further separation due to the chemical potential. Such a joint separation exists naturally. However, this situations is hard to be described with the reference line language, as the separation on the real axis can only be captured by a line of infinitesimal length.
Give these difficulties, it might be desirable to employ a reference-free formalism to describe the spectral gap, as in many earlier literature before the invention of the reference points/lines. Indeed, the separation of the bands themselves already contains all information of a gap in a many-body system.

In sum, reference points and lines are about the location of the spectrum. They are usually independent of band gaps, separation gaps and nodal structures. To characterize the latter cases, the band gaps and separation gaps presented in this work are the natural choice. As a comparison, the one-band Hatano-Nelson model \cite{PhysRevLett.77.570} was considered trivial in a multi-band sense. There is only one band, therefore the cost of energy to exciting to other bands is undefined. Its topology is instead characterized by a fixed-reference winding.

\section{Configuration spaces and general linear groups}\label{ap_hpgl}
Here, we give additional introduction to some mathematical concepts used in the main text.

The configuration space is the central concept for eigenvalue topology. It is constructed as lists of distinct elements. Taking a set, or space $X$, we construct a list of $n$ pairwise distinct elements in $X$: $(x_1,x_2,\dots,x_n )$ where all $x_i\in X$ and $x_i\ne x_j $ for any pair $i\ne j$. The configuration space $\textrm{Conf}_n(X)$ is made up of all such lists:
\begin{equation}
    \textrm{Conf}_n(X)=\ \{(x_1,x_2,\dots,x_n )\in X^n\vert x_i\ne x_j,\, \forall i\ne j\}.
\end{equation}
We look at a simple example. Take $X=\mathbb R, n=2$. Then the space $\mathbb R^2$ is the two-dimensional plane. The configuration space $\textrm{Conf}_2(\mathbb R)$ is made up of points $(x_1,x_2)\in \mathbb R^2$ such that $x_1\ne x_2$. This is to delete the diagonal $x_1=x_2$ in the two-dimensional plane. So we have $\textrm{Conf}_2(\mathbb R)=\mathbb R^2-\{(x,x)\vert x\in \mathbb R\}$. Larger-$n$ situations are constructed similarly by removing some ``diagonals'' in $\mathbb R^n$. The configuration space, as a subspace of $X^n$, inherits a natural topology from $X^n$ ($\mathbb R^2$ in the example), which is employed to find out its homotopy properties.
There is a natural action of $S_n$ (the permutation group) on the configuration space, by permuting the order among $x_i$.

Next we list some information for $GL_n(\mathbb R)$ and $GL_m(\mathbb C)$ that is useful for calculations in the main text. 

The polar decomposition $A=UP$ ($U\in U(m)$, $P\in Mat^+(m)$, the space of positive definite Hermitian matrices) gives rise to a homeomorphism between $GL_m(\mathbb C)$ and $U(m)\times Mat^+(m)$. Since $Mat^+(m)$ is contractible,  we have a homotopy equivalence:
\begin{equation}
    GL_m(\mathbb C) \simeq U(m).
\end{equation}
Similarly, the real general linear group satisfies
\begin{equation}
    GL_n(\mathbb R) \simeq O(n).
\end{equation}

\renewcommand{\arraystretch}{1.5}
\begin{table}[t]

\begin{tabular}{| c|c|c|c|}
 \hline
   $GL_n(\mathbb R)\simeq O(n)$ & $n=1$& $n=2$&$n>2$\\
 \hline
  $\pi_0$ &   \multicolumn{3}{c|}{ $\mathbb Z_2$} \\
\hline
  $\pi_1$ &  $\{1\}$ & $\mathbb Z$&$\mathbb Z_2$\\
\hline
 $\pi_2$ & \multicolumn{3}{c|}{  $\{1\}$}\\
 \hline
\end{tabular}

\begin{tabular}{| c|c|c|c|}
 \hline
   $GL_m(\mathbb C)\simeq U(m)$ & $m=1$& $m=2$&$m>2$\\
 \hline
  $\pi_0$ &   \multicolumn{3}{c|}{ $\{1\}$} \\
\hline
  $\pi_1$ & \multicolumn{3}{c|}{ $\mathbb Z$}  \\
\hline
 $\pi_2$ & \multicolumn{3}{c|}{  $\{1\}$}\\
 \hline
\end{tabular}
\caption{Homotopy groups of real general linear groups and complex general linear groups. The real general linear group has two connected components that are homeomorphic to each other, so the higher ($>0$) homotopy groups tell the topological information of both components.}\label{tb_htggl}
\end{table}

The homotopy groups are summarized in Table~\ref{tb_htggl}. We briefly explain some entries in the following. The orthogonal group
$O(n)$ has two connected components $O^+(n)$ and $O^-(n)$, distinguished by the sign of the determinant. In particular, $O(1)$ consists of only two points.
$U(m)$, instead, is connected.

The group $O(2)\cong U(1)$ is homeomorphic to a circle $S^1$. 
The group $O(3)$ has a double cover $SU(2)$, and $SU(2)$ is homeomorphic to $S^3$.
The fundamental group of $O(n)$ or $GL_n(\mathbb R)$ becomes stable for $n\geq 3$. The unitary group
$U(m)$ for any $m\geq 1$ is homeomorphic to  $SU(m)\times U(1)$ as a manifold (not as a Lie group).
Its fundamental group is $\mathbb Z$, coming from the $U(1)$ part, representing how many cycles the determinant can rotate around 0.

The group $GL_n(\mathbb R)$ can be embedded in $GL_{n+1}(\mathbb R)$ in a natural way by letting $GL_n(\mathbb R)$ act on the first $n$ coordinates of $\mathbb R^{n+1}$.
The induced maps $\pi_1(GL_n(\mathbb R))\to \pi_1(GL_{n+1}(\mathbb R))$ are as follows:
\begin{equation}
\begin{aligned}
n=2&:~~\text{mod 2 map from $\mathbb Z$ to $\mathbb Z_2$};\\
n\geq 3&:~~\text{isomorphism of $\mathbb Z_2$}.
\end{aligned}
\end{equation}

The group $GL_m(\mathbb C)$ can also be naturally embedded in $GL_{2m}(\mathbb R)$.
Writing $A\in GL_m(\mathbb C)$ as $A=B+iC$ where $B, C\in GL_{m}(\mathbb R)$, then its image in $GL_{2m}(\mathbb R)$ is simply:
\begin{equation}
    A'=\begin{pmatrix}
        B&C\\
        -C&B
    \end{pmatrix}.
\end{equation}
The image of this embedding can be identified with real, invertible matrices that commute with $\big(\begin{smallmatrix}
    &\mathbf1_m\\
    -\mathbf1_m&0
\end{smallmatrix}\big)$.
The induced maps $\pi_1(GL_m(\mathbb C))\to \pi_1(GL_{2m}(\mathbb R))$ are as follows:
\begin{equation}
\begin{aligned}
m=1&:~~\text{isomorphism of $\mathbb Z$};\\
m\geq 2&:~~\text{mod 2 map from $\mathbb Z$ to $\mathbb Z_2$}.
\end{aligned}
\end{equation}

\section{Product structure of eigenvalue and eigenvector topology}\label{ap_pdtetv}

In this section, we give a solid proof for the direct product structure of the eigenvector and the eigenvalue topology for gapped \PT-symmetric Hamiltonians.
Namely, we prove Eq.~\eqref{eq-pi1isproduct} (copied here):
\begin{equation}
    \pi_1(X^{(m,n)})=\mathbb B_m \times (\mathbb Z_2)^{n-1}~~ (m>0,n>0).
\end{equation}
Here, $X^{(m,n)}=(\mathcal E\times M^{(m,n)})/S_m$ and $M^{(m,n)}=GL_{n+2m}(\mathbb R)/[(\mathbb R^{\times})^n\times (\mathbb C^{\times})^m]$. 

First, from the fibration $M^{(m,n)}\to X^{(m,n)}\to \mathcal E/S_m$, we have an exact sequence:
\begin{align}
    \cdots\to \pi_2(X^{(m,n)})\to \pi_2(\mathcal E/S_m)\to\pi_1(M^{(m,n)}) \nonumber\\
    \to \pi_1(X^{(m,n)}) \to \pi_1(\mathcal E/S_m) \to\pi_0(M^{(m,n)})\cdots.
\end{align}
Since $\pi_2(X^{(m,n)})\to \pi_2(\mathcal E/S_m)$ is surjective (it is the same as $\pi_2(\mathcal E\times M)\to \pi_2(\mathcal E)$) and $\pi_0(M^{(m,n)})=0$ by connectedness, we have:
\begin{equation}\label{eq-prodproof1}
    0\to\pi_1(M^{(m,n)}) \to \pi_1(X^{(m,n)}) \to \pi_1(\mathcal E/S_m) \to 0.
\end{equation}
Second, from the covering space $M^{(m,n)}\to M^{(m,n)}/S_m$ induced from the $S_m$-action, we have:
\begin{equation}\label{eq-prodproof2}
    0\to \pi_1(M^{(m,n)})\to  \pi_1(M^{(m,n)}/S_m)\to S_m\to  0.
\end{equation}
The second map in Eq.~\eqref{eq-prodproof1} and that in Eq.~\eqref{eq-prodproof2} are related by the natural projection $X^{(m,n)}\to M^{(m,n)}/S_m$, meaning that we have the following commuting diagram:
\begin{equation}
\begin{tikzcd}[row sep=small, column sep=1.5ex]
0\arrow[r] & \pi_1(M^{(m,n)})\arrow[d, equal] \arrow[r, "f"] & \pi_1(X^{(m,n)})\arrow[d, "p"] \arrow[r] & \pi_1(\mathcal E/S_m)\arrow[r] & 0\\
0\arrow[r] & \pi_1(M^{(m,n)})\arrow[r,"pf"] & \pi_1(M^{(m,n)}/S_m)\arrow[r] & S_m\arrow[r] & 0
 \end{tikzcd}.\label{eq_cds}
\end{equation}

In our problem, we claim that Eq.~\eqref{eq-prodproof2} actually splits. 
In other words, 
\begin{equation}\label{eq-prodproof3}
    \pi_1(M^{(m,n)}/S_m)\cong \pi_1(M^{(m,n)})\times S_m. 
\end{equation}
This is simply saying that the permutation $S_m$ on the complex eigenvalues does not mess up with $\pi_1(M^{(m,n)})$, which only receives contribution from real eigenvectors.
To formally see that, we show that there is a retraction of $pf$ in Eq.~\eqref{eq-prodproof2}.
Indeed, there is a map $M^{(m,n)}/S_m\to GL_{n+2m}(\mathbb R)/((\mathbb R^{\times})^n\times GL^+_{2m}(\mathbb R))$, and an isomorphism $\pi_1(GL_{n+2m}(\mathbb R)/((\mathbb R^{\times})^n\times GL^+_{2m}(\mathbb R)))\cong \pi_1(M^{(m,n)})$, from which we can define a map $r: \pi_1(M^{(m,n)}/S_m)\to \pi_1(M^{(m,n)})$.
It is straightforward to check that $r\circ pf=id_{\pi_1(M^{(m,n)})}$, hence $r$ is a retraction.

The commuting diagram Eq.~\eqref{eq_cds} together with Eq.~\eqref{eq-prodproof3} implies that Eq.~\eqref{eq-prodproof1} also splits:
\begin{equation}
    \pi_1(X^{(m,n)})\cong \pi_1(\mathcal E/S_m)\times \pi_1(M^{(m,n)}).
\end{equation}
Indeed, we can just define a map $r': \pi_1(X^{(m,n)})\to\pi_1(M^{(m,n)})$ as $r'=r\circ p$. 
That $r'$ is a retraction follows from $r'f=rpf=id_{\pi_1(M^{(m,n)})}$.  This concludes the proof.

\section{Braiding models}
\label{appendix:braiding-models}
In this appendix, we provide a constructive method for \PT-symmetric operators with arbitrary eigenvalue braiding, before showcasing two simple models that were obtained via this method.

\subsection{Constructing \PT-symmetric models featuring arbitrary braids}
The following procedure is composed of two steps. 
We first use established methods such as the one outlined in Ref.~\onlinecite{PhysRevLett.126.010401}, to construct a generic operator that features the desired eigenvalue braid.
We then construct a \PT-symmetric operator from this, by explicitly using the required symmetry of eigenvalues and eigenvectors.
In the following, we outline this second step.

A \PT-symmetric, i.e. real, and non-degenerate operator \(H\) can be deconstructed uniquely into three commuting parts,
\begin{equation}
    H= H_- + H_0 + H_+,
\end{equation}
where \(H_+\) has eigenvalues in the complex upper half-plane, \(H_0\) has real eigenvalues, and \(H_-\) has eigenvalues in the complex lower half-plane. 
Any two distinct components project onto different spaces, i.e., their product is zero.
These components individually need not be real, in fact we will find that \(H_+\) cannot be real.
Since the eigenvalues, as well as eigenvectors, are permuted the same way under complex conjugation, \(H_-\) is in fact the complex conjugate of \(H_+\).

For models with only non-real eigenvalues, we may disregard the matrix \(H_0\), and immediately see a constructive method for \(H\):
Take an arbitrary \(m\)-band Hamiltonian \(h_m(k)\) that features the desired braid, and construct \(H_+\) from it via
\begin{equation}
    H_+(k) = T \left((h_m(k) + i  c  \mathbf{1}_m) \oplus \mathbf0^m \right) T^{-1},
\end{equation}
where \(\mathbf1_m, \mathbf0_m\) are identity and trivial linear maps of \(m\) dimensions, respectively, and \(T\) is an appropriately chosen basis transformation.
We ensure non-degenerate eigenvalue braid structure by shifting the eigenvalues of \(h_m\) sufficiently far into the upper half complex plane using \(c>0\) such that they remain non-real throughout.
We then extend the vector space to \(2m\) dimensions by adding \(0_m\).
Finally, we choose the transformation matrix \(T\) such that the image of the \(m\) eigenvectors of \(h_m\), together with their complex conjugate, form a basis of the \(2m\)-dimensional space.
One simple choice is
\begin{equation}
    T = \mqty(\mathbf1_m & \mathfrak{i} \mathbf1_m\\ \mathfrak{i} \mathbf1_m& \mathbf1_m).
\end{equation}

Thus one obtains an operator \(H_+\) such that \(H = H_++H_- = 2 \Re(H_+)\), its real part, is a \PT-symmetric operator with the desired braid structure in the upper half plane, and the corresponding (mirrored / inverse) braid in the lower half plane.

\begin{figure*}
    \def\svgwidth{\linewidth}
    %--
    
    \begingroup%
		  \makeatletter%
		  \providecommand\color[2][]{%
		    \errmessage{(Inkscape) Color is used for the text in Inkscape, but the package 'color.sty' is not loaded}%
		    \renewcommand\color[2][]{}%
		  }%
		  \providecommand\transparent[1]{%
		    \errmessage{(Inkscape) Transparency is used (non-zero) for the text in Inkscape, but the package 'transparent.sty' is not loaded}%
		    \renewcommand\transparent[1]{}%
		  }%
		  \providecommand\rotatebox[2]{#2}%
		  \newcommand*\fsize{\dimexpr\f@size pt\relax}%
		  \newcommand*\lineheight[1]{\fontsize{\fsize}{#1\fsize}\selectfont}%
		  \ifx\svgwidth\undefined%
		    \setlength{\unitlength}{510.00003052bp}%
		    \ifx\svgscale\undefined%
		      \relax%
		    \else%
		      \setlength{\unitlength}{\unitlength * \real{\svgscale}}%
		    \fi%
		  \else%
		    \setlength{\unitlength}{\svgwidth}%
		  \fi%
		  \global\let\svgwidth\undefined%
		  \global\let\svgscale\undefined%
		  \makeatother%
		  \begin{picture}(1,0.31372544)%
		    \lineheight{1}%
		    \setlength\tabcolsep{0pt}%
		    \put(0,0){\includegraphics[width=\unitlength,page=1]{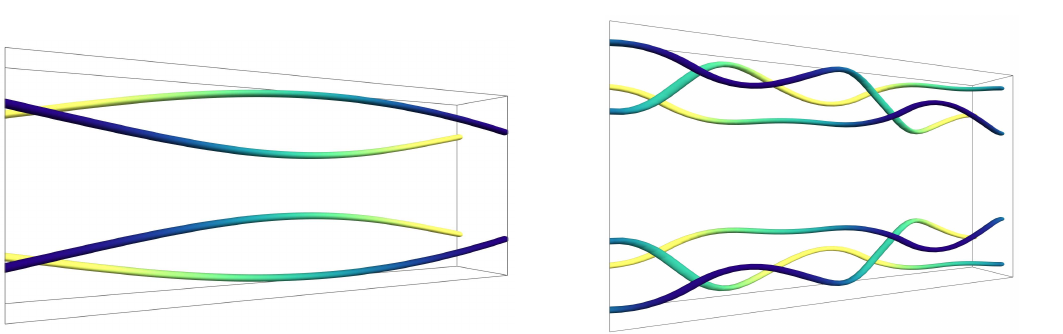}}%
		    \put(0.25892365,0.25275089){\color[rgb]{0,0,0}\makebox(0,0)[t]{\lineheight{1.25}\smash{\begin{tabular}[t]{c}\(k\)\end{tabular}}}}%
		    \put(0.46363383,0.23323188){\color[rgb]{0,0,0}\makebox(0,0)[t]{\lineheight{1.25}\smash{\begin{tabular}[t]{c}\(2\pi\)\end{tabular}}}}%
		    \put(0.02018788,0.27065768){\color[rgb]{0,0,0}\makebox(0,0)[t]{\lineheight{1.25}\smash{\begin{tabular}[t]{c}\(0\)\end{tabular}}}}%
		    \put(0.47967452,0.13431375){\color[rgb]{0,0,0}\makebox(0,0)[lt]{\lineheight{1.25}\smash{\begin{tabular}[t]{l}\(0\)\end{tabular}}}}%
		    \put(0.47967452,0.18284316){\color[rgb]{0,0,0}\makebox(0,0)[lt]{\lineheight{1.25}\smash{\begin{tabular}[t]{l}\(1\)\end{tabular}}}}%
		    \put(0.47967452,0.08578434){\color[rgb]{0,0,0}\makebox(0,0)[lt]{\lineheight{1.25}\smash{\begin{tabular}[t]{l}\(-1\)\end{tabular}}}}%
		    \put(0.47655175,0.03527655){\color[rgb]{0,0,0}\makebox(0,0)[rt]{\lineheight{1.25}\smash{\begin{tabular}[t]{r}\(-\frac12\)\end{tabular}}}}%
		    \put(0.42604009,0.047568){\color[rgb]{0,0,0}\makebox(0,0)[rt]{\lineheight{1.25}\smash{\begin{tabular}[t]{r}\(\frac12\)\end{tabular}}}}%
		    \put(0.47605269,0.13534029){\color[rgb]{0,0,0}\makebox(0,0)[rt]{\lineheight{1.25}\smash{\begin{tabular}[t]{r}Im\(E\)\end{tabular}}}}%
		    \put(0.48818387,0.04763855){\color[rgb]{0,0,0}\makebox(0,0)[lt]{\lineheight{1.25}\smash{\begin{tabular}[t]{l}Re\(E\)\end{tabular}}}}%
		    \put(0.78755209,0.27574164){\color[rgb]{0,0,0}\makebox(0,0)[t]{\lineheight{1.25}\smash{\begin{tabular}[t]{c}\(k\)\end{tabular}}}}%
		    \put(0.93165974,0.25470669){\color[rgb]{0,0,0}\makebox(0,0)[t]{\lineheight{1.25}\smash{\begin{tabular}[t]{c}\(2\pi\)\end{tabular}}}}%
		    \put(0.58288311,0.29702232){\color[rgb]{0,0,0}\makebox(0,0)[t]{\lineheight{1.25}\smash{\begin{tabular}[t]{c}\(0\)\end{tabular}}}}%
		    \put(0.95593949,0.14504368){\color[rgb]{0,0,0}\makebox(0,0)[lt]{\lineheight{1.25}\smash{\begin{tabular}[t]{l}\(0\)\end{tabular}}}}%
		    \put(0.95593949,0.18707956){\color[rgb]{0,0,0}\makebox(0,0)[lt]{\lineheight{1.25}\smash{\begin{tabular}[t]{l}\(2\)\end{tabular}}}}%
		    \put(0.95593949,0.1030078){\color[rgb]{0,0,0}\makebox(0,0)[lt]{\lineheight{1.25}\smash{\begin{tabular}[t]{l}\(-2\)\end{tabular}}}}%
		    \put(0.95164537,0.03879845){\color[rgb]{0,0,0}\makebox(0,0)[rt]{\lineheight{1.25}\smash{\begin{tabular}[t]{r}\(-1\)\end{tabular}}}}%
		    \put(0.91404913,0.04806527){\color[rgb]{0,0,0}\makebox(0,0)[rt]{\lineheight{1.25}\smash{\begin{tabular}[t]{r}\(1\)\end{tabular}}}}%
		    \put(0.95103338,0.27836856){\color[rgb]{0,0,0}\makebox(0,0)[t]{\lineheight{1.25}\smash{\begin{tabular}[t]{c}Im\(E\)\end{tabular}}}}%
		    \put(0.96447607,0.04660602){\color[rgb]{0,0,0}\makebox(0,0)[lt]{\lineheight{1.25}\smash{\begin{tabular}[t]{l}Re\(E\)\end{tabular}}}}%
		    \put(0.02405552,0.29902131){\color[rgb]{0,0,0}\makebox(0,0)[rt]{\lineheight{1.25}\smash{\begin{tabular}[t]{r}(a)\end{tabular}}}}%
		    \put(0.54811758,0.29901959){\color[rgb]{0,0,0}\makebox(0,0)[rt]{\lineheight{1.25}\smash{\begin{tabular}[t]{r}(b)\end{tabular}}}}%
		    \put(0,0){\includegraphics[width=\unitlength,page=2]{fig12.pdf}}%
		  \end{picture}%
		\endgroup%

     %--
    \caption{Braids of the two models given in Eq.~\eqref{eq_z-braid} and after Eq.~\eqref{eq_3-braid}.
    The strands are colored according to the imaginary part of the corresponding eigenvalue to distinguish over-crossing from under-crossing strands.
    One shows a four-band model with a braid corresponding to \(\sigma_1\in\B_2\), the other shows a non-trivial three-strand braid corresponding to \(\sigma_2^{-1}\sigma_1\in\B_3\).
    }
    \label{fig:pt-braids}
\end{figure*}

We now apply this method to construct PT-symmetric models with illustrative braids, which we show in Fig.~\ref{fig:pt-braids}.

\subsection{\texorpdfstring{\(\mathbb{Z}\text{-braid}, N=4, m=2\)}{Z-braid, N=4, m=2}}

The lowest number of bands for which eigenvalue braids occur in \PT-symmetric models is \(m=2,n=0\), see Table~\ref{tab:homotopy-groups}.

We apply our constructive method to the textbook example of a braiding two-band operator, 
\begin{equation}
    h_2(k) = \mqty(0&1\\\exp(i k) & 0),
    \label{eq:z-braid-model-nonsym}
\end{equation}
which features a single counterclockwise exchange of eigenvalues as \(k\to k+2\pi\).
As imaginary shift away from the real axis we choose \(c=2\), and obtain 
\begin{equation}
H(k)=\mqty( 
    0 & \frac{1}{2} & 1 & 0 \\
    \frac{\cos (k)}{2} & 0 & \frac{\sin (k)}{2} & 1 \\
    -1 & 0 & 0 & \frac{1}{2} \\
    -\frac{\sin (k)}{2} & -1 & \frac{\cos (k)}{2} & 0
).
\label{eq_z-braid}
\end{equation}

The braid group on two strands is Abelian. 
In the following we introduce a model with a more generic braid group.

\subsection{\texorpdfstring{\(\mathbb{B}_3\text{-braid}, N=6, m=3\)}{B3 braid, N=6, m=3}}
The lowest number of bands to obtain non-Abelian braiding is \(N=6\), for models of \(m=3\) pairs of complex conjugate eigenvalues, which feature a \(\mathbb{B}_3\) group structure.
As a non-trivial braid with interesting non-Abelian structure we choose the boundary braid of the model in Ref.~\cite{konig2022braid}.
We take 
\begin{equation}
    h_3=\mqty(
    1-e^{i k_x} & -i \left(1+e^{i k_x}\right) & 0 \\
 -i \left(1+e^{i k_x}\right) & e^{i k_x}-e^{i k_y} & -i \left(1+e^{i
   k_y}\right) \\
 0 & -i \left(1+e^{i k_y}\right) & -1+e^{i k_y}
    )
    \label{eq_3-braid}
\end{equation}
and obtain the braid in question by setting \(k_x= \pi\cos(k), k_y = \pi\sin(k)\).
This is the model used for the illustrative braids in Fig.~\ref{fig_nphs}.

\section{Second homotopy groups of band-gapped matrices}\label{ap_2hmtpbg}

In this Appendix, we give the second homotopy groups of the band-gapped matrices. As before, we need to discuss the results based on the number $m$ pairs of complex eigenvalues and number $n$ of real eigenvalues.

To obtain the higher homotopy groups of $X^{(m,n)}$, we can neglect the discrete group $S_m$ in Eq.~\eqref{eq_rdndsp}, since such an action only changes $\pi_0$ and $\pi_1$. Then we only look the following space
\begin{equation}
    \mathbb R^n\times \mathrm{Conf}_m(\mathbb C)\times \frac{GL_{n+2m}(\mathbb R)}{(\mathbb R^{\times})^n\times (\mathbb C^{\times})^m}.
\end{equation}
The first two spaces $\mathbb R^n$and $\mathrm{Conf}_m(\mathbb C)$ do not contribute to the second homotopy group. So we only need to study the eigenvector part
\begin{equation}
    M^{(m,n)}=\frac{GL_{n+2m}(\mathbb R)}{(\mathbb R^{\times})^n\times (\mathbb C^{\times})^m}.
\end{equation}
The second homotopy groups of $M^{(m,n)}$ is obtained from the following exact sequence:
\begin{equation}
    0\to\pi_2(M^{(m,n)})\to \pi_1((\mathbb C^{\times})^m)\to \pi_1(GL_{n+2m}(\mathbb R)).
\end{equation}
This tells that $\pi_2(M^{(m,n)})$ is isomorphic to the kernel of the map $\pi_1((\mathbb C^{\times})^m)\to \pi_1(GL_{n+2m}(\mathbb R))$.

If $m=0$, we get $\pi_1((\mathbb C^{\times})^m)=0$, so is the kernel.
If $m=1$ and $n=0$, the above map is an isomorphim of $\mathbb Z$.
If $m\geq 2$ or if $m=1$ and $n\geq 1$, then the map can be represented as $\mathbb Z^m\to\mathbb Z_2$, such that $(k_1,k_2,\dots,k_m)\in \mathbb Z^m$ is mapped to $\sum k_i\textrm{ mod }2$. 
The kernel is the  sequences of $m$ integers whose sum is even. It is a subgroup of $\mathbb Z^m$ that is isomorphic to $\mathbb Z^m$.  In fact, a basis of this group can be chosen as $(1,-1,0,\dots,0),(1,1,0,\dots,0),\allowbreak(0,1,1,0,\dots,0),\dots,(0,0,\dots,1,1)$. 

To summarize, 
\begin{equation}
    \pi_2(M^{(m,n)})\cong \begin{cases}
            0,~~(m=0; \text{~or~}m=1, n=0)\\
        \mathbb Z^m,~~(\text{otherwise})\\
    \end{cases}.
\end{equation}

\section{Spectral flattening of separation-gapped non-Hermitian matrices}\label{ap_sft}

In this appendix, we show that the spectral flattening in Sec.~\ref{sc_mspfl} is continuous for \PT-symmetric non-Hermitian separation-gapped matrices.

Recall that for any such separation-gapped matrix $A$, not necessarily Hermitian, we define $sf(A)$ as the spectral flattened version of $A$ by taking all eigenvalues in the upper(and lower)-half-plane to be flatten to $i$ (and $-i$, respectively):
\begin{equation}
    sf(A)=i\sum_{\lambda\in\text{spec}_+(A)}P_\lambda-i\sum_{\lambda\in\text{spec}_-(A)}P_\lambda.
\end{equation}

To show that $sf(\cdot)$ is continuous, it suffices to show that $P_\pm=\sum_{\lambda\in\text{spec}_+}P_\lambda$ is continuous.
First, notice that for any separation-gapped $A$, not necessarily diagonalizable, we have \cite{kato2013perturbation}:
\begin{equation}\label{eq:complexint}
    P_+(A)=\frac{1}{2\pi i}\int_\mathcal{C}(z-A)^{-1}dz,
\end{equation}
where $\mathcal{C}$ is an arbitrary complex contour on the upper-half-plane that surrounds $\text{spec}_+(A)$.
Second, due to the continuity of eigenvalues \cite{kato2013perturbation}, for any fixed separation-gapped $A$ and fixed contour $\mathcal{C}$, there exists a $\delta>0$ such that spec$_+(B)$ is also enclosed in $\mathcal{C}$ as long as $\norm{B-A}<\delta$ and hence Eq.~\eqref{eq:complexint} also holds for any such $B$.

In general, for any two invertible matrices $X$ and $Y$, we have:
\begin{equation}
    X^{-1}(Y-X)Y^{-1}=X^{-1}-Y^{-1}.
\end{equation}
Therefore
\begin{align}
    \norm{Y^{-1}}&\leq\norm{X^{-1}}+\norm{X^{-1}(Y-X)Y^{-1}}\nonumber\\ &\leq
    \norm{X^{-1}}+\norm{X^{-1}}\norm{Y-X}\norm{Y^{-1}},
\end{align}
which implies (assuming $\norm{X^{-1}}\norm{Y-X}<1$)
\begin{equation}
    \norm{Y^{-1}}
    \leq \frac{\norm{X^{-1}}}{1-\norm{X^{-1}}\norm{Y-X}}
\end{equation}
and
\begin{align}
    \norm{X^{-1}-Y^{-1}}&\leq\norm{X^{-1}}\norm{Y-X}\norm{Y^{-1}}\nonumber\\ &\leq
    \frac{\norm{X^{-1}}^2\norm{Y-X}}{1-\norm{X^{-1}}\norm{Y-X}}.
\end{align}

Applying this inequality to $X=z-A$ and $Y=z-B$ and assuming $\norm{(z-A)^{-1}}\delta<1$ (here $\delta$ is an upper bound for $\norm{B-A}$):
\begin{equation}\label{eq:estimateDiff}
    \norm{(z-A)^{-1}-(z-B)^{-1}}\leq\frac{\norm{(z-A)^{-1}}^2\delta}{1-\norm{(z-A))^{-1}}\delta}.
\end{equation}
The expression $\norm{(z-A)^{-1}}$ as a continuous function of $z$ on the closed contour $\mathcal{C}$ has an upper bound. So the right hand side of Eq.~\eqref{eq:estimateDiff} can be made arbitrarily small.
The proof is done by applying Eq.~\eqref{eq:complexint} to $A$ and $B$.

By now we have proved that the spectral flattening is applicable to separation between real bands and complex bands. With a more careful examination of the above proof, we remark that we can actually generalize the proof to a broader form of separation-gapped matrices. The most important steps in the proof are the linear flattening and bounded spectrum. So we can let the spectral separation condition to be: the spectrum of each partition set $J_\alpha$ is contained in an open bounded convex subset $\mathcal E_\alpha$ of $\mathbb C$.  All $\mathcal E_\alpha$ are disjoint, $\bar{\mathcal E_\alpha}\cap \bar{\mathcal E_j}=\varnothing$, where $\bar{\mathcal E_\alpha}$ is the closure of $\mathcal E_\alpha$. The convex subset allows us to linearly flatten the spectrum and the boundedness will make sure that all eigenvalues can be enclosed in a proper contour.

\section{Chern number and Euler number from Berry connections}\label{ap_cherneuler}
The Chern class and Euler class are topological invariants for complex vector bundles and orientable real vector bundles. 
They are fully determined by the topology of the vector bundles and is independent of the geometric information (connection, curvature, metric). 
However, in physics literature, they usually arise from Berry connections and Berry curvatures. 
In this appendix, we provide more information that is relevant for our paper.

\subsection{Chern number: equivalence between different conventions}\label{ap_cnbc}

The Chern-Weil theory \cite{tu2017differential} tells us that the Chern class can be calculated using formulas such as Eq.~\eqref{eq-Berryintegral} starting from any connection, as long as it is globally well-defined. 
In physics literature, there several possible choices such for the connection.
We list the definition in terms of the ``connection matrices" here:
\begin{equation}
\begin{aligned}
    \mathcal A^{LR}_{ij,a}(\mathbf k)&= i\langle u^L_i(\mathbf k)|\partial_au^R_j(\mathbf k)\rangle, 
    \\ \mathcal A^{RR}_{ij,a}(\mathbf k)&= i\langle u^R_i(\mathbf k)|\partial_au^R_j(\mathbf k)\rangle, \\
    \mathcal A^{RL}_{ij,a}(\mathbf k)&= i\langle u^R_i(\mathbf k)|\partial_au^L_j(\mathbf k)\rangle, \\
    \mathcal A^{LL}_{ij,a}(\mathbf k)&= i\langle u^L_i(\mathbf k)|\partial_au^L_j(\mathbf k)\rangle.\label{eq_bobc}
\end{aligned}
\end{equation}

In the main text, we discussed the ``LR" convention, corresponding to the following connection:
\begin{equation}\label{eq-appdefconnection}
   \nabla_a|u^R_j(\mathbf k)\rangle=
   P\partial_a|u^R_j(\mathbf k)\rangle,
\end{equation}
where $P=\sum_{ij} \ket{u^R_i}\bra{u^L_j}$ corresponds to the eigendecomposition of $H$ (a not-necessarily-orthogonal decomposition of a vector space $X=Y\oplus Z$ defines a projection $P$ onto $Y$ by $P|_Y=id_Y$ and $P|_Z=0$ ).

The RR convention, however, does \emph{not} give rise to a globally well-defined connection.
Consider a linear transformation $V(\mathbf k)$ on the eigenvectors, then the ``connection matrix" transforms as
\begin{equation}
    \mathcal A^{RR}_{a}(\mathbf k)\to V^{\dagger}\mathcal A^{RR}_{a}(\mathbf k)V+iV^{\dagger}(\mathbf k)\partial_a V(\mathbf k).\label{eq_trmRR}
\end{equation}
This is not the correct transformation rule that a connection should have \cite{tu2017differential}.
This means if we write the Berry curvature from different gauges of the eigenvectors, the result may not agree. The necessity of different gauge choices always happens in topological systems, where we cannot write the Bloch wave functions globally in the BZ \cite{DJThouless_1984,bott1982differential}. Instead, we have to write the wave function on several patches of the BZ, namely the trivialization open sets, and glue them carefully via gauge transformations. In non-Hermitian settings, the eigenmodes can be braided around the BZ, which enforces nontrivial $GL(\mathbf R)$ transformations.

The exception is when the transition $V(\mathbf k)$ across different patches of the BZ Eq.~\eqref{eq_trmRR} is unitary everywhere. In this situation, the transformation Eq.~\eqref{eq_trmRR} coincides with Eq.~\eqref{eq_bctr}. The RR connection matrix is then smoothly glued together and gives rise to a globally-defined connection. An example is a single Chern band \cite{PhysRevLett.120.146402}. There, the matrix $V(\mathbf k)$ is simply a phase after requiring  $\braket{u^R}{u^R}=1$.

Another way to fix this problem is to define (non-canonically, namely, depending on the choice of local frames) RR connection locally, and then use a partition of unity to ``glue them" into a global connection \cite{tu2017differential}, similar to the proof of existence of a connection for a vector bundle over a manifold. 
However, this is in general not practical, since the above define connection can be very complicated in the overlapping regions of different open sets.

The RL convention is a globally well-defined connection on a different vector bundle. 
Here the vector bundle is defined by the left eigenvectors instead of the right ones. 
Nevertheless, this vector bundle shares the same Chern number as the LR convention.
For a proof, notice that:
\begin{align}
     \mathcal A^{LR}_{ij,a}(\mathbf k)&= i\langle u^L_i(\mathbf k)|\partial_au^R_j(\mathbf k)\rangle=-i\langle \partial_au^L_i(\mathbf k)|u^R_j(\mathbf k)\rangle \nonumber\\
     &=-i\langle u^R_j(\mathbf k)|\partial_a u^L_i(\mathbf k)\rangle^\ast=\mathcal A^{RL\ast}_{ji,a}(\mathbf k).
\end{align}
and hence
\begin{align}
    \Omega^{LR}_{ij,xy}(\mathbf k)=\Omega^{RL\ast}_{ji,xy}(\mathbf k).
\end{align}
For the Chern number, only the real diagonal part of the Berry curvature contributes. The LR and the RL choices share the same real diagonal part according to the equation above. This completes the proof.

The relation between the LL convention and the LR convention is the same as that between RR and LR.

In summary, regarding the choices of left/right eigenvectors,
\begin{itemize}
    \item the LR and RL convention always give globally well-defined connection and give rise to the same Chern number;
    \item the RR and LL conventions do not give a global-defined connection automatically; however, in some situations they can be made global. This usually happens when there is no braid or degeneracy in the bands that we are studying. In this situation, they also give rise to the same Chern number as the LR and RL convention.
\end{itemize}

\subsection{Euler number: compatibility between the metric and the connection}\label{ap_eulcl}

Unlike the Chern class, to express the Euler class through Berry connection and Berry curvature, we should pick up a connection compatible with a Riemannian metric on the real vector bundle and orthogonalize the basis; arbitrary choices of connection do not lead to the Euler class in general \cite{tu2017differential,milnor1974characteristic}.

A Riemannian metric on a real vector bundle is a way to associate a positively definite bilinear form for each fiber, the Hilbert space at each $\mathbf k$, (not to be confused with the Riemannian metric on the base space). 
A connection is said to be compatible with a metric if 
\begin{equation}
    \partial_a(u,v)=(\nabla_a u,v)+( u,\nabla_a v)
\end{equation}
for any vector fields $u$ and $v$. 
For our question, a natural metric on the subbundle is the one induced from the standard inner product on $\mathbb R^N$: $(u',u)=\langle u'|u\rangle$.
However, if we define the connection on the subbundle using the LR convention as in the main text, or equivalently through Eq.~\eqref{eq-appdefconnection}, it is in general not compatible with the above metric.
This is because the projection $P$ is usually not orthogonal with respective to the standard inner product on $\mathbb R^N$.

In order to overcome this, let us choose $P$ to be orthogonal projection onto the subspaces.
For a concrete calculation,
we apply the following Gram-Schmidt process to the eigenvectors
\begin{equation}\label{eq_gsort}
\begin{aligned}
    |\tilde u_1(\mathbf k)\rangle&=\frac{|u_1^R(\mathbf k)\rangle}{\norm{\ket{u_1^R(\mathbf k)}}},\\
    |\tilde u_2(\mathbf k)\rangle&=\pm
    \frac{|u_2^R(\mathbf k)\rangle-|\tilde u_1(\mathbf k)\rangle\langle \tilde u_1(\mathbf k)\rangle| u_2^R(\mathbf k)\rangle}
    {\norm{|u_2^R(\mathbf k)\rangle-|\tilde u_1(\mathbf k)\rangle\langle \tilde u_1(\mathbf k)\rangle| u_2^R(\mathbf k)\rangle}}.
\end{aligned}
\end{equation}
Note that the vector $|\tilde u_2(\mathbf k)\rangle$ is in general not an eigenvector of the system. 
In the second equation, we choose the sign properly so that $(\ket{\tilde u_1(\mathbf k)},\ket{\tilde u_2(\mathbf k)})$ has the correct orientation, i.e., when we move to different patches in the BZ, the transition function $V$ has positive determinant.
Then we just choose $P=\sum_{i=1,2} \ket{\tilde u_i(\mathbf k)}\bra{\tilde u_i(\mathbf k)}$.
In other words, we define
\begin{equation}
    \overline\nabla_a|\tilde u_j(\mathbf k)\rangle=-i\sum_{j'=1,2}\tilde{\mathcal A}_{jj',a}(\mathbf k)|\tilde u_j(\mathbf k)\rangle,\quad j=1\textrm{ or }2,
\end{equation}
where
\begin{equation}
    \tilde{\mathcal A}_{jj',a}(\mathbf k)=i\langle \tilde u_j(\mathbf k)|\partial_a \tilde u_{j'}(\mathbf k)\rangle.
\end{equation}

Now we are able to proceed to writing the Euler class in terms of the curvature of $\tilde{\mathcal A}$. 
The curvature is skew symmetric and the Euler class is its off-diagonal component:
\begin{equation}
    \frac{1}{2\pi i}\widetilde\Omega_{12,xy}(\mathbf k),\quad \mathbf k\in U_\alpha
\end{equation}

The above equation is already a globally well-defined 2-form, although to perform the above calculation we need to work on a trivializable open set.
As always, we can perform the calculation for a family of trivializable open set that covers the whole base space (just to be careful with the orientation, see comments below Eq.~\eqref{eq_gsort}).
For examples, if we study the Euler class on a two-dimensional sphere, we can choose the two trivialization open sets to be the upper and the lower hemispheres. Then we fix the signs of $ |\tilde u_2(\mathbf k)\rangle$ such that the transition function of the bases at the equator has a positive determinant.

\bibliography{refPTH}
 
\end{document}